\documentclass[prd,12pt,longbibliography
,groupedaddress, nofootinbib,notitlepage
]{revtex4-1}

\usepackage{amsmath, amssymb, cool, txfonts}
\usepackage{natbib}
\usepackage{graphicx}
\usepackage{bbm}
\usepackage{epsfig}
\usepackage{epstopdf}
\usepackage{latexsym}
\usepackage{subfigure}
\usepackage{booktabs}
\usepackage{subfigure}
\pdfoutput=1
\makeatletter
\def\@fpheader{\relax}
\makeatother

\def\A{{\cal A}}

\def\J{{\cal J}}

\def\be{\begin{equation}}
\def\ee{\end{equation}}
\def\bea{\begin{eqnarray}}
\def\eea{\end{eqnarray}}
\def\ba{\begin{array}}
\def\ea{\end{array}}
\def\ben{\begin{enumerate}}
\def\een{\end{enumerate}}

\newcommand{\dsl}{\pa \kern-0.5em /} 
\def\a{\alpha}

\def\g{\gamma}

\def\l{\lambda}

\def\pa {\partial}

\begin{document}


\title{Correlation functions of the Bjorken flow in the holographic Schwinger-Keldysh approach}



\author{Toshali Mitra}
\email[]{toshali.mitra@apctp.org}
\affiliation{Asia Pacific Center for Theoretical Physics, Pohang 37673, Korea}
\affiliation{The Institute of Mathematical Sciences, Chennai 600113, India} 
\affiliation{Homi Bhabha National Institute, Training School Complex, AnushaktiNagar, Mumbai 400094,
India}

\author{Avik Banerjee}
\email[]{avikphys02@gmail.com}
\affiliation{Laboratoire de Physique de l'\'{E}cole Normale Supérieure, ENS, Universit\'{e} PSL, CNRS, Sorbonne Universit\'{e}, Universit\'{e} de Paris, F-75005 Paris, France}
\affiliation{Center for Strings, Gravitation and Cosmology, Department of Physics, Indian Institute of Technology Madras, Chennai 600036, India}

\author{Ayan Mukhopadhyay}
\email[]{ayan.mukhopadhyay@pucv.cl}
\affiliation{Instituto de F\'{\i}sica,
Pontificia Universidad Cat\'{o}lica de Valpara\'{\i}so,
Avenida Universidad 330, Valpara\'{\i}so, Chile}
\affiliation{Center for Strings, Gravitation and Cosmology, Department of Physics, Indian Institute of Technology Madras, Chennai 600036, India}

\date{\today}

\begin{abstract}
One of the outstanding problems in the holographic approach to many-body physics is the explicit computation of correlation functions in nonequilibrium states. We provide a new and simple proof that the horizon cap prescription of Crossley-Glorioso-Liu for implementing the thermal Schwinger-Keldysh contour in the bulk is consistent with the Kubo-Martin-Schwinger periodicity and the ingoing boundary condition for the retarded propagator at any arbitrary frequency and momentum. The generalization to the hydrodynamic Bjorken flow is achieved by a Weyl rescaling in which the dual black hole's event horizon attains a constant surface gravity and area at late time although the directions longitudinal and transverse to the flow expands and contract respectively. The dual state's temperature and entropy density thus become constants (instead of the perfect fluid expansion) although no time-translation symmetry emerges at late time. Undoing the Weyl rescaling, the correlation functions can be computed systematically in a large proper time expansion in inverse powers of the average of the two reparametrized proper time arguments. The horizon cap has to be pinned to the nonequilibrium event horizon so that regularity and consistency conditions are satisfied. Consequently, in the limit of perfect fluid expansion, the Schwinger-Keldysh correlation functions with space-time reparametrized arguments are simply thermal at an appropriate temperature. A generalized bilocal thermal structure holds to all orders. We argue that the Stokes data (which are functions rather than constants) for the hydrodynamic correlation functions can decode the quantum fluctuations behind the horizon cap pinned to the evolving event horizon, and thus the initial data.
\end{abstract}

\maketitle

\tableofcontents

\section{Introduction} \label{sec1}
\subsection{Motivation and aims}
One of the outstanding issues in the holographic duality, that maps strongly interacting quantum systems to semi-classical gravity in one higher dimension, is to understand the dictionary in real time. The applications of the holographic approach to many-body physics are especially limited without explicit and implementable prescriptions for computing out-of-equilibrium correlation functions. Even in the weak-coupling limit, these correlation functions are fundamental tools for studying decoherence and thermalization, e.g. to understand how the commutator and the anti-commutator evolve to satisfy the fluctuation-dissipation relation leading to the emergence of the Kubo-Martin-Schwinger periodicity at an appropriate temperature, and how the occupation numbers of quasi-particles equilibrate or evolve to new fixed points \cite{Berges:2004yj,Sieberer_2016,Heyl:2017blm,Mori:2017qhg}. These correlation functions are actually indispensable for understanding dynamics far from equilibrium in the strong coupling limit where the system cannot be described by quasi-particles. Although one can compute the one-point functions such as the energy-momentum tensor in real time using the correspondence between time-dependent geometries with regular horizons and states in the dual theory, with the remarkable fluid-gravity correspondence \cite{Policastro:2001yc,Bhattacharyya:2007vjd,Baier:2007ix,Rangamani:2009xk} providing a primary example, and numerical relativity \cite{Chesler:2013lia} providing a powerful tool, an explicit computation of a generating functional for hydrodynamic Schwinger-Keldysh correlation functions has not been achieved yet.\footnote{A limited number of observables can still be computed analytically or numerically. The equal time two-point functions can be computed via the geodesic approximation when the operator has a large scaling dimension, even out of equilibrium. The out-of-equilibrium retarded correlation function can also be computed by implementing linear causal response appropriately -- see \cite{Banerjee:2016ray} for a general prescription. Furthermore, equal-time Green's function can be computed for operators with large anomalous scaling dimensions in the geodesic approximation and has been used to understand thermalization \cite{Balasubramanian:2011ur}.} 

The object of interest is  the generating functional
\begin{align}
\exp(iW[J_1, J_2]) = {\rm Tr}\left(\hat{\rho}_0 ~T_c \exp\left(-i\oint {\rm d}t \int{\rm d}^{d-1}x \,\,J(t,\mathbf{x}) \hat{O}(t,\mathbf{x}) \right)\right) 
\end{align}
in the dual field theory, where $\hat{\rho}_0$ denotes the initial density matrix, $\oint$ the closed time Schwinger-Keldysh contour composed of the forward and backward arms --- $\oint = \int_{-\infty}^\infty  +\int^{-\infty}_\infty$, where the source $J(t,\mathbf{x})$ is specified such that it is $J_1(t,\mathbf{x})$ and $J_2(t,\mathbf{x})$ on the forward and backward arms of the contour, respectively, and $T_c$ denotes contour ordering. Formally, we can rewrite the density-matrix as
\begin{equation}
    \hat{\rho}_0 = \int [\mathcal{D}\phi_1][\mathcal{D}\phi_2]\,\, \rho_0(\phi_1,\phi_2)\,\vert \phi_1\rangle\langle\phi_2 \vert
\end{equation}
in terms of a basis $\vert \phi \rangle$ of field configurations, and construct the kernel 
\begin{equation}
\mathcal{K}(\phi_1,\phi_2; J_1, J_2) =  \langle\phi_2\vert T_c \exp\left(-i\oint {\rm d}t \int{\rm d}^{d-1}x \,\,J(t,\mathbf{x})\hat{O}(t,\mathbf{x})\right)\vert \phi_1\rangle.
\end{equation}
Then the functional $W[J_1,J_1]$ can be obtained from
\begin{equation}
    \exp(iW[J_1, J_2]) = \int [\mathcal{D}\phi_1][\mathcal{D}\phi_2]\,\, \rho_0(\phi_1,\phi_2) \mathcal{K}(\phi_1,\phi_2; J_1, J_2).
\end{equation}
When $\hat\rho_0$ is the thermal density matrix, this computation simplifies drastically. One needs to just add an appendage of length $T^{-1}$ at the end of the closed real-time contour parallel to the negative imaginary axis, and impose periodic boundary conditions on the full contour implementing Kubo-Martin-Schwinger (KMS) periodicity \cite{le_bellac_1996}. \\

\begin{figure}[h!]
    \centering
    \includegraphics[scale=0.32]{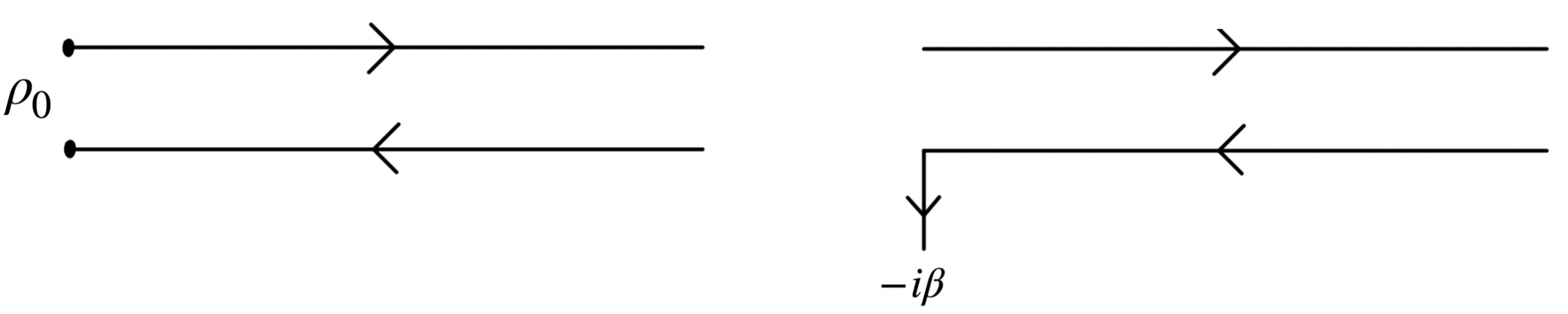}
    \caption{Left: The general Schwinger-Keldysh contour for an arbitrary initial density matrix $\hat\rho_0$. Field configurations need to be specified at the two ends of the closed time contour which are represented in bold. Right: In thermal equilibrium this simplifies. One needs an appendage to the contour extending along the negative imaginary axis by $\beta = T^{-1}$ and then impose periodic boundary conditions for the full contour.}
    \label{fig:SK-FT}
\end{figure}

One can also expect a similar simplification for the Bjorken flow which provides the simplest example of the evolution of an expanding system on the forward light cone (see Sec \ref{sec:BjRev} for details). The state is assumed to have boost invariance, and also translational and rotational invariance along the transverse plane so that the energy-momentum tensor can be expressed only in terms of the energy density $ T_{\tau\tau}=\epsilon(\tau)$ via Ward identities, where $\tau = \sqrt{t^2 - z^2}$ is the proper time of an observer co-moving with the flow and $z$ being the longitudinal coordinate along which the expansion happens. (Here we use units where $c=1$.) At late time, $\epsilon(\tau)$ is described by hydrodynamics and thus it reaches a perfect fluid expansion, so that
\begin{equation}\label{Eq:Bjorken-ep0}
    \epsilon (\tau) \approx \epsilon_0 \left(\frac{\tau_0}{\tau}\right)^{\frac{d}{d-1}}.
\end{equation}
In the hydrodynamic regime, it can be described by a single constant parameter, namely
\begin{equation}
    \mu := \epsilon_0 \tau_0^{\frac{d}{d-1}}.
\end{equation}
The full hydrodynamic series for $\epsilon(\tau)$ in powers of $\tau^{-\frac{d-2}{d-1}}$ (essentially a derivative expansion) is given in terms of $\mu$ (which is determined by initial conditions) and the transport coefficients which are determined by the fundamental microscopic theory. In this case, it is natural to ask whether there can be a simpler computation of the Schwinger-Keldysh partition function of $W[J_1,J_2]$ in the hydrodynamic limit, since just like the thermal case, the state can be essentially captured by a single parameter. 

More generally, we would expect that general methods for computing $W[J_1,J_2]$ would exist in the hydrodynamic regime where the energy-momentum tensor and conserved currents are described only by the hydrodynamic variables, namely the four-velocity $u^\mu(t,\mathbf{x})$, the energy density $\epsilon(t,\mathbf{x})$ (or equivalently the temperature $T(t,\mathbf{x})$), etc., and we would not require the knowledge of the detailed (off-diagonal) matrix elements $\rho(\phi_1,\phi_2)$ of the state or the kernel $\mathcal{K}(\phi_1,\phi_2; J_1, J_2)$ explicitly. In fact, $W[J_1,J_2]$ is related to the generalization of the thermodynamic free energy to hydrodynamics via Legendre transform, and the latter especially in the context of macroscopic space-time configurations of conserved currents is also known as the large deviation functional \cite{TOUCHETTE20091} which can be computed in many models studied in classical nonequilibrium statistical mechanics.\footnote{The large deviation functional gives the probability for a macroscopic space-time profile of a conserved charge or current density which does not necessarily satisfy the hydrodynamic equations. In a quantum system, the off-diagonal matrix elements in the basis of macroscopic field configurations for the conserved charges and currents eventually decohere, but the decoherence would be of interest. See \cite{Bernard_2021} for a recent discussion on the possibility of a quantum generalization of large deviation functional methods.} 

The primary aim of this work is to show how the explicit computation of $W[J_1,J_2]$ can be achieved by holographic methods in the hydrodynamic limit of the Bjorken flow. We will also present concrete steps for understanding how to go beyond the hydrodynamic limit and \textit{recover} the initial state. 

\subsection{A brief historical review}
The first major advance in understanding thermal real-time correlation functions in holography was the Son-Starinets prescription for computing the retarded correlation function, according to which the ingoing boundary condition at the horizon implements the causal linear response in the classical gravity (large $N$ and infinite strong coupling) approximation \cite{Son:2002sd}. Using the Chesler and Yaffe method for causal time evolution in the bulk \cite{Chesler:2013lia}, this approach was suitably generalized to compute the out-of-equilibrium retarded correlation function in holography \cite{Banerjee:2016ray}. The first concrete implementation of the Schwinger-Keldysh contour in holography is due to Son and Herzog utilizing the eternal black hole geometry \cite{Herzog:2002pc}. The two boundaries of the eternal black hole were shown to provide the forward and backward arms of the Schwinger-Keldysh closed time contour with the backward arm displaced by $-i\beta/2$ (note $\beta = T^{-1}$) along the imaginary axis. 

The most concrete prescription for real-time gauge-gravity duality for general initial states is due to Skenderis and van Rees \cite{Skenderis:2008dh,Skenderis:2008dg}. This however requires detailed understanding of the state in terms of semiclassical field configurations of dual gravity, and is best defined for states which can be constructed using Euclidean path integrals. In this prescription, one explicitly constructs the bulk geometry corresponding to the boundary Schwinger-Keldysh contour with specified sources, and extends data on the field theory contour into the bulk in an appropriate manner. It is, however, not easy to apply this approach to realistic computations for generic initial states. Furthermore, as mentioned above, we would expect a simpler approach in the hydrodynamic regime. We will compare this approach with ours in Sec \ref{sec:IC}.

Our method is based on  generalizing the recently proposed horizon cap prescription due to Crossley, Glorioso and Liu (CGL)  \cite{Glorioso:2018mmw} for the static black brane dual to the thermal state. Here the Schwinger-Keldysh contour is realized by a \textit{horizon cap}, in which the ingoing Eddington-Finkelstein radial coordinate goes around the horizon in the complex plane in a little circle of radius $\epsilon$ (not to be confused with the energy density) before going back to the real axis and reaching the second boundary. Thus the two arms of the Schwinger-Keldysh contour at the two boundaries are connected continuously in the bulk through the bulk radial contour. The horizon cap implements the appropriate analytic continuation of bulk fields from one arm of the bulk geometry to the other with the sources $J_1$ and $J_2$ specified independently at the two boundaries.

\subsection{Summary of the results}

Our first result is the demonstration that the CGL horizon cap prescription reproduces both the KMS periodicity and the ingoing boundary condition prescription of Son and Starinets (needed to obtain the retarded correlation function from causal response) in thermal equilibrium. Earlier in \cite{Glorioso:2018mmw}, both of these were demonstrated only up to quadratic orders in the frequency. Our proof relies on a simple and novel matrix factorization of thermal correlation functions which is reproduced by the holographic method for extracting the correlation function of a scalar operator in the field theory via the on-shell action of the dual bulk scalar field whose mass is determined by the scaling dimension of the operator.\footnote{The most important ingredient for realization of the matrix factorization is that only the terms which are obtained from the product of an ingoing mode with an outgoing mode can contribute to the quadratic on-shell action of the free bulk scalar field minimally coupled to gravity.}


Our primary tool for extending the CGL prescription into the (conformal) hydrodynamic Bjorken regime is Weyl rescaling that maps the asymptotic perfect fluid expansion to a flow in which the temperature and entropy density become \textit{constant} at late time in a non-trivial background metric \textit{without any time-like Killing vector}.  This Weyl transformation can be lifted to a bulk diffeomorphism \cite{Henningson:1998gx,deHaro:2000vlm} in which the dual black hole's event and apparent horizon coincide at a fixed radial location at a very late proper time. The area and surface gravity of the horizons, and therefore the entropy density and the temperature of the dual fluid, remain constant at late time, but the horizons shrink in the directions transverse to the flow and expand in the longitudinal direction. In the absence of any time-like Killing vector, the late-time behavior of the Weyl transformed Bjorken flow is \textit{not} thermal although the temperature and energy density become constant.

The requirement that the dual geometry has a regular future black hole horizon necessitates viscous corrections to the perfect fluid Bjorken flow, which implies the same for the Weyl transformed version mentioned above since the bulk dual of the latter is obtained simply via bulk diffeomorphism. Furthermore, bulk regularity determines the precise values of \textit{all} transport coefficients of the holographic theory order by order in the derivative expansion (equivalent to the large proper time expansion) with the Bjorken flow giving a special case of the fluid-gravity correspondence \cite{Bhattacharyya:2007vjd}. The large proper time expansion of the holographic Bjorken flow has been worked out to very high orders \cite{Heller:2013fn}. We discuss the bulk dual of the Bjorken flow and its Weyl transformed version in Section \ref{sec2} in detail.

We want to emphasize that the Weyl rescaling (and hence the dual bulk diffeomorphism) is determined purely by the late perfect fluid flow regime that is mapped to that of a constant temperature in a non-trivial background metric in the field theory (which is also the boundary metric of the dual black brane geometry after the bulk diffeomorphism) as mentioned above. The Weyl transformation  and consequently the Weyl rescaled background metric of the field theory do \textit{not} receive any correction at first and higher orders in the derivative (i.e. large proper time) expansion. 

As a result of the bulk diffeomorphism which implements the Weyl transformation of the dual Bjorken flow, the Klein-Gordon equation for a bulk scalar field with arbitrary mass can be mapped to that of a static black brane at late proper time.  In this map, the frequency needs to be appropriately scaled, and the momenta in the static black brane geometry are identified with \textit{co-moving} longitudinal and transverse momenta.\footnote{Note that the map to the static black brane holds only at the leading order in the large proper time expansion. The co-moving momenta however are also defined at the leading order itself via the Weyl transformed background metric of the field theory (boundary metric of the bulk geometry).} For vanishing momenta, this reproduces the result of Janik and Peschanski \cite{Janik:2006gp}.

The utility of this map to the static black brane is to establish the horizon cap prescription in the asymptotic perfect fluid limit. It follows that the Schwinger-Keldysh correlation functions of the operator dual to the bulk scalar field can be mapped to thermal correlation functions after suitable space-time reparametrizations in the asymptotic perfect fluid limit of the Bjorken flow (i.e. when both the proper time arguments of the two-point correlation functions are sufficiently large). The non-thermal nature of the Schwinger-Keldysh correlation functions can thus be \textit{absorbed} into these space-time reparametrizations up to overall (proper time-dependent) Weyl factors. 

We also show that the horizon cap prescription extends to all orders in the late proper time expansion. Firstly, at first and higher orders, the Klein-Gordon equation of the bulk scalar takes the same form as at the leading order but with source terms, and can be systematically solved at each order in the large proper time expansion such that the leading behavior of the ingoing and outgoing modes are exactly the same as at the zeroth order (that can be mapped to the static black brane) implying that the standard analytic continuation at the horizon cap can be performed at each order. Remarkably, this is possible \textit{only} if the horizon cap is pinned to the time-dependent event horizon (and not the apparent or other dynamical horizons which do not coincide with the event horizon beyond the zeroth order).

Second, the requirement of the near-horizon behavior of the ingoing and outgoing modes at the time-dependent event horizon does not completely determine all the order-by-order corrections. The undetermined coefficients are those that give the leading behavior of the ingoing modes at the horizon at first and higher orders in the proper time expansion. However, when the on-shell action of the scalar field is used to extract the Schwinger-Keldysh correlation functions, we find that they are consistent with all field theory identities provided one of these identities which hold for arbitrary nonequilibrium states provided one of these identities is used to determine the unfixed coefficients that give the leading near-horizon behavior of the ingoing modes. 

Therefore, we establish the horizon cap prescription can be unambiguously extended to determine the Schwinger-Keldysh correlation functions of the entire hydrodynamic tail of the Bjorken flow at any arbitrary order in the large proper time expansion. Aside from satisfying all field theory identities, we find that the retarded propagator given by the ingoing modes reproduces the normalizable bulk solutions at complex frequencies (which map to quasi-normal modes of the static black brane \cite{Janik:2006gp}) with vanishing sources to all orders although the relevant phase factors at first and higher orders are determined as functions of the frequency via the near-horizon behavior of the outgoing modes at real frequencies. The latter implies non-trivially that one can obtain the normalizable bulk solutions dual to the transients (nonequilibrium generalization of the collective excitations of the system) to all orders in the large proper time expansion from the retarded correlation function generalizing how quasinormal modes are obtained from the poles of the retarded propagator in thermal equilibrium. To all orders, the Schwinger-Keldysh correlation functions satisfy a matrix factorized form implying a bilocal thermal structure.

Let us briefly provide some explicit expressions. The $d$-dimensional Bjorken flow in the field theory is conveniently expressed in the Milne coordinates, $\tau =\sqrt{t^2-z^2}$ (the proper time with $z$ being the longitudinal Minkowski coordinate), the rapidity $\zeta = {\rm arctanh}(z/t)$ and the transverse coordinates $\vec{x}_\perp$.

It is convenient to define
\begin{equation}\label{Eq:sigma-zeta-def}
   \sigma = \tau^{\frac{d-2}{d-1}} \tau_0^{\frac{1}{d-1}}, \quad \hat\zeta = \zeta\tau_0.
\end{equation}
with $\tau_0$ being an arbitrarily chosen fixed proper time. Furthermore, for the arguments $(\sigma_1, \zeta_1, \vec{x}_{\perp 1})$ and $(\sigma_2, \zeta_2, \vec{x}_{\perp 2})$ of the two-point correlation function let us define
\begin{align}
    \overline{\sigma} = \frac{1}{2}(\sigma_1 +\sigma_2), \quad \sigma_r = \sigma_1 -\sigma_2,\quad  \widetilde{\zeta}_r = (\hat{\zeta}_1- \hat{\zeta}_2)\frac{\overline\sigma}{\tau_0}, \quad \widetilde{x_\perp}_r = \vert \vec{x}_{\perp 1} -\vec{x}_{\perp 2}\vert \left(\frac{\tau_0}{\overline{\sigma}}\right)^{\frac{1}{d-2}}.
\end{align}

As discussed above, at late proper time, the Schwinger-Keldysh correlation functions $G(\sigma_1, \sigma_2, \hat\zeta_1 - \hat\zeta_2, \vert\vec{x}_{\perp1}-\vec{x}_{\perp2}\vert)$ of the Bjorken flow can be mapped to thermal correlation functions $G_\beta$ at a specific temperature $T=\beta^{-1}$ as follows
\begin{equation}\label{Eq:Gbeta-pf}
 G(\sigma_1, \sigma_2, \hat\zeta_1 - \hat\zeta_2, \vert\vec{x}_{\perp1}-\vec{x}_{\perp2}\vert) \rightarrow   \left(\frac{\tau_0}{\overline{\sigma}}\right)^{\frac{2\Delta_O}{d-2}}{G}_{\beta} \left(\frac{d-1}{d-2}\sigma_r, \sqrt{\widetilde{\zeta}_r^2 +\widetilde{x_\perp}_r ^2}\right).
\end{equation}
This limit implies that both $\sigma_1$ and $\sigma_2$ are large. Above $\Delta_O$ is the scaling dimension of the operator (which is determined by the mass of the dual bulk scalar field). Also both $G$ and $G_\beta$ are $2\times 2$ matrices with indices determining whether the arguments $(\sigma_1, \zeta_1, \vec{x}_{\perp 1})$ and $(\sigma_2, \zeta_2, \vec{x}_{\perp 2})$ are in the forward or backward legs of the Schwinger-Keldysh time-contour. Finally, the temperature $T$ of the thermal is determined by the (only free) parameter $\epsilon_0$ of the Bjorken flow (see Eq. \eqref{Eq:Bjorken-ep0}) as follows
\begin{equation}\label{Eq:T0-etc}
    \beta = T^{-1} = \frac{4\pi \varepsilon_0^{1/d}}{d}, \quad \varepsilon_0 = \frac{16\pi G_N}{d-1} \epsilon_0,
\end{equation}
where the $d+1$-dimensional gravitational constant $G_N$ is given by the rank of the gauge group of the dual theory (as for instance, in $\mathcal{N}=4$ super Yang-Mills theory with $SU(N)$ gauge group in $4$ dimensions, $G_N^{-1} = 2N^2/\pi$).\footnote{Apparently, the thermal form on the right hand side of Eq. \ref{Eq:T0-etc} depends on the choice of $\tau_0$. However, a shift in $\tau_0$ also changes the parameter $\epsilon_0$ appropriately so that the thermal form in Eq. \ref{Eq:T0-etc} is unambiguous. This is explained in Section \ref{Sec:PFSK} explicitly.} Remarkably, the map to the thermal form involving space-time reparametrization implies the emergence of $SO(d-1)$ rotational symmetry, and also time translation symmetry although these are absent in the original coordinates $\tau$ and $\zeta$. 

As mentioned above, we can systematically include viscous and higher-order corrections to the correlation functions and obtain
\begin{equation}\label{Eq:corr-hydro-series}
    G(\sigma_1, \sigma_2, \hat\zeta_1 - \hat\zeta_2, \vert\vec{x}_{\perp1}-\vec{x}_{\perp2}\vert) =  \left(\frac{\tau_0}{\overline{\sigma}}\right)^{\frac{2\Delta_O}{d-2}} \sum_{n=0}^\infty \frac{1}{\overline\sigma^n \varepsilon_0^{n/d}}G_n\left(T_0\sigma_r, T_0\widetilde{\zeta}_r, T_0\widetilde{x_\perp}_r\right)
\end{equation}
where $G_0$ coincides with the thermal correlation function $G_\beta$ given by \eqref{Eq:Gbeta-pf}. 

We also note that our result that the horizon cap for the holographic hydrodynamic Bjorken flow should be pinned to the nonequilibrium event horizon captures the causal nature of the Schwinger-Dyson equations for the real-time (out-of-equilibrium) correlation functions in the dual field theory.\footnote{The latter is manifest when written in terms of the coupled evolution of the statistical and spectral functions \cite{Berges:2004yj}.}  

The series in Eq. \eqref{Eq:corr-hydro-series} is not expected to be convergent, and therefore requires a trans-series completion with appropriate Stokes data (distinct from the Stokes data for the expectation value of the energy density which defines the bulk geometry) which should be actually functions of $\sigma_r$, $\widetilde\zeta_r$ and $\widetilde{x_\perp}_r$. We discuss their physical role in deciphering the information of the initial state which is lost in hydrodynamization, and also how they can be used to decode the interior of the event horizon.
\subsection{Plan}
The paper is organized as follows. In Section \ref{sec:CGL}, we introduce the Crossley-Glorioiso-Liu (CGL) horizon cap prescription for the thermal Schwinger-Keldysh correlation functions in holography. We prove that the prescription indeed reproduces the KMS periodicity so that they are given just in terms of the retarded correlation function, and that the latter is exactly what we obtain from the Son-Starinets prescription. As mentioned, we use a new matrix factorization of thermal correlation functions. In Section \ref{sec2}, we review the Bjorken flow and its holographic dual. We also introduce the Weyl rescaling of the Bjorken flow along with the dual bulk diffeomorphism such that the final state has a fixed temperature and entropy density. As mentioned, although the event horizon has a constant surface gravity and area at late time, it stretches and expands in the directions longitudinal and transverse to the flow respectively. Additionally, we discuss the proper residual gauge transformation corresponding to radial reparametrization.

In Section \ref{sec:ScalarField}, we study the probe bulk scalar field in the gravitational background dual to the hydrodynamic Bjorken flow and show how we can preserve the analytic structure of the horizon cap to all orders in the proper time expansion. Crucially, we find that it requires the horizon cap to be pinned to the nonequilibrium event horizon. In Section \ref{sec:CFs}, we use these results to extract the real-time correlation functions of the hydrodynamic Bjorken flow. After presenting the result for the perfect fluid limit in terms of a thermal propagator with space-time reparametrizations, we show how we systematically obtain the corrections in a proper time expansion. We also discuss many non-trivial consistency checks of our results. In Section \ref{sec:IC}, we present a discussion on how a trans-series completion of this expansion can lead to seeing the quantum fluctuations behind the nonequilibrium event horizon, and matching with initial data lost during hydrodynamization.

Finally, we conclude in Section \ref{sec:Con} with an outlook.

\section{The CGL horizon cap of the thermal black brane}\label{sec:CGL}

{The Crossley-Glorioso-Liu (CGL) horizon cap prescription is a simple proposal for the holographic realization of the Schwinger-Keldysh contour at thermal equilibrium \cite{Glorioso:2018mmw}. The thermal nature of the correlation functions obtained from this prescription, including their consistency with the Kubo-Martin-Schwinger (KMS) periodicity, has been explicitly verified up to quadratic order in the small frequency expansion in \cite{Glorioso:2018mmw}. In \cite{Glorioso:2018mmw}, it has also been verified that the retarded correlation function is implied by the ingoing boundary condition, as demanded by the Son-Starinets prescription \cite{Son:2002sd} up to the quadratic order in frequency. These were sufficient to obtain a rudimentary effective theory of diffusion and dissipative hydrodynamics from holography \cite{Ghosh:2020lel, Crossley:2015evo,Glorioso:2017fpd,Glorioso:2017lcn,Jensen:2018hse,Liu:2018kfw,Glorioso:2018mmw,deBoer:2018qqm,Chakrabarty:2019aeu,He:2021jna,Jana:2020vyx,He:2022jnc}. Here, we present an elegant proof that the CGL horizon cap indeed gives thermal correlation functions satisfying KMS periodicity, and that it also implies the Son-Starinets prescription for the retarded correlation function, at \textit{any arbitrary frequency and momenta}.\footnote{The real-time {prescription \cite{Skenderis:2008dg,Skenderis:2008dh}} of van Rees and Skenderis leads to the ingoing boundary condition {as shown in \cite{vanRees:2009rw}} in thermal equilibrium. Our methods discussed here discuss a natural generalization away from equilibrium for out-of-equilibrium, especially hydrodynamic states.} {For other approaches, see \cite{Jana:2020vyx}.}}

The generating functional for the thermal Schwinger-Keldysh correlation functions in a quantum field theory is\footnote{Unless specified, we will always put the backward arm of the Schwinger-Keldysh time contour infinitesimally below the real axis. Also, we will often omit explicit mention of the appendage of the contour along the imaginary axis which creates the thermal state in the infinite past.}
\begin{equation}\label{Eq:SKCorrs-Defn}
    e^{W[J_1, J_2]} = {\rm Tr}\left[ \hat\rho_\beta T_c e^{i\int {\rm d}t\int{\rm d}^{d-1}x\,(\hat{O}_1(\mathbf{x},t)J_1(\mathbf{x},t)-\hat{O}_2(\mathbf{x},t)J_2(\mathbf{x},t)) }\right]
\end{equation}
with $\rho_\beta$ denoting the thermal density matrix, $1$ and $2$ denoting the forward and backward arms of the contour, and $T_c$ denoting (time) contour ordering. The contour ordering implies that (with $\langle \cdot \rangle \equiv {\rm Tr}(\hat\rho_\beta \cdot)$)
\begin{align}\label{Eq:FT-Thermal}
& G_{11}(t -t', \mathbf{x}-\mathbf{x}') = \frac{\delta^2 W}{\delta J_1(t,\mathbf{x}) \delta J_1(t',\mathbf{x}') } = -i\langle T(\hat{O}(t,\mathbf{x}) \hat{O}(t',\mathbf{x}'))\rangle, \\\nonumber
& -G_{12}(t -t', \mathbf{x}-\mathbf{x}') = \frac{\delta^2 W}{\delta J_1(t,\mathbf{x}) \delta J_2(t',\mathbf{x}') } = i\langle \hat{O}(t',\mathbf{x}') \hat{O}(t,\mathbf{x}))\rangle, \\\nonumber
& -G_{21}(t -t', \mathbf{x}-\mathbf{x}') = \frac{\delta^2 W}{\delta J_2(t,\mathbf{x}) \delta J_1(t',\mathbf{x}') } = i\langle \hat{O}(t,\mathbf{x}) \hat{O}(t',\mathbf{x}'))\rangle,\\\nonumber
& G_{22}(t -t', \mathbf{x}-\mathbf{x}') = \frac{\delta^2 W}{\delta J_2(t,\mathbf{x}) \delta J_2(t',\mathbf{x}') } = -i\langle \overline{T}(\hat{O}(t,\mathbf{x}) \hat{O}(t',\mathbf{x}'))\rangle.
\end{align}
Succinctly we can write the above as
\begin{align}\label{Eq:FT-Thermal-2}
G_{ij}(x_1, x_2)= \frac{\delta^2 W}{\delta J_i(x_1) \delta J_j(x_2) }(-)^{i+j}
\end{align}
with $(i,j)= (1,2)$. The above holds even out of equilibrium with $\rho_\beta$ in \eqref{Eq:SKCorrs-Defn} replaced by an arbitrary initial state $\rho_0$.

It can readily be shown that the KMS periodicity (arising from the ${\rm Tr}$ in \eqref{Eq:SKCorrs-Defn} after extending the contour along the negative imaginary axis by $\beta$ as shown in Fig. \ref{fig:SK-FT}) implies that the thermal correlation functions in Fourier space (with $i$ and $j$ standing for the $1$ (forward) or $2$ (backward) arms of the contour) defined as
\begin{equation}
    G_{ij}(\omega,\mathbf{k}) = \int {\rm d}t~ {\rm d^{d-1}}x\, e^{i\omega(t-t')}e^{-i\mathbf{k}\cdot(\mathbf{x}-\mathbf{x}')}G_{ij}(t -t', \mathbf{x}-\mathbf{x}')
\end{equation}
assume the form
\begin{eqnarray}\label{Eq:G-Thermal}
 G_{ij} &=&\begin{pmatrix}
G_R(\omega, \mathbf{k})(1 + n(\omega)) - G_A(\omega, \mathbf{k})n(\omega)\, &\, - (G_R(\omega, \mathbf{k})- G_A(\omega, \mathbf{k})n(\omega)\\
- (G_R(\omega, \mathbf{k})- G_A(\omega, \mathbf{k})(1+n(\omega))\, &\,   G_R(\omega, \mathbf{k})n(\omega)-G_A(\omega, \mathbf{k})(1 + n(\omega))
\end{pmatrix}\nonumber\\
\end{eqnarray}
where
\begin{equation}
    G_R(t -t', \mathbf{x}-\mathbf{x}') = -i \theta(t-t')\langle [\hat{O}(t,\mathbf{x}) ,\hat{O}(t',\mathbf{x}')] \rangle 
\end{equation}
is the retarded propagator,  
\begin{equation}
    G_A(t -t', \mathbf{x}-\mathbf{x}') = -i \theta(t'-t)\langle [\hat{O}(t',\mathbf{x}') ,\hat{O}(t,\mathbf{x})] \rangle
\end{equation}
is the advanced propagator, and $n(\omega) = 1/(e^{\beta\omega}-1)$ is the Bose-Einstein distribution function. It is easy to see from these definitions that
\begin{equation}
    G_A^*(\omega,\mathbf{k}) = G_R(\omega,\mathbf{k})~.
\end{equation}

The crucial element of the proof of why the CGL prescription works is a simple and general factorization property of thermal correlation functions in field theory (irrespective of whether the theory is holographic or not). The Schwinger-Keldysh thermal correlation functions \eqref{Eq:G-Thermal} obtained by differentiating the real-time partition function at a temperature $T = \beta^{-1}$ can be factorized as shown below
\begin{eqnarray}\label{Eq:ThMF}
G_{ij} (\omega, \mathbf{k})
=\sigma_3\cdot  \begin{pmatrix}
A(\omega, \mathbf{k}) \, &\, B(\omega, \mathbf{k})\\
A(\omega, \mathbf{k}) \, &\, B(\omega, \mathbf{k})e^{\beta\omega}
\end{pmatrix}
\cdot
 \begin{pmatrix}
a(\omega, \mathbf{k}) \, &\, b(\omega, \mathbf{k})\\
a(\omega, \mathbf{k}) \, &\, b(\omega, \mathbf{k})e^{\beta\omega}
\end{pmatrix}^{-1},
\end{eqnarray}
in which $\sigma_3 = {\rm diag}(1,-1)$ is the third Pauli matrix, and
\begin{equation}
    G_R(\omega, \mathbf{k})  = \frac{A(\omega, \mathbf{k})}{a(\omega, \mathbf{k})}=  \frac{B^*(\omega, \mathbf{k})}{b^*(\omega, \mathbf{k})}=G_A^*(\omega, \mathbf{k}).
\end{equation}
Clearly $A\rightarrow \lambda A$, $B\rightarrow \tilde{\lambda} B$, $a\rightarrow \lambda a$ and ${b}\rightarrow \tilde{\lambda} b$ gives the same thermal matrix, so the factorization is unique up to the multiplicative complex functions $\lambda(\omega,\mathbf{k})$ and $\tilde\lambda(\omega,\mathbf{k})$.

\begin{figure}[htp]
  \centering
 \includegraphics[width=15.0cm, height=4.6cm,keepaspectratio]{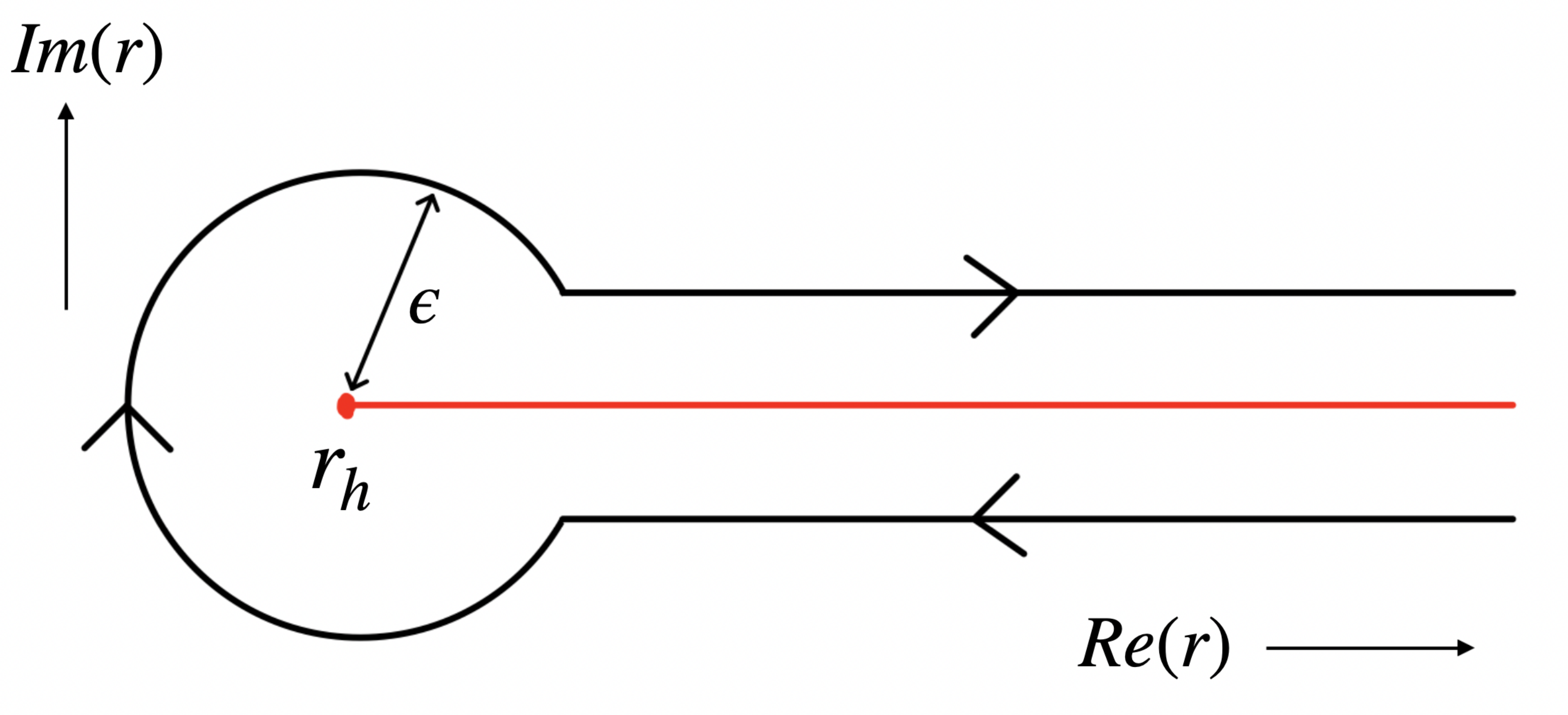} \label{Fig:ThermalHC1}
 \includegraphics[width=15.0cm,height=4.6cm,keepaspectratio]{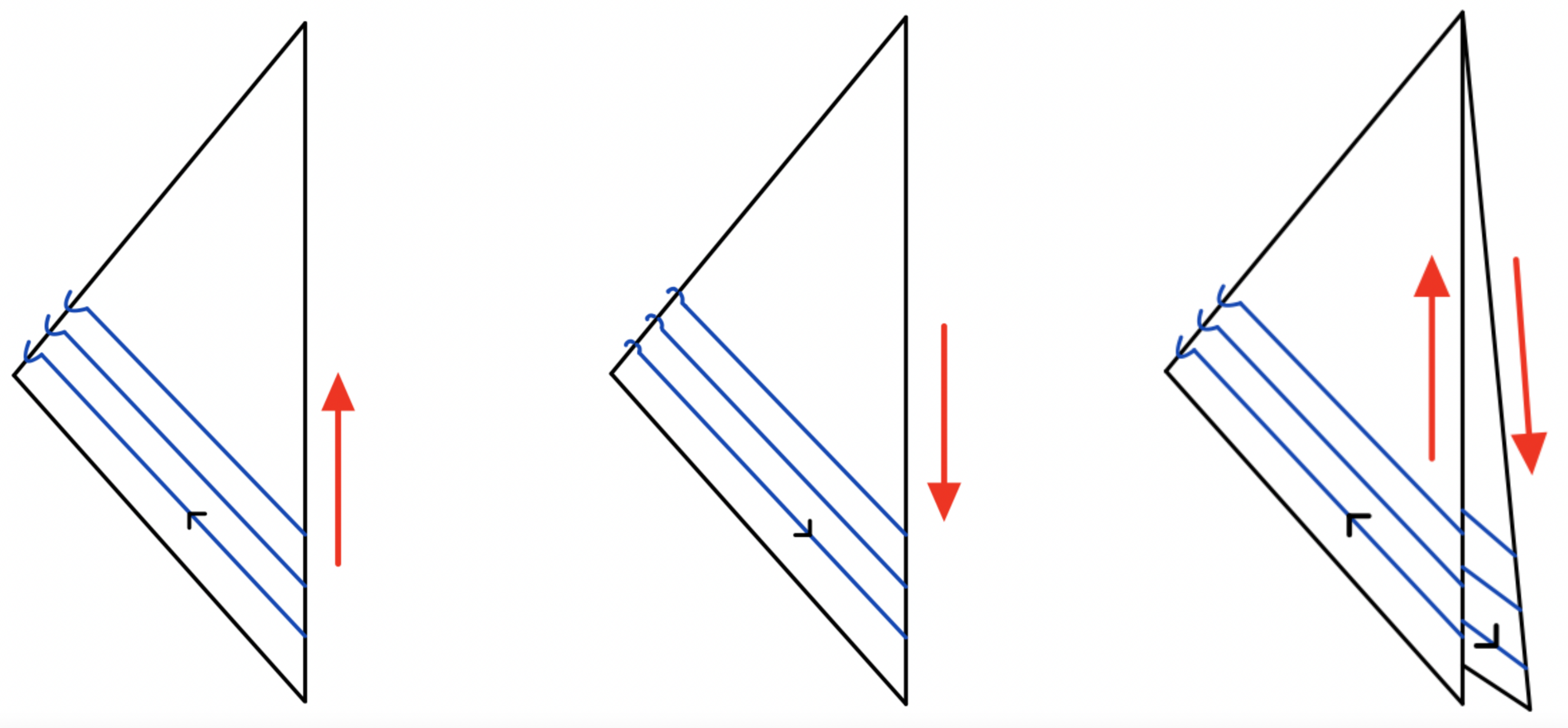} \label{Fig:ThermalHC2}
\caption{\textbf{Top:} The radial contour in the complexified two-sheeted black brane geometry on a constant time hypersurface. The radial coordinate goes around the horizon $r = r_h$ in the complex plane forming the \textit{horizon cap}, and connects the two arms of the  Schwinger-Keldysh contour at the two boundaries at a constant Lorentzian time. The analytic continuation along the {horizon cap} gives a well-defined Dirichlet value problem for the bulk fields in terms of their sources specified at the two boundaries. \textbf{Bottom:} Shows the time evolution of the radial contour (shown in blue lines) in the full complexified space-time ending at the two Lorentzian arms at the two boundaries. The reversal of the direction of time in the second copy implies that the full complexified space-time has a single orientation after the two copies are smoothly glued with the horizon cap. }\label{Fig:ThermalHC}
\end{figure}

The CGL horizon cap glues two copies of the black brane geometry, whose boundaries represent the forward and backward arms of the Schwinger-Keldysh time contour respectively, at the horizon as shown in Fig. \ref{Fig:ThermalHC}. For reasons to become clear later, this prescription is easily implemented in the ingoing Eddington-Finkelstein (EF) coordinates. The ingoing EF radial coordinates of the two geometries, representing the forward ($1$) and backward ($2$) arms of the time contour respectively, are displaced along the imaginary axis by $\mp \epsilon$ (i.e. $r_1 \rightarrow r_1 - i\epsilon$ and $r_2 \rightarrow r_2 + i\epsilon$). The smooth gluing is achieved by the encircling of the complexified radial coordinate around the horizon $r= r_h$ \textit{clockwise} along a circle of radius $\mathcal{O}(\epsilon)$ as it is analytically continued from the (first) copy dual to the forward contour to the (second) copy dual to the backward contour. The direction of time in the second copy has to be reversed so that full complexified bulk geometry has a single orientation. Therefore, the analytic continuation of the radial coordinate automatically necessitates the closed Schwinger-Keldysh time contour. 

Explicitly, the $AdS_{d+1}$ static black brane geometry in the ingoing Eddington-Finkelstein coordinates is 
\begin{eqnarray}\label{Eq:SBB}
{\rm d}s^2 = - \frac{2}{r^2}{\rm d}r {\rm d}t -\frac{1}{r^2}\left(1 - r^d\varepsilon_0 \right){\rm d}t^2 + \frac{1}{r^2} ({\rm d}x_1^2 +\cdots +{\rm d}x_{d-1}^2).
\end{eqnarray}
where $r$ is the bulk radial coordinate, $t$ is the Eddington-Finkelstein time and the horizon is at $r = r_h = \varepsilon_0^{-1/d}$. The on-shell action for bulk fields in this geometry is identified with the generating functional of connected real-time correlation functions of the dual operators at the boundary. A bulk scalar field configuration can be written in the form
\begin{eqnarray}
\Phi(r,t, \mathbf{x}) = \int \frac{{\rm d}\omega}{2\pi} \frac{{\rm d}^{d-1}k}{(2\pi)^{d-1}} \, e^{-i\omega t}e^{i\mathbf{k}\cdot\mathbf{x}}\Phi(r,\omega, \mathbf{k}).
\end{eqnarray}
On-shell, $\Phi(r,\omega, \mathbf{k})$ is a sum of two linearly independent solutions $\phi_{\rm in}(r,\omega, \mathbf{k})$ and $\phi_{\rm out}(r,\omega, \mathbf{k})$ which are ingoing and outgoing at the horizon respectively. Therefore, 
\begin{eqnarray}
\Phi(r,\omega, \mathbf{k}) = \phi_{\rm in}(r,\omega, \mathbf{k}) p(\omega,\mathbf{k})+\phi_{\rm out}(r,\omega, \mathbf{k}) q(\omega,\mathbf{k})
\end{eqnarray}
generally with $p(\omega,\mathbf{k})$ and $q(\omega,\mathbf{k})$ representing the arbitrary Fourier coefficients of the solutions which are ingoing and outgoing at the horizon respectively. The latter thus provide a basis of solutions for given $\omega$ and $\mathbf{k}$, and can be uniquely defined via the following conditions
\begin{equation}\label{Eq:in-out-static}
    \phi_{\rm in}(r_h,\omega, \mathbf{k}) = 1, \qquad\lim_{r\rightarrow r_h} \phi_{\rm out}(r,\omega, \mathbf{k})(r_h-r)^{-\frac{i\beta\omega}{2\pi}} = 1,
\end{equation}
where
\begin{equation}\label{Eq:THrh}
    \beta = T^{-1} = \frac{4\pi r_h}{d}
\end{equation}
is the inverse Hawking temperature of the black brane, and $r_h = \varepsilon_0^{-1/d}$ is the radial location of the horizon. Indeed, near the horizon ($r\approx r_h$), 
\begin{equation}
    \Phi(r,\omega, \mathbf{k}) \approx p(\omega,\mathbf{k})+(r_h-r)^{\frac{i\beta\omega}{2\pi}} q(\omega,\mathbf{k})
\end{equation}
as should follow from the universal validity of the geometrical optics approximation at the horizon. The CGL horizon cap prescription for the analytic continuation of the radial coordinate from one copy of the bulk space-time to another then implies that the Fourier coefficients of the on-shell solutions in the two copies are related by
\begin{equation}
   p_2(\omega,\mathbf{k}) =p_1(\omega,\mathbf{k}), \qquad q_2(\omega,\mathbf{k}) = e^{\beta\omega}q_1(\omega,\mathbf{k}),
\end{equation}
with $1$ and $2$ denoting the copies ending on the forward and backward arms of the time contour respectively at their boundaries. The on-shell solution in the full geometry can therefore be written in the following matrix form:
\begin{eqnarray}
\begin{pmatrix}
\Phi_1(r,t, \mathbf{x}) \\ \Phi_2(r,t, \mathbf{x})
\end{pmatrix}
= \int \frac{{\rm d}\omega}{2\pi} \frac{{\rm d}^{d-1}k}{(2\pi)^{d-1}} \, e^{-i\omega t}e^{i\mathbf{k}\cdot\mathbf{x}} \mathcal{M}(r,\omega, \mathbf{k})\cdot \begin{pmatrix}
p(\omega, \mathbf{k}) \\ q(\omega, \mathbf{k})
\end{pmatrix}
\end{eqnarray}
with the matrix
\begin{equation}
    \mathcal{M}(r,\omega, \mathbf{k}) = \begin{pmatrix}
     \phi_{\rm in}(r,\omega, \mathbf{k})~~~~~\phi_{\rm out}(r,\omega, \mathbf{k})\\\phi_{\rm in}(r,\omega, \mathbf{k})~~~~~e^{\beta\omegaup}\phi_{\rm out}(r,\omega, \mathbf{k})
    \end{pmatrix}
\end{equation}
providing a basis of solutions for the entire complexified space-time comprising of the two copies smoothly glued at the horizon. The sources $J_1(\omega, \mathbf{k})$ and $J_2(\omega, \mathbf{k})$ specified at the two boundaries (see below) implement the Dirichlet boundary conditions that determine $p(\omega, \mathbf{k})$ and $q(\omega, \mathbf{k})$ uniquely for real frequencies and momenta, and thus yielding a unique bulk field configuration in the full complexified space-time.

According to the holographic dictionary, the generating functional for the connected correlation functions is identified with the on-shell action for the scalar field dual to the operator $\hat{O}$, on the full complexified space-time, i.e.
\begin{equation}\label{Eq:W-S}
    W[J_1, J_2]= iS_{\rm on-shell}.
\end{equation}
Assuming minimal coupling to gravity, the on-shell quadratic action for the bulk scalar field $\Phi$ dual to a scalar operator takes the form 
\begin{eqnarray}\label{Eq:Ssplit1}
S_{on-shell} = S_{in-in} + S_{in-out} + S_{out-out},
\end{eqnarray}
The first piece $S_{in-in}$ is quadratic in the ingoing mode. Since the ingoing mode is analytic at the horizon, the contributions from the two arms cancel each other out (as the solutions are the same on the two arms) while the circle around the horizon does not contribute as well. Therefore, $S_{in-in} =0$. Note if we keep the ingoing mode alone, then $J_1 = J_2$. In the field theory, $W[J_1 =J_2] =0$ because the partition function $e^W$ with the same unitary evolution forward and backward in time, equals unity. Therefore, $S_{in-in} =0$ is consistent with field theory.

The second piece $S_{in-out}$, which is the sum of cross-terms between the in and outgoing modes, has a branch point at the horizon. Integrating over the two arms amounts to integrating around a branch cut, and results in the two boundary terms on-shell, i.e.
\begin{eqnarray}\label{Eq:Ssplit2}
S_{in-out} = S_{bdy1} + S_{bdy2}.
\end{eqnarray}

The third piece $S_{out-out}$ has a possibility of a pole at the horizon, i.e. $(r_h-r)^{-1}$ terms in the Lagrangian density arising from the radial derivative acting on the non-analytic piece $(r_h-r)^{\frac{i\beta\omega}{2\pi}}$, which we denote collectively as $S_\epsilon$. Essentially $S_\epsilon$ gets contributions from the following two terms:
\begin{eqnarray}\label{Eq:Sepsilon}
    S_\epsilon \propto \int {\rm d}\omegaup \int {\rm d}^dk  \oint_{\epsilon}\, {\rm d}r \sqrt{-G}\Bigg(G^{rr}\partial_r \phi_{out}^*\partial_r \phi_{out} +  G^{rt}\left(\partial_r \phi_{out}^*\partial_t \phi_{out}+ \partial_r \phi_{out}\partial_t \phi_{out}^*\right)\Bigg).
\end{eqnarray}
Remarkably, the poles originating from these two terms cancel each other out (note $\beta$ is given by \eqref{Eq:THrh}) resulting in $S_\epsilon =0$. The remaining terms quadratic in the outgoing mode are analytic, so that $S_{out-out}$ is also the sum of two boundary terms. These two boundary terms cancel each other out as in the ingoing case. On the gravitational side, the easy way to see this is by first writing the contributions from the forward and backward arms of the contour separately. The boundary contributions from one arm in the integrand would be proportional to $$\phi_b(-\omega, k)(\partial_r \phi)_b(\omega, k) + \cdots$$with $\phi_b$ and $(\partial_r \phi)_b$ denoting the boundary values of $\phi$ and its radial derivative respectively, and  $\cdots$ include counter-terms too. The contributions from the forward and backward parts of the contour come with opposite signs. If we consider the terms quadratic in the outgoing mode, then $\phi_b(\omega, k)$ picks up a factor of $e^{\beta\omega}$ while $\phi_b(-\omega, k)$ picks up a factor of $e^{-\beta\omega}$ via analytic continuation through the horizon cap, and the product of these factors is unity. Therefore, the contributions from the forward and backward contours cancel out leading to $S_{out-out}=0$.

It is useful to see this also from the field theory perspective. If we keep the outgoing modes only, then $J_2(\omega,\mathbf{k}) = J_1(\omega,\mathbf{k})e^{\beta\omega}$. In any field theory  \cite{Liu:2018kfw} $$W[J_1(t,\mathbf{x}), J_2(t,\mathbf{x})] = W_T[J_2(t-i\beta,\mathbf{x}), J_1(t,\mathbf{x})], \,\,\, {\rm i.e.}\,\,\,W[J_1(t,\mathbf{x}), J_2(t + i\beta,\mathbf{x})] = W_T[J_2(t,\mathbf{x}), J_1(t,\mathbf{x})],$$
where $W_T$ stands for the time-reversed process in which we specify with the same density matrix in the future instead of the initial time.\footnote{Succinctly, $W[J_1, J_2] = {\rm Tr}(\rho_0 \tilde{U}[J_2]U[J_1])$ where $U[J_1]$ is forward evolution with source $J_1$ and $\tilde{U}[J_2]$ is backward evolution with source $J_2$. Similarly, $W_T[J_1, J_2] = {\rm Tr}(\tilde{U}[J_2]\rho_0 U[J_1]) ={\rm Tr}(\rho_0 U[J_1]\tilde{U}[J_2]).$ } For $J_1(t,\mathbf{x})= J_2(t,\mathbf{x})= J(\omega, \mathbf{k})$, this amounts to $$
    W_T[J(\omega, \mathbf{k}), J(\omega, \mathbf{k}))] = W[J(\omega, \mathbf{k}), J(\omega, \mathbf{k})e^{\beta\omega}].$$The LHS of the above equation vanishes because once again the forward and backward evolution with the same source are inverses of each other (there is no operator insertion in the past now although the state is specified in the future). Therefore, the RHS of the above equation should vanish too, implying that
\begin{equation}
    W[J_2(\omega, \mathbf{k}) =J_1(\omega, \mathbf{k}) e^{\beta\omega}] = 0.
\end{equation}
Thus, $S_{out-out} =0$ is consistent with field theory. See also footnote 12 for a more straightforward verification that the thermal correlators in the dual theory originate from $S_{in-out}$ alone.

\textit{The upshot is that we obtain only two boundary contributions from the cross-term between the ingoing and outgoing modes, so that}
\begin{eqnarray}\label{Eq:Ssplit}
S_{on-shell} \equiv S_{in-out}= S_{bdy1} + S_{bdy2}
\end{eqnarray}
where $S_{bdy1}$ and $S_{bdy2}$ are the contributions from the two boundaries after taking into account counter-terms necessary for holographic renormalization \cite{Skenderis:2002wp}. This implies that
\begin{equation}\label{Eq:OnShell}
    S_{\rm on-shell}[J_1,J_2] = \int {\rm d}t\int{\rm d}^{d-1}x\,(\langle O_1(t, \mathbf{x})\rangle J_1(t, \mathbf{x})-\langle O_2(t, \mathbf{x})\rangle J_2(t, \mathbf{x}))
\end{equation}
\textit{with} 
\begin{eqnarray}\label{Eq:J-Oeq}
    \begin{pmatrix}
J_1(\omega, \mathbf{k}) \\ J_2(\omega, \mathbf{k})\end{pmatrix} &=& \mathcal{S}(\omega, \mathbf{k})\cdot \begin{pmatrix}
p(\omega, \mathbf{k}) \\ q(\omega, \mathbf{k})
\end{pmatrix}, \nonumber\\
\begin{pmatrix}
\langle O_1(\omega, \mathbf{k}) \rangle \\ \langle O_2(\omega, \mathbf{k}) \rangle\end{pmatrix} &=& \mathcal{R}(\omega, \mathbf{k})\cdot \begin{pmatrix}
p(\omega, \mathbf{k}) \\ q(\omega, \mathbf{k}) 
\end{pmatrix}\nonumber\\&=&
(\mathcal{R}\cdot \mathcal{S}^{-1})(\omega, \mathbf{k})\cdot\begin{pmatrix}
J_1(\omega, \mathbf{k}) \\ J_2(\omega, \mathbf{k})
\end{pmatrix},
\end{eqnarray}
The matrices $\mathcal{R}$ and $\mathcal{S}$ are defined as follows. Let the asymptotic ($r\approx 0$) expansions of the ingoing and outgoing modes be\footnote{The scaling dimension $\Delta$ is related to the mass via $\Delta = \frac{d}{2}+ \sqrt{\frac{d^2}{4}+m^2L^2}$ with $L$ being the AdS radius.}
\begin{eqnarray}
\phi_{\rm in}(r,\omega,\mathbf{k}) &=& r^{d-\Delta}(a_0(\omega,\mathbf{k}) + \cdots ) + r^\Delta (A_0(\omega,\mathbf{k})+\cdots), \nonumber\\
\phi_{\rm out}(r,\omegaup,\mathbf{k}) &=& r^{d-\Delta}(b_0(\omega,\mathbf{k}) + \cdots ) + r^\Delta (B_0(\omega,\mathbf{k})+\cdots).
\end{eqnarray}
Then
\begin{equation}\label{Eq:Seq}
    \mathcal{S}(\omega, \mathbf{k}) = \begin{pmatrix}
    a_0(\omega,\mathbf{k}) ~~~~ b_0(\omega,\mathbf{k})\\a_0(\omega,\mathbf{k}) ~~~~ e^{\beta\omega}b_0(\omega,\mathbf{k}) 
    \end{pmatrix}= \lim_{r\rightarrow0}r^{\Delta -d}\mathcal{M}(r,\omega, \mathbf{k})
\end{equation}
and\footnote{Note that, asymptotically, $\mathcal{M}(r,\omega, \mathbf{k}) =r^{d-\Delta}(\mathcal{S}(\omega, \mathbf{k}) +\cdots) + r^\Delta (\mathcal{R}(\omega, \mathbf{k}) +\cdots)$.}
\begin{equation}\label{Eq:Req}
    \mathcal{R}(\omega, \mathbf{k}) =  (2\Delta -d)\begin{pmatrix}
   A_0(\omega,\mathbf{k}) ~~~~ B_0(\omega,\mathbf{k})\\A_0(\omega,\mathbf{k}) ~~~~ e^{\beta\omega}B_0(\omega,\mathbf{k})
    \end{pmatrix} + \cdots.
\end{equation}
The $\cdots$ above stands for (state-independent) contact terms which we ignore. Denoting
\begin{equation}\label{Eq:hatG}
    \widehat{G}(\omega,\mathbf{k})= (\sigma_3\cdot\mathcal{R}\cdot\mathcal{S}^{-1})(\omega,\mathbf{k})
\end{equation}
(with $\sigma_3 = {\rm diag}(1, -1)$) we find from \eqref{Eq:OnShell}, \eqref{Eq:J-Oeq}, \eqref{Eq:Seq}, \eqref{Eq:Req} and \eqref{Eq:hatG} that\footnote{$J_i(\omega,\mathbf{k})= \int {\rm d}t\, {\rm d}^{d-1}x\, e^{i\omega t}e^{-i\mathbf{k}\cdot\mathbf{x}} J_i(t,\mathbf{x})$.}
\begin{eqnarray}\label{Eq:Onshell-Action-JO-redux}
S_{\rm on-shell}[J_1, J_2] & =& \int \frac{{\rm d}\omega}{2\pi} \frac{{\rm d}^{d-1}k}{(2\pi)^{d-1}} (J_1(-\omega,-\mathbf{k})\langle O_1(\omega,\mathbf{k})\rangle-J_2(-\omega,-\mathbf{k})\langle O_2(\omega,\mathbf{k})\rangle)\nonumber\\ 
&=& \int \frac{{\rm d}\omega}{2\pi} \frac{{\rm d}^{d-1}k}{(2\pi)^{d-1}}  (J_1(-\omega,-\mathbf{k})\,\, J_2(-\omega,-\mathbf{k}))\cdot\widehat{G}(\omega,\mathbf{k})\cdot \begin{pmatrix}
J_1(\omega,\mathbf{k})\\ J_2(\omega,\mathbf{k})
\end{pmatrix}.
\end{eqnarray}
Therefore, the identification \eqref{Eq:W-S} together with \eqref{Eq:FT-Thermal} implies that,\footnote{The reader can check that substituting $p(\omega,\mathbf{k}) + q(\omega,\mathbf{k}) = J_1(\omega,\mathbf{k})$ and $p(\omega,\mathbf{k}) +q(\omega,\mathbf{k})e^{\beta\omegaup}= J_2(\omega,\mathbf{k})$ in the on-shell action \eqref{Eq:Onshell-Action-JO-redux}, and using the thermal form of the propagators below, that indeed only cross-terms between $p$ and $q$, i.e. the in and outgoing modes appear. There are no contributions from terms quadratic in $p$ or in $q$, implying that $S_{in-in} = S_{out-out} =0$ as claimed above, and also $S_{on-shell} = S_{in-out}$.}
\begin{eqnarray}\label{Eq:Gravity-Thermal}
G_{ij}(\omega,\mathbf{k}) = \frac{\partial^2 S_{\rm on-shell}}{\partial J_i(-\omega,-\mathbf{k}) \partial J_j(\omega,\mathbf{k}) } (2\pi)^d=\widehat{G}(\omega,\mathbf{k}),
\end{eqnarray}
From the matrix factorization of thermal correlation functions given by \eqref{Eq:ThMF}, we readily find from \eqref{Eq:Seq}, \eqref{Eq:Req} and \eqref{Eq:hatG} that the correlation functions obtained by differentiating the on-shell gravitational action are thermal, i.e. assume the form \eqref{Eq:G-Thermal} provided\footnote{It is obvious that to map to the factorization in \eqref{Eq:ThMF}, we have to set $A = (2\Delta -d)A_0$, $B= (2\Delta -d)B_0$, $a = a_0$ and $b=b_0$.}
\begin{equation}
    G_R(\omegaup, \mathbf{k}) = (2\Delta -d)\frac{A_0(\omegaup,\mathbf{k})}{a_0(\omegaup,\mathbf{k})}, \qquad G_A(\omegaup, \mathbf{k}) =(2\Delta -d) \frac{B_0(\omegaup,\mathbf{k})}{b_0(\omegaup,\mathbf{k})}.
\end{equation}
Remarkably, the above are exactly the Son-Starinets prescriptions \cite{Son:2002sd} for the retarded and advanced propagators according to which they are obtained from the ingoing and outgoing boundary conditions at the horizon respectively. Furthermore, since the outgoing mode is time reverse of the ingoing mode (which is not manifest in the Eddington-Finkelstein gauge but can be evident from transforming to Schwarzchild-like coordinates),\footnote{Note that the notion of in/outgoing modes are gauge-invariant up to overall multiplicative factors, but these cancel out in the ratio of the normalizable to the non-normalizable modes. This is why the Son-Starinets prescription is also gauge-invariant.} we should have
\begin{equation}
    \frac{B_0^*(\omegaup,\mathbf{k})}{b_0^*(\omegaup,\mathbf{k})}= \frac{A_0(\omegaup,\mathbf{k})}{a_0(\omegaup,\mathbf{k})}, \quad {\rm i.e.}\quad  G_A^*(\omegaup, \mathbf{k}) = G_R(\omegaup, \mathbf{k}) \quad {\rm holds}.
\end{equation}

We therefore conclude that the CGL horizon cap prescription reproduces the Son-Starinets prescription for the retarded propagator together with KMS periodicity and the thermal structure of the correlation functions at any frequency and momentum. A similar approach was adopted earlier by  Son and Herzog by identifying the two sides of the eternal black hole with the forward and backward arms of the Schwinger-Keldysh contour \cite{Son:2009vu}. However, in this case, the backward part of the time contour needs to be shifted by $\beta/2$ along the negative imaginary axis. The main advantage of the CGL prescription is that we do not need an eternal black hole geometry for its implementation suggesting that its nonequilibrium generalization would be generically more feasible. Furthermore, it is also not clear if out-of-equilibrium correlation functions can be analytically continued in their time arguments as required by the Son and Herzog implementation of the Schwinger-Keldysh contour. Also, it should be possible to define integration over bulk vertices and bulk quantum loops in the CGL prescription as well via the analytic structure of the complexified space-time with the horizon cap. However, this is outside the scope of the present work, and therefore we do not further discuss about this issue. Finally, we note that the arguments presented here are simpler compared with \cite{Glorioso:2018mmw} since we do not employ an expansion about $\omega = 0$ which obscures the analytic continuation at the horizon cap by producing $(\log\omega)^n$ terms.

\section{Bjorken flow, Weyl rescaling and the holographic dual}\label{sec2}
\subsection{Bjorken flow and its Weyl rescaling}\label{sec:BjRev}
\begin{figure}[h!]
    \centering
    \includegraphics[scale=0.5]{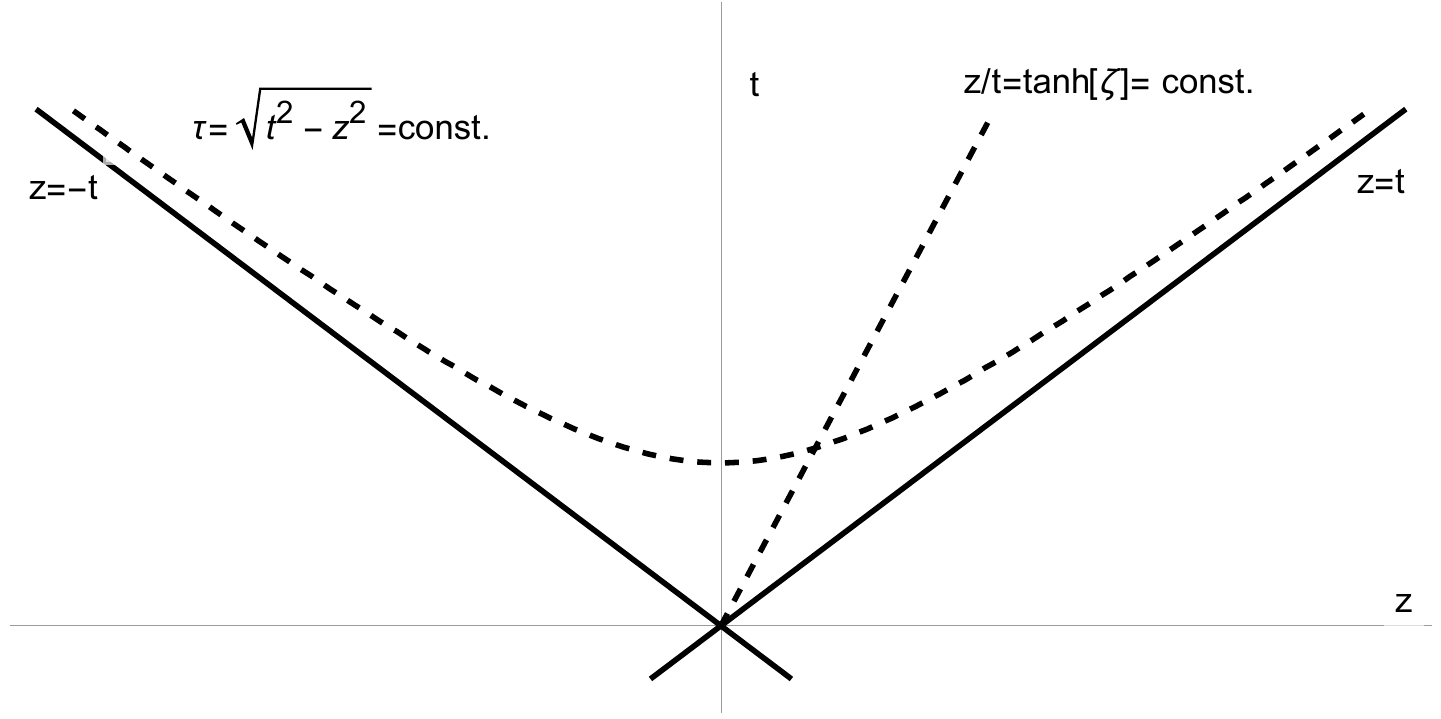}
    \caption{The schematic diagram of the Bjorken flow illustrates the evolution of an expanding system on the forward light cone. Here $z$ is the longitudinal direction along which the system expands. The transverse directions have been suppressed. Initial data is specified on a constant $\tau$ hyperboloid.}
    \label{fig:bjorken_diagram}
\end{figure}
Bjorken flow \cite{Bjorken:1982qr} is a simple model describing the expanding plasma produced by heavy ion collisions. This model is based on the assumptions of boost invariance, and translational and rotational symmetries in the transverse directions of an expanding system. The evolution occurs inside a forward light cone as shown in Fig. \ref{fig:bjorken_diagram}.
It is convenient to describe the Bjorken flow in the Milne proper time $\tau$ and the rapidity $\zeta$ which are related to the Minkowski (lab frame) time $t$ and the longitudinal  coordinate $z$ (along which the system is expanding) as
$$ t= \tau \cosh{\zeta} \hspace{1cm} \text{and} \hspace{1cm} z= \tau \sinh{\zeta}.$$
The transverse coordinates  $x_\perp$ are the same in both Milne and Minkowski coordinate systems. In the Milne coordinates, the Minkowski metric takes the form
\begin{equation}\label{Eq:Milne}
{\rm d}s^2 = -{\rm d}\tau^2 +\tau^2{\rm d}\zeta^2 + {\rm d}s_\perp^2~,
\end{equation}
where ${\rm d}s_\perp$ is the line element in the transverse plane.

The symmetries of the Bjorken flow imply that the expectation value of any operator depends only on the proper time $\tau$. Thus an ansatz for the expectation value of the energy-momentum tensor in a $d$-dimensional theory, which is also consistent with the transverse translational and rotational symmetries of the Bjorken flow, can take the form
\begin{equation}\label{Eq:TBjorken}
\langle T_{\mu\nu}\rangle = {\rm diag}(\epsilon(\tau),\,\tau^{2} p_L(\tau),\,\underbrace{p_T(\tau),\,\cdots,\,p_T(\tau)}_{(d-2) \,\, {\rm times}}),\end{equation}
in the Milne coordinates for $d>2$. Clearly, $\epsilon$, $p_L$ and $p_T$ denote the energy density, longitudinal and transverse pressures respectively. The local conservation of energy and momentum $ \nabla_\mu T^{\mu\nu} =0$ implies that
\begin{equation}\label{Eq:pL}
   p_L(\tau) = - \epsilon(\tau) - \tau \partial_\tau\epsilon(\tau),
\end{equation}
and the conformal Ward identity $T^\mu_{\,\,\mu}=0$ imposes
\begin{equation}\label{Eq:pT}
    p_T(\tau) = \frac{2}{d-2} \left(\epsilon(\tau) + \frac{1}{2}\tau \partial_\tau \epsilon(\tau)\right).
\end{equation}
Consequently, the evolution of the energy-momentum tensor is determined by $\epsilon(\tau)$ in a conformal field theory.


At large proper time $\tau$, the Bjorken flow admits a hydrodynamic description \cite{Jeon:2015dfa}. To explicitly map the energy-momentum tensor \eqref{Eq:TBjorken} to that of a fluid, we need to set the flow velocity as $$u^\mu =(1, \vec{0}),$$i.e. ${\rm d}\tau$ is co-moving with the flow in the Milne coordinates. In a conformal system, the large proper time expansion of $\epsilon(\tau)$ is given by a single parameter, namely 
\begin{equation}\label{Eq:mu}
    \mu :=\epsilon_0\tau_0^{\frac{d}{d-1}},
\end{equation}
which is determined by the initial conditions \textemdash ~$\epsilon_0$ is a constant energy density, and $\tau_0$ can be chosen to be the value of $\tau$ where we intialize. The large proper time expansion of $\epsilon(\tau)$ takes the form
\begin{equation}\label{Eq:epsilonexp}
   T^\tau_{\,\,\tau}=-\epsilon(\tau) =-\mu\tau^{-\frac{d}{d-1}} \left(1+\sum_{n=1}^\infty\lambda_n\, \mu^{-\frac{n}{d}}\tau^{-n\frac{d-2}{d-1}} \right),
\end{equation}
where $\lambda_n$ are (state-independent) constants that are determined by the transport coefficients of the microscopic theory. As for instance, $\lambda_1$ is related to the shear viscosity $\eta$ as
\begin{equation}\label{Eq:lambda1}
    \lambda_1 = -\frac{\eta(\epsilon)}{\epsilon^{\frac{d-1}{d}}},
\end{equation}
which should indeed be a constant in a conformal theory. The leading term of the expansion $\propto \tau^{-d/(d-1)}$ gives an exact solution of the Euler equations, and thus represents the expansion of a conformal perfect fluid.

 In what follows, we will need a Weyl transformation of the Bjorken flow. In a conformal theory, the hydrodynamic equations are Weyl covariant \cite{Rangamani:2009xk}. We are ignoring the Weyl anomaly for the moment, but we will explicitly mention it later. Under a Weyl transformation which transforms the metric and the energy-momentum tensor as
 \begin{equation}
     {\rm d}s^2 \rightarrow \widetilde{{\rm d}s}^2 = \Omega(x)^2 {\rm d}s^2, \quad T_{\mu\nu}\rightarrow \widetilde{T}_{\mu\nu}= \Omega(x)^{-d+2}T_{\mu\nu},
 \end{equation}
 the new solutions of the hydrodynamic equations are given by
 \begin{equation}
     u^\mu \rightarrow \tilde{u}^\mu = \Omega(x)^{-1} u^\mu, \quad \epsilon \rightarrow \tilde{\epsilon} = \Omega(x)^{-d}\epsilon,
 \end{equation}
 in any conformal theory. Consider the combined operation of the time reparametrization
 \begin{equation}\label{Eq:TimeRepa}
     \sigma = \tau^{\frac{d-2}{d-1}} \tau_0^{\frac{1}{d-1}}
 \end{equation}
 and the Weyl scaling with
 \begin{equation}\label{Eq:Omega}
     \Omega(\sigma) = \left(\frac{\tau_0}{\sigma}\right)^{\frac{1}{d-2}},
 \end{equation}
under which the Milne metric \eqref{Eq:Milne} transforms to (with $\hat{\zeta} = \zeta\tau_0$)
 \begin{equation}\label{Eq:Weylmetric}
     \widetilde{{\rm d}s}^2 = - \frac{(d-1)^2}{(d-2)^2}{\rm d}\sigma^2 + \frac{\sigma^2}{\tau_0^2} {\rm d}{\hat{\zeta}}^2 + \left(\frac{\sigma}{\tau_0}\right)^{-\frac{2}{d-2}} {\rm d}s_\perp^2,
 \end{equation}
 and the energy-momentum tensor given by \eqref{Eq:TBjorken}, \eqref{Eq:pL} and \eqref{Eq:pT} transforms to 
 \begin{eqnarray}\label{Eq:TBjorkenWeyl}
 \widetilde{T}_{\sigma\sigma} &=&\Omega^{-d+2}\tau'(\sigma)^2\epsilon(\tau(\sigma))\nonumber\\&=&\frac{(d-1)^2}{(d-2)^2}  \left(\frac{\sigma}{\tau_0}\right)^{\frac{d}{d-2}}\epsilon(\tau(\sigma))\nonumber\\ &=&\frac{(d-1)^2}{(d-2)^2} \widetilde\epsilon(\sigma), \quad \left({\rm Note}\,\, \widetilde\epsilon(\sigma) := \left(\frac{\sigma}{\tau_0}\right)^{\frac{d}{d-2}}\epsilon(\tau(\sigma)) \right)\nonumber\\
 \widetilde{T}_{\hat\zeta\hat\zeta} &=& \Omega^{-d+2}\left(\frac{\tau(\sigma)}{\tau_0}\right)^2 p_L(\tau(\sigma))
 \nonumber\\&=& -\left(\frac{\sigma}{\tau_0}\right)^{\frac{3d-4}{d-2}}\Big(\epsilon(\tau(\sigma))+\tau(\sigma)\epsilon'(\tau(\sigma))\Big)\nonumber\\&=& \left(\frac{\sigma}{\tau_0}\right)^2\frac{1}{d-1}\Big(\widetilde\epsilon(\sigma)-(d-2)\sigma\tilde\epsilon'(\sigma)\Big),\nonumber\\
 \widetilde{T}_{ii} &=& \Omega^{-d+2} P_T(\tau(\sigma))
 \nonumber\\&=& \frac{2}{d-2}\frac{\sigma}{\tau_0}\Big(\epsilon(\tau(\sigma))+\frac{1}{2}\tau(\sigma)\epsilon'(\tau(\sigma))\Big)\nonumber\\&=&
 \left(\frac{\sigma}{\tau_0}\right)^{-\frac{2}{d-2}}\frac{1}{d-1}\Big(\widetilde\epsilon(\sigma)+\sigma\tilde\epsilon'(\sigma)\Big),
 \end{eqnarray}
with $\widetilde{T}_{ii}$ denoting the diagonal transverse components and $'$ denoting the derivative w.r.t. the argument of the corresponding function. It follows that in the hydrodynamic limit, the Bjorken expansion \eqref{Eq:epsilonexp} takes the resultant form
 \begin{equation}\label{Eq:epsilonWeylexp}
   \widetilde{T}^\sigma_{\,\,\sigma} \equiv -\widetilde{\epsilon}(\sigma)=-\left(\frac{\sigma}{\tau_0}\right)^{\frac{d}{d-2}}\epsilon(\tau(\sigma)) =- \epsilon_0 \left(1+\sum_{n=1}^\infty\lambda_n\, \epsilon_0^{-\frac{n}{d}}\sigma^{-n} \right).
 \end{equation}
 The Weyl scaled metric \eqref{Eq:Weylmetric} has the property that 
 \begin{equation}\label{Eq:sqrt-tilde-g}
     \sqrt{-\widetilde{g}} = \frac{d-1}{d-2}~,
 \end{equation}
 is a constant, and the spatial volume factor is unity,  same as in the Minkowski coordinates. However, the longitudinal volume expands, while the transverse volume contracts with the evolution.  Also note that for the Weyl scaled Bjorken flow \eqref{Eq:epsilonWeylexp}, we have
 \begin{equation}
     \lim_{\sigma\rightarrow\infty}\widetilde{\epsilon}(\sigma) = \epsilon_0.
 \end{equation}
Therefore, instead of a perfect fluid expansion, the flow attains a constant temperature, energy and entropy densities at late time although no time-like Killing vector exists in the background metric. The latter feature leads to viscous and higher-order corrections. The large (reparametrised) proper time expansion is determined by $\epsilon_0$, the final thermal value of the energy density, while $\tau_0$ appears in the Weyl scaling factor $\Omega$ as explicit in Eq. \eqref{Eq:Omega}.

The Weyl scaling depends explicitly on $\tau_0$. However, note that for $\tau_0\rightarrow \xi \tau_0$, we obtain from \eqref{Eq:TimeRepa} that $\sigma \rightarrow \xi^{1/(d-1)} \sigma$. Thus the Weyl factor given by \eqref{Eq:Omega} scales as $\Omega\rightarrow \xi^{1/(d-1)}\Omega$, implying that $\widetilde{{\rm d}s}^2 \rightarrow \xi^{2/(d-1)} \widetilde{{\rm d}s}^2$ and $\epsilon_0 \rightarrow \xi^{-1/(d-1)}\epsilon_0.$ Therefore, the dimensionless variables $\epsilon_0^{-1/d}\sigma^{-1}$ (which provides the proper time expansion parameter) and $\sigma^n \widetilde{T}^\sigma_{\,\,\sigma} $ are invariant under $\tau_0\rightarrow \xi \tau_0$, and are thus independent of $\tau_0$.

We also note that the Weyl transformation studied here is itself not corrected at first and higher orders in the large and proper time expansion.

\subsection{Gravitational setup}\label{sec2.1}
 The gravitational dual of the Bjorken flow \cite{Shuryak:2005ia,Janik:2005zt,Janik:2006ft,Kinoshita:2008dq,Beuf:2009cx,Heller:2013fn} has been extensively studied in the literature with the late time evolution providing a primary example of the fluid/gravity correspondence \cite{Policastro:2001yc,Bhattacharyya:2007vjd,Baier:2007ix,Rangamani:2009xk} where large order resummation of the hydrodynamic series \cite{Heller:2013fn} has been explicitly carried out revealing the hydrodynamization \cite{Heller:2015dha} of a far-from-equilibrium state. When a state hydrodynamizes, the energy-momentum tensor can be described as an optimally truncated (divergent and asymptotic) hydrodynamic series even when it is far from equilibrium \cite{Heller:2013fn,Heller:2015dha}. In the context of the Bjorken flow, the evolution of the energy density approaches a hydrodynamic attractor \cite{Heller:2013fn,Heller:2015dha}. This is a generic property of a many-body relativistic system irrespective of whether its degrees of freedom interact weakly or strongly (see \cite{Soloviev:2021lhs} for a recent review).

Here we will review the gravitational dual of the Bjorken flow in the hydrodynamic limit and then describe its Weyl transformation in detail. This Weyl transformation is what has been described in the previous subsection. In the bulk it is implemented by an appropriate diffeomorphism. As a result of this transformation, the state reaches a constant temperature and entropy density at late proper time instead of attaining perfect fluid expansion. The dual black hole also attains a horizon with constant surface gravity and area. However, even at late proper time there is no time-like Killing vector -- the directions longitudinal and transverse to the flow expand and contract respectively such that the horizon area remains constant at late proper time. Along with the Weyl transformation of the metric and the energy-momentum tensor described in the previous subsection, the holographic dual also produces the Weyl anomaly. The Weyl transformation will be an important tool in implementing the horizon cap prescription out of equilibrium.

Additionally, we will focus on the residual gauge freedom which allows us to fix the nonequilibrium event or apparent horizon at a fixed radial location. We will see that it is crucial to pin the horizon cap at the nonequilibrium event horizon for regularity, and therefore this gauge freedom will play an important role. We will explicitly show that this gauge freedom does not affect the dual metric or the dual energy-momentum tensor (and is thus a proper gauge transformation).

The holographic dual of the Bjorken flow in a $d$-dimensional conformal theory is a $(d+1)$-dimensional geometry which satisfies the Einstein's equations with a negative cosmological constant:
\begin{equation}\label{Eq:Einstein}
    R_{MN} - \frac{1}{2}R G_{MN} - \frac{d(d-1)}{2 L^2}G_{MN} =0.
\end{equation}
In what follows, we will set $L =1$ for convenience. In addition to the field theory coordinates, we need an extra radial coordinate to describe the dual geometry. The state of the conformal theory dual to a specific solution of \eqref{Eq:Einstein}, lives at the boundary ($\rho=0$) in the boundary metric, which is defined as
\begin{equation}\label{Eq:bmetric}
    g^{\rm b}_{\mu\nu} = \lim_{\rho\rightarrow 0}\rho^2 G_{\mu\nu},
\end{equation}
where $a$ and $b$ stand for the field theory indices. Since we are considering the Bjorken flow in the Milne metric \eqref{Eq:Milne}, $g^{\rm b}_{\mu\nu}$ should coincide with it. Similarly, if we consider the Weyl scaled version of the Bjorken flow, the boundary metric should coincide with \eqref{Eq:Weylmetric}.

Before considering the Bjorken flow, it is useful to first understand the vacuum solution, which is pure (maximally symmetric) $AdS_{d+1}$ space-time with the desired boundary metric. In the ingoing Eddington-Finkelstein gauge, the vacuum state in the Milne metric \eqref{Eq:Milne} is thus dual to 
\begin{eqnarray}\label{Eq:Vacuum}
{\rm d}s^2 = - \frac{2}{r^2} {\rm d}r {\rm d}\tau -\frac{1}{r^2}{\rm d}\tau^2 +\left(1 +\frac{\tau}{r}\right)^2 {\rm d}\zeta^2 + \frac{1}{r^2}{\rm d}s_\perp^2~,
\end{eqnarray}
where $r$ is the radial coordinate. Similarly, the vacuum in Weyl scaled metric \eqref{Eq:Weylmetric} is dual to 
\begin{eqnarray}\label{Eq:VacuumWeyl}
{\rm d}s^2 &=& - \frac{2}{v^2}\frac{d-1}{d-2} {\rm d}v {\rm d}\sigma -\frac{1}{v^2}\left(\frac{(d-1)^2}{(d-2)^2}+\frac{2(d-1)v}{
(d-2)^2\sigma}\right){\rm d}\sigma^2 +\frac{1}{\tau_0^2}\left(1 +\frac{\sigma}{v}\right)^2 {\rm d}{\hat{\zeta}}^2 \nonumber\\&& + \frac{1}{v^2}\left(\frac{\sigma}{\tau_0}\right)^{-\frac{2}{d-2}}{\rm d}s_\perp^2,
\end{eqnarray}
where $v$ is the radial coordinate. These bulk metrics \eqref{Eq:Vacuum} and \eqref{Eq:VacuumWeyl} are related by the diffeomorphism
\begin{equation}\label{Eq:Diffeo}
   \tau = \tau_0^{-\frac{1}{d-2}}\sigma^{\frac{d-1}{d-2}},\quad r = v\left(\frac{\sigma}{\tau_0}\right)^{\frac{1}{d-2}}.
\end{equation}
For both cases, \eqref{Eq:Vacuum} and \eqref{Eq:VacuumWeyl}, we obtain the boundary metrics \eqref{Eq:Milne} and \eqref{Eq:Weylmetric} from \eqref{Eq:bmetric}, after replacing $\rho$ with $r$ and $v$, respectively. Any Weyl transformation at the boundary is dual to a bulk diffeomorphism. Since the boundary metrics \eqref{Eq:Milne} and \eqref{Eq:Weylmetric} are related by a Weyl transformation, \eqref{Eq:Diffeo} is simply a specific instance of this general feature of holographic duality. Note that $\tau$ and $\sigma$ are related exactly by the time reparametrization \eqref{Eq:TimeRepa} at the boundary.\footnote{Diffeomorphisms such as \eqref{Eq:Diffeo} which implement global transformations on the dual state are called improper diffeomorphisms which are always part of residual gauge freedom after gauge fixing in the bulk. The latter can also have additional proper diffeomorphisms which do not affect the dual physical quantities.} 

Holographic renormalization \cite{Henningson:1998gx,Balasubramanian:1999re,deHaro:2000vlm,Skenderis:2002wp} provides the framework for extracting the $\langle T_{\mu\nu}\rangle$ corresponding to the state in the field theory dual to a specific asymptotically $AdS_{d+1}$ bulk geometry. The procedure essentially amounts to covariantly regularizing the Brown-York tensor on a cut-off hypersurface with local counterterms built out of the induced metric, and then taking this surface to the boundary. For the bulk geometry \eqref{Eq:Vacuum}, $\langle T_{\mu\nu}\rangle =0$ in the dual vacuum state living in the flat Milne metric \eqref{Eq:Milne} at the boundary. On the other hand, for the vacuum state living on the Weyl transformed Milne metric \eqref{Eq:Weylmetric} which is dual to the bulk geometry \eqref{Eq:VacuumWeyl}, $\langle T_{\mu\nu}\rangle =0$ only if $d$ is odd. For even $d$, holographic renormalization reproduces the Weyl anomaly of the dual field theory. In the case of $d=4$, we obtain (using minimal subtraction scheme)
\begin{eqnarray}
\langle \widetilde{T}_{\mu\nu}\rangle = \frac{1}{8\pi G_N}\mathcal{A}_{\mu\nu} 
\end{eqnarray}
with 
\begin{eqnarray}\label{Eq:AnomalousT}
\mathcal{A}_{\mu\nu} &=& \frac{1}{16}\left( \frac{4}{3}\widetilde{R}_{\mu\nu}\widetilde{R}-2 \widetilde{R}_{\mu\rho}\widetilde{R}^\rho_{\,\,\nu}-\tilde{g}_{\mu\nu}\left(\frac{1}{2}\widetilde{R}^2 -\widetilde{R}_{\rho\sigma}\widetilde{R}^{\rho\sigma}\right)\right),\nonumber\\
&=&{\rm diag}\left(-\frac{1}{32\sigma^4},-\frac{11}{648\sigma^2\tau_0^2},\frac{25\tau_0}{648\sigma^5},\frac{25\tau_0}{648\sigma^5}\right)~,
\end{eqnarray}\label{Eq:Amn}
where $\tilde{g}$ denotes the Weyl rescaled background metric \eqref{Eq:Weylmetric}, $\widetilde{R}_{\mu\nu}$ is the Ricci tensor built out of it, etc. It is easy to verify that
\begin{equation}
    \widetilde\nabla_\mu \widetilde{T}^\mu_{\,\,\nu} =0,
\end{equation}
i.e. energy and momentum are conserved in the Weyl rescaled background metric \eqref{Eq:Weylmetric} (with $\widetilde\nabla$ being the covariant derivative built out of it), and 
\begin{eqnarray}\label{Eq:WeylAn}
\widetilde{T}^\mu_{\,\,\mu} =- \frac{1}{8\pi G_N}\left(\frac{1}{24}\widetilde{R}^2 -\frac{1}{8}\widetilde{R}_{\rho\sigma}\widetilde{R}^{\rho\sigma}\right) =\frac{1}{8\pi G_N} \frac{2}{27\sigma^4}.
\end{eqnarray}
Using 
\begin{equation}\label{Eq:NGN}
    \frac{1}{G_N} =  \frac{2N^2}{\pi},
\end{equation}
we can readily find that \eqref{Eq:WeylAn} reproduces the Weyl anomaly of $SU(N)$ $\mathcal{N}=4$ super-symmetric Yang-Mills theory \cite{Henningson:1998gx,Balasubramanian:1999re}.

An asymptotically $AdS_{d+1}$ metric dual to a Bjorken flow on the flat Milne metric \eqref{Eq:Milne} at the boundary, is a solution to the vacuum Einstein's equations \eqref{Eq:Einstein} which takes the form
\begin{eqnarray} \label{Eq:geom}
ds^2 =  -  \frac{2}{r^2}  {\rm d}r  {\rm d}\tau- \frac{A(r,\tau)}{r^2}  {\rm d} \tau^2 + \left(1+ \frac{ \tau}{r}\right)^2 e^{L(r,\tau)}  {\rm d}\zeta^2 +\frac{ e^{K(r,\tau)}}{r^2}  {\rm d}s_{\perp}^2 
\end{eqnarray}
with the following Dirichlet asymptotic boundary conditions 
\begin{equation}\label{Eq:DirichletMetric}
  A(r,\tau) \rightarrow 1, \hspace*{0.6cm} K(r,\tau) \rightarrow 0 \hspace{0.6cm} \text{and} \hspace{0.6cm} L(r,\tau) \rightarrow 0 \hspace{0.6cm} \text{as}\quad r\rightarrow 0.
\end{equation}
These boundary conditions ensure that the boundary metric \eqref{Eq:bmetric} (with $r$ being the radial coordinate in place of the generic $\rho$) coincides with the Milne metric \eqref{Eq:Milne}.

The Einstein equations \eqref{Eq:Einstein} can be readily solved in the late-time expansion, as functions of the scaling variable
\begin{equation}\label{Eq:s}
    s = r \left(\frac{\tau_0}{\tau}\right)^{\frac{1}{d-1}},
\end{equation}
and with the expansion parameter being
\begin{equation}\label{Eq:exp-parameter}
\left(\tilde{\mu}^{1/d} \tau^{\frac{d-2}{d-1}}\right)^{-1}
\end{equation}
where $\tilde\mu := \varepsilon_0 \tau_0^{\frac{d}{d-1}}$ is a constant which will be related to the single parameter $\mu$ of the Bjorken flow defined in \eqref{Eq:mu} (or equivalently to $\epsilon_0$) below. Explicitly,
\begin{eqnarray}\label{Eq:AsympGen}
A(r,\tau) &=& 1 - \varepsilon_0 s^d + \sum_{i=1}^\infty \left(\tilde{\mu}^{1/d} \tau^{\frac{d-2}{d-1}}\right)^{-i} a_{(i)}(\varepsilon_0^{\frac{1}{d}}s),\nonumber\\
L(r,\tau) &=&  \sum_{i=1}^\infty\left(\tilde{\mu}^{1/d} \tau^{\frac{d-2}{d-1}}\right)^{-i} l_{(i)}(\varepsilon_0^{\frac{1}{d}}s),\nonumber\\
K(r,\tau) &=&  \sum_{i=1}^\infty \left(\tilde{\mu}^{1/d} \tau^{\frac{d-2}{d-1}}\right)^{-i} k_{(i)}(\varepsilon_0^{\frac{1}{d}}s).
\end{eqnarray}
The functions $a_{(i)}$, $l_{(i)}$ and $k_{(i)}$ satisfy ordinary differential equations with source terms at each order. We require that these functions do not blow up at the perturbative horizon which is at $s_h = \varepsilon_0^{-1/d}$, i.e. at
\begin{equation}\label{Eq:Bhor}
r_h =\varepsilon_0^{-\frac{1}{d}}\left(\frac{\tau}{\tau_0}\right)^{\frac{1}{d-1}} 
\end{equation}
Together with the Dirichlet boundary conditions \eqref{Eq:DirichletMetric}, these finiteness conditions ensure that we obtain solutions which are free of naked singularities in the perturbative expansion and which are unique up to terms which are determined by a single coefficient \cite{Heller:2013fn}. This coefficient captures the residual gauge freedom of the ingoing Eddington-Finkelstein coordinates which is the reparametrization of the radial coordinate $r$ (without spoiling the manifest translational and rotational symmetries along the transverse directions). Usually, this residual gauge freedom is fixed by setting the radial location of the apparent or event horizon at \eqref{Eq:Bhor} to all orders in the late proper time expansion \cite{Chesler:2013lia}. However, in what follows, we will show that the residual gauge freedom should actually be fixed by the regularity of the horizon cap. Note that although the requirement that the horizon cap should be pinned to the evolving event horizon for regularity is a gauge-invariant statement, the residual gauge freedom will be crucial to implement the prescription in the Eddington-Finekelstein coordinates. We therefore keep this gauge freedom unfixed here and later show how it is fixed by the regularity of the horizon cap such that the latter is pinned to the evolving event horizon (which lies at a fixed radial location after the residual gauge fixing).

It is also crucial to emphasize that the residual gauge freedom involving the reparametrization of the radial coordinate is a proper diffeomorphism, i.e. it leaves \textit{both} the boundary metric (which is the flat Milne background \eqref{Eq:Milne}) and also the $\langle T_{\mu\nu}\rangle$ of the dual Bjorken flow extracted from holographic renormalization is \textit{invariant}. It's useful to see this explicitly. For illustration, let's consider the $AdS_5$ case. The asymptotic expansions take the form:
\begin{eqnarray}\label{renormalization:asymptotic}
A(r,\tau) &=& 1 + r a_1(\tau) + r^2\left(\frac{a_1(\tau)^2}{4}- a_1'(\tau)\right) + r^4 a_4(\tau) + \cdots,\nonumber\\
K(r,\tau) &=& r a_1(\tau) - r^2\frac{a_1(\tau)^2}{4} +r^3 \frac{a_1(\tau)^3}{12}+ r^4 k_4(\tau) + \cdots,\nonumber\\
L(r,\tau) &=& r a_1(\tau) - r^2\left(\frac{a_1(\tau)^2}{4}+\frac{ a_1(\tau)}{\tau}\right) \nonumber\\&& +r^3 \left(\frac{a_1(\tau)^3}{12}+\frac{a_1(\tau)^2}{2\tau}+\frac{ a_1(\tau)}{\tau^2}\right)+ r^4 l_4(\tau)+ \cdots.
\end{eqnarray}
Above, the function $a_1(\tau)$ is related to the residual gauge freedom, and can be chosen arbitrarily. Furthermore, the constraints of Einstein's equations \eqref{Eq:Einstein} impose
\begin{eqnarray}\label{Einstein:constraint}
a_4'(\tau) &=& - \frac{a_1(\tau)^4}{12\tau}  - \frac{4}{3\tau} a_4(\tau) - \frac{8}{3\tau} k_4(\tau), \nonumber\\
l_4(\tau) + 2 k_4(\tau) &=& - \frac{a_1(\tau)}{\tau^3} -\frac{3}{4}\frac{a_1(\tau)^2}{\tau^2}-\frac{1}{4}\frac{a_1(\tau)^3}{\tau}-\frac{3}{32}a_1(\tau)^4.
\end{eqnarray}
Using these constraints, one can find via holographic renormalization that
\begin{eqnarray}\label{Eq:HolT5}
    \langle T_{\mu\nu} \rangle &=& \frac{3}{16\pi G_N}\times \\\nonumber&&
    \begin{pmatrix}
    - a_4(\tau) ~ & 0~ & 0~ & 0\\ 0 ~& \tau^2\left(a_4(\tau)+ \tau a_4'(\tau)\right) ~& 0~&0 \\0~&0 ~& -a_4(\tau)- \frac{1}{2}\tau a_4'(\tau) ~& 0\\0~&0~&0 ~& -a_4(\tau)- \frac{1}{2}\tau a_4'(\tau)
    \end{pmatrix}.
\end{eqnarray}
Firstly, the above result is exact to \textit{all} orders in the late proper time expansion. Second, we readily find that $ \langle T_{\mu\nu}(\tau) \rangle $ is determined by $a_4(\tau)$ alone and is independent of the arbitrary function $a_1(\tau)$ capturing the residual gauge freedom in the asymptotic expansion \textit{after} utilizing the constraints \eqref{Einstein:constraint} in the renormalized Brown-York stress tensor. Thus $ \langle T_{\mu\nu}(\tau) \rangle $ is invariant under the residual gauge transformation. Furthermore, comparing \eqref{Eq:HolT5} with \eqref{Eq:TBjorken}, \eqref{Eq:pL} and \eqref{Eq:pT} (for $d=4$) we find that $ \langle T_{\mu\nu}(\tau) \rangle $ takes the general form of the energy-momentum tensor of Bjorken flow with the identification
\begin{equation}
    \epsilon(\tau) = - \frac{3}{16\pi G_N} a_4(\tau).
\end{equation}
One can repeat the same exercise in arbitrary dimensions ($d>2$) and show that $ \langle T_{\mu\nu}(\tau) \rangle $ obtained from holographic renormalization takes the general form given by \eqref{Eq:TBjorken}, \eqref{Eq:pL} and \eqref{Eq:pT} with the identification
\begin{equation}\label{Eq:ep-gen}
    \epsilon(\tau) = - \frac{d-1}{16\pi G_N} a_d(\tau),
\end{equation}
where $a_d(\tau)$ is the coefficient of $r^d$ in the asymptotic expansions of $A(r,\tau)$.

Furthermore, extracting $a_d(\tau)$ from \eqref{Eq:AsympGen} we obtain that at the leading order in the late  proper time expansion
\begin{equation}\label{Eq:epsilon-d-asymp}
    \epsilon(\tau) \approx \frac{d-1}{16\pi G_N}\varepsilon_0 \left(\frac{\tau_0}{\tau}\right)^{\frac{d}{d-1}}=\frac{d-1}{16\pi G_N} r_h(\tau)^{-d}= \frac{d-1}{16\pi G_N}\left(\frac{4\pi T(\tau)}{d}\right)^d,
\end{equation}
where we have used \eqref{Eq:Bhor}, and also \eqref{Eq:THrh} to define an instantaneous Hawking temperature $T(\tau)$ given by
\begin{equation}\label{Eq:T-d-asymp}
    T(\tau) = \frac{d}{4\pi}\varepsilon_0^{\frac{1}{d}}\left(\frac{\tau_0}{\tau}\right)^{\frac{1}{d-1}}.
\end{equation}
Once again comparing with the general (hydrodynamic) late proper time expansion \eqref{Eq:epsilonexp}, we find that
\begin{equation}
    \epsilon_0 = \frac{d-1}{16\pi G_N}\varepsilon_0.
\end{equation}
For the case of $AdS_5$, the identification \eqref{Eq:NGN} implies that at late proper time
\begin{equation}\label{Eq:epsilon-4-asymp}
     T(\tau) \approx \frac{1}{\pi} \varepsilon_0^{1/4}\left(\frac{\tau_0}{\tau}\right)^{\frac{1}{3}} , \qquad\quad\epsilon(\tau) \approx \frac{3}{8}\pi^2 N^2 T(\tau)^4 = \frac{3}{8\pi^2} N^2 \varepsilon_0 \left(\frac{\tau_0}{\tau}\right)^{\frac{4}{3}}.
\end{equation}
For any $d$, $\epsilon(\tau)$ is given by a perfect fluid expansion given by \eqref{Eq:epsilon-d-asymp} at late time with
\begin{equation}
    p_L(\tau) \approx p_T(\tau) \approx \frac{1}{d-1}\epsilon(\tau) \approx \frac{1}{16\pi G_N} r_h(\tau)^{-d}.
\end{equation}
Thus, at late proper time, the energy density $\epsilon(\tau)$ and the pressures $p_L(\tau)$ and $p_T(\tau)$ are thus given by the thermal equation of state obtained from a static black brane geometry, but with a time-dependent temperature \eqref{Eq:T-d-asymp} which satisfy the Euler equations.

To construct a regular horizon cap, it is useful to change coordinates from $r$ and $\tau$ to $v$ and $\sigma$ following \eqref{Eq:Diffeo} as in the case of the space-time dual to the vacuum state. Note, it follows from \eqref{Eq:s} that $v$ and $s$ are the same. In these new coordinates, the metric \eqref{Eq:geom} takes the form:
\begin{eqnarray}\label{Eq:Metric-Bjorken-vsigma}
   {\rm d}s^2 &=& - \frac{2}{v^2}\frac{d-1}{d-2} {\rm d}v {\rm d}\sigma -\frac{1}{v^2}\left(\frac{(d-1)^2}{(d-2)^2}A(v,\sigma)+\frac{2(d-1)v}{
(d-2)^2\sigma}\right){\rm d}\sigma^2 \nonumber\\&&+\frac{1}{\tau_0^2}\left(1 +\frac{\sigma}{v}\right)^2 e^{L(v,\sigma)}{\rm d}{\hat{\zeta}}^2 + \frac{1}{v^2}\left(\frac{\sigma}{\tau_0}\right)^{-\frac{2}{d-2}}e^{K(v,\sigma)}{\rm d}s_\perp^2, 
\end{eqnarray}
and is dual to Bjorken flow on the Weyl rescaled background metric \eqref{Eq:Weylmetric} at the boundary. We want to emphasize that just like the Weyl transformation, the dual bulk diffeomorphism which achieves the above form of the metric is given by \eqref{Eq:Diffeo} exactly, and is therefore not corrected at first and higher orders in the large proper time expansion.

Holographic renormalization and the constraints of the Einstein's equations \eqref{Eq:Einstein} imply that the energy-momentum tensor of the dual Bjorken flow takes the form
\begin{equation}
    \langle \widetilde{T}_{\mu\nu}\rangle =  T^W_{\mu\nu} + \frac{1}{8 \pi G_N} \mathcal{A}_{\mu\nu},
\end{equation}
where $T^W_{\mu\nu}$ takes the general Bjorken form with non-vanishing components given by \eqref{Eq:TBjorkenWeyl} in which 
\begin{equation}
    \tilde{\epsilon}(\sigma) = - \frac{d-1}{16 \pi G_N} \frac{(d-1)^2}{(d-2)^2}a_d(\sigma),
\end{equation}
and $\mathcal{A}_{\mu\nu}$ is the Weyl anomaly appearing for even $d$. Comparing with \eqref{Eq:ep-gen}, we indeed verify that the bulk diffeomorphism \eqref{Eq:Diffeo} implements the Weyl transformation and time reparametrization in the dual theory (with the Weyl factor given by \eqref{Eq:Omega}) and also reproduces its Weyl anomaly. Particularly, for $d=4$, we recall that $\mathcal{A}_{\mu\nu}$ is simply given by \eqref{Eq:AnomalousT}. The anomalous term is state-independent (and is always the same as in the Weyl transformed vacuum state). We again note that $\langle \widetilde{T}_{\mu\nu}\rangle$ is invariant under the residual gauge symmetry since it is independent of $a_1(\tau)$ after we implement the gravitational constraints following the previous discussion.

Obviously, the late proper time expansion \eqref{Eq:AsympGen} takes the form 
\begin{eqnarray}\label{Eq:AsympGen2}
A(v,\sigma) &=& 1 - \varepsilon_0 v^d + \sum_{i=1}^\infty \left( \varepsilon_0^{1/d}\sigma\right)^{-i} a_{(i)}(\varepsilon_0^{1/d}v),\nonumber\\
L(v,\sigma) &=&  \sum_{i=1}^\infty \left( \varepsilon_0^{1/d}\sigma\right)^{-i}  l_{(i)}(\varepsilon_0^{1/d}v),\nonumber\\
K(v,\sigma) &=&  \sum_{i=1}^\infty \left( \varepsilon_0^{1/d}\sigma\right)^{-i}  k_{(i)}(\varepsilon_0^{1/d}v).
\end{eqnarray}
Explicitly, for $d=4$, 
\begin{eqnarray}\label{Eq:a1l1k1}
a_{(1)}(x) &=& \alpha_1\frac{x(1+x^4)}{3}+ \frac{2x^4(1 +x)}{3},\nonumber\\
k_{(1)}(x) &=&\alpha_1\frac{x}{3} + \frac{1}{2}g(x),\nonumber\\
l_{(1)}(x) &=&\alpha_1\frac{x}{3}-g(x),
\end{eqnarray}
where
\begin{equation}\label{Eq:g}
    g(x) = \frac{4}{3}x -\frac{1}{3}\ln (x^2 +1)-\frac{2}{3}\ln (x +1) - \frac{2}{3} \arctan x  .
\end{equation}
Above $\alpha_1$ is the dimensionless parameter associated with the residual gauge freedom. At any order in the late proper time expansion, the terms multiplying $\alpha_1$ in \eqref{Eq:a1l1k1} remain the same, however, we should replace $\alpha_1$ by $\alpha_n$ at the $n$-th order. It is also easy to see that $g(x)$ is finite at $x=1$ implying that the metric is regular (with no naked singularity) at the perturbative horizon
\begin{equation}
 v_h = \varepsilon_0^{-1/d} + \mathcal{O}(\sigma^{-1}).
\end{equation}
At late proper time the dual black brane has a constant surface gravity and area although the directions longitudinal to the flow keep expanding and those transverse to the flow keep contracting.

The above metric reproduces the late proper time expansion of $\widetilde{\epsilon}(\sigma)$ which takes the form \eqref{Eq:epsilonWeylexp} with $\epsilon_0$ given by \eqref{Eq:ep-gen} and $\lambda_n$ taking specific values for a given $d$. Particularly, for any $d$, we obtain
\begin{equation}\label{Eq:lambda1-hol}
    \lambda_1 = -  \frac{1}{(d-1)^{\frac{d-1}{d}}(16\pi G_N)^{1/d}}.
\end{equation}
It is easy to verify from \eqref{Eq:lambda1} and the equation of state (see \eqref{Eq:epsilon-d-asymp})
\begin{equation}\label{Eq:beta-weyl}
  T=\beta^{-1} = \frac{d}{4\pi}\varepsilon_0^{1/d},\quad \epsilon = \frac{d-1}{16\pi G_N}\varepsilon_0 = \frac{d-1}{16\pi G_N}\left(\frac{4\pi T}{d}\right)^d
\end{equation}
that \eqref{Eq:lambda1-hol} implies via \eqref{Eq:lambda1} that
\begin{equation}
    \frac{\eta}{s} = \frac{1}{4\pi}~,
\end{equation}
for any $d>2$. More details of the perturbative expansion are in Appendix \ref{sec:BJ2}.

\section{The bulk scalar field and the horizon cap of the Bjorken flow}\label{sec:ScalarField}
The key to obtaining the real-time correlation functions is solving the dynamics of the scalar field in the gravitational background dual to the Bjorken flow. The starting point, however, is to construct the analog of the bulk Schwinger-Keldysh contour with the horizon cap for the gravitational background itself. This is straightforward. The metric dual to the Weyl scaled Bjorken flow (given by Eqs. \eqref{Eq:Metric-Bjorken-vsigma} and \eqref{Eq:AsympGen2}) reaches a constant horizon temperature at late time although the boundary metric has time-dependent spatial components. Since we would be working perturbatively in the late proper time expansion, we will fix the horizon cap at the constant late-time value $v = \varepsilon_0^{-1/d}$ to all orders in the perturbative late proper time expansion while keeping the residual gauge freedom of radial reparametrization unfixed as mentioned above. The metric is analytic to all orders at the horizon cap. Therefore, there is no modification to the metric on the other arm of the bulk space-time as it is reached via the complexified $v$ contour encircling the horizon as shown in Fig. \ref{Fig:ThermalHC} (as emphasized earlier, there is no analytic continuation in $\sigma$ and other coordinates). Exactly the same gravitational background is valid on both arms of the complex $v$ contour. It is also easy to see that the on-shell Einstein-Hilbert action on the two arms cancel each other out implying that the dual (non-)equilibrium partition function in the absence of additional sources (and with the same boundary metric on the two arms) is exactly zero, as should be the case.\footnote{In the field theory, this is equivalent to the statement that $W[J_1= J_2 =0] =0$.}

There is a crucial subtlety to this rather simple construction. We should worry about the residual gauge symmetries at first and higher orders in the late proper time expansion. In what follows, we will show that the analytic behavior of the sourced scalar field retains its equilibrium nature to all orders in the late proper time expansion at the horizon cap, provided the residual gauge symmetry is fixed in a unique way at each order. This addresses an issue which would have arisen if we had fixed the residual gauge symmetries to keep the apparent or the event horizon at $v = \varepsilon_0^{-1/d}$ to all orders in the late proper time expansion. However, the location of these two horizons differ at second and higher orders in the proper time expansion. We find that the gauge fixing which implements the regularity of the horizon cap is exactly the same which fixes the event (but not the apparent) horizon at $v_h =\varepsilon_0^{1/d}$ up to third order in the proper time expansion in the case of $AdS_4$ and $AdS_5$. Although we do not have an analytic proof that this feature will continue to hold at higher orders, we expect it to be the case as we explicitly find that the gauge fixing is independent of the mass of the bulk scalar field, and it should also hold for fermion, vector and higher rank tensor fields.

At the outset, we repeat our emphasis that the Weyl rescaling (and hence the dual bulk diffeomorphism) is determined purely by the late proper time perfect flow regime that is mapped to that of a constant temperature in a non-trivial background metric in the field theory (which is also the boundary metric of the dual black brane geometry after the bulk diffeomorphism). This should be evident from our construction of the bulk dual of the Weyl-transformed Bjorkan flow (in the $v$ and $\sigma$ coordinates) in the previous section. Both these Weyl transformation and the background metric of the field theory do \textit{not} receive any correction at first and higher orders in the derivative (i.e. large proper time) expansion. Nevertheless, with such a Weyl transformation (bulk diffeomorphism) we will be able to ensure that the Klein-Gordon equation of the bulk scalar takes the same form  at first and higher orders as that at the leading order but with source terms, and that \textit{the leading near-horizon behavior is exactly the same as that in the static black brane geometry (in the $v$ and $\sigma$ coordinates) to all orders in the large proper time expansion} when the horizon cap is pinned to the evolving event horizon. This will be instrumental in establishing the horizon cap prescription in the hydrodynamic tail of the Bjorken flow.

To see the main advantage of working in the $v$ and $\sigma$ coordinates, note that the explicit form of the leading order metric (as evident from Eqs. \eqref{Eq:Metric-Bjorken-vsigma} and \eqref{Eq:AsympGen2}) is:
\begin{eqnarray}\label{Eq:Metric-Bjorken-vsigma-zero}
   {\rm d}s^2 &=& - \frac{2}{v^2}\frac{d-1}{d-2} {\rm d}v {\rm d}\sigma -\frac{1}{v^2}\left(\frac{(d-1)^2}{(d-2)^2}(1- v^d \varepsilon_0+ \mathcal{O}(\sigma^{-1}))+\frac{2(d-1)v}{
(d-2)^2\sigma}\right){\rm d}\sigma^2 \nonumber\\&&+\frac{1}{\tau_0^2}\left(1 +\frac{\sigma}{v}\right)^2 (1+ \mathcal{O}(\sigma^{-1})){\rm d}{\hat{\zeta}}^2 + \frac{1}{v^2}\left(\frac{\sigma}{\tau_0}\right)^{-\frac{2}{d-2}}(1+ \mathcal{O}(\sigma^{-1})){\rm d}s_\perp^2.
\end{eqnarray}

{It is useful to define the \textit{comoving momenta} which depend on $\sigma$}
\begin{equation}\label{Eq:kLkTredef}
    \kappa_L = k_L \frac{\tau_0}{\sigma}, \quad \vec{\kappa}_T = \vec{k}_T\left(\frac{\tau_0}{\sigma}\right)^{-\frac{1}{d-2}}.
\end{equation}
such that it takes care of the longitudinal expansion and transverse contraction of the boundary metric. A natural ansatz for the scalar field consistent with boost invariance of the background geometry is:
\begin{equation}\label{Eq:Anstaz0n}
    \Phi (v, \sigma, \hat\zeta, \vec{x}_\perp) \approx e^{i\left(-\frac{d-1}{d-2}\omegaup\sigma + k_L \hat\zeta + \vec{k}_T\cdot \vec{x}_\perp\right)} f(v,\omegaup,\kappa_L, \vec{\kappa}_T).
\end{equation}
where remarkably there is no explicit $\sigma$ dependence in $f$ (only implicitly through $\kappa_L$ and $\vec{\kappa}_T$) while in the phase we have the usual fixed conjugate momenta $k_L$ and $\vec{k}_T$. This is exact at leading order and can be further corrected to obtain a systematic expansion of the equations of motion in powers of $\sigma^{-1}$ as shown below. Particularly, the Klein Gordon equation $(\Box - m^2)\Phi =0 $ for the bulk scalar field at the leading order $\sigma^0$ is then\footnote{To obtain this we should substitute $k_L$ and $\vec{k}_T$ in \eqref{Eq:Anstaz0n} by the redefined momenta \eqref{Eq:kLkTredef} and remember to differentiate $f$ w.r.t. $\sigma$ as the redefined momenta \eqref{Eq:kLkTredef} depend on $\sigma$.}
\begin{align}\label{Eq:ZerothKG}
    \mathcal{D}f = \mathcal{O}(\sigma^{-1})
\end{align}
where 
\begin{equation}\label{Eq:D}
    \mathcal{D} = v^2( 1- v^d \varepsilon_0) \partial_v^2 -v(d-1 - 2 i v\omegaup +v^d\varepsilon_0 )\partial_v - (m^2 + v(v(\kappa_L^2 +{\kappa}_T^2) + i(d-1)\omegaup)).
\end{equation}
Remarkably, the left hand side of \eqref{Eq:ZerothKG} is just the Klein Gordon equation for the massive bulk scalar in the $AdS_{d+1}$ static black brane geometry \eqref{Eq:SBB} at temperature $\beta^{-1}$ given by \eqref{Eq:beta-weyl}, with $v$ substituting for $r$, and $\omegaup$, $\kappa_L$ and $\vec{\kappa}_T$ identified with the canonical frequency and momenta. The implication is that $f$ will have exactly the same solutions at the leading order as in the static black brane geometry and therefore the same analytic structure at the horizon cap at the leading order. For the homogeneous ($k_L = \vec{k}_T = 0$) and massless case, this result reduces to the observation made by Janik and Peschanski \cite{Janik:2006gp} in the context of the transients of the Bjorken flow. 

Note that the dependence of $f$ in Eq. \eqref{Eq:Anstaz0n} on the co-moving momenta implies that we do not have separation of variables, but this should not be expected as the background at late time corresponds to an expanding boost-invariant perfect fluid, which has no time-like Killing vector. Nevertheless, the map to the Laplacian of a static black hole geometry is possible at leading order because the boost invariant perfect fluid is given only by a time-dependent temperature.

The ansatz \eqref{Eq:Anstaz0n} can be corrected to incorporate the viscous and all higher order corrections to the gravitational Bjorken flow background systematically while ensuring the regularity of the horizon cap after fixing the residual gauge freedom perturbatively in late proper time expansion. {This modified ansatz involves an expansion in $\sigma^{-1}$ with coefficients which are functions of $v$, $\omegaup$ and the co-moving momenta, such that the equations of motion can be solved systematically in $\sigma^{-1}$ expansion as well. Obviously, this implies that we also take into account the implicit dependence of the co-moving momenta on $\sigma$ while obtaining the equations of motion order by order in $\sigma^{-1}$.}

Explicitly, the ansatz for the bulk scalar field in the full complexified space-time with $1$ and $2$ labeling the sheets of the bulk space-time ending at the forward and backward arms of the Schwinger-Keldysh contour respectively at their boundaries is
\begin{eqnarray}\label{Eq:Ansatz1}
\begin{pmatrix}
\Phi_1 (v, \sigma, \hat\zeta, \vec{x}_\perp)\\\Phi_2 (v, \sigma, \hat\zeta, \vec{x}_\perp)
\end{pmatrix} &=& \int{\rm d}\omegaup {\rm d}k_L{\rm d}^{d-2}k_T \,\, e^{i\left(k_L \hat\zeta + \vec{k}_T\cdot \vec{x}_\perp\right)}e^{-i\frac{d-1}{d-2}\omegaup\sigma}\left(\frac{\sigma}{\tau_0}\right)^{i\gamma_0(\omegaup/\varepsilon_0^{1/d})}\nonumber\\ &&
\sum_{n=0}^\infty \left(\varepsilon_0^{1/d}\sigma\right)^{-n}\mathcal{M}_n(v, \omegaup, \kappa_L, \vec{\kappa}_T)\cdot 
\begin{pmatrix}
p(\omegaup, k_L, k_T)\\q(\omegaup, k_L, k_T)
\end{pmatrix},
\end{eqnarray}
where 
\begin{equation}\label{Eq:Mnc}
    \mathcal{M}_n(v, \omegaup, \kappa_L, \vec{\kappa}_T) = \begin{pmatrix}
    \phi_{n,in}(v, \omegaup, \kappa_L, \vec{\kappa}_T)~&\phi_{n,out}(v, \omegaup, \kappa_L, \vec{\kappa}_T)\\\phi_{n,in}(v, \omegaup, \kappa_L, \vec{\kappa}_T)~&e^{\beta\omegaup}\phi_{n,out}(v, \omegaup, \kappa_L, \vec{\kappa}_T)
    \end{pmatrix}.
\end{equation}
Above, in \eqref{Eq:Ansatz1}, a non-analytic term proportional to $ \sigma^{i\gamma_0}$ has been introduced with $\gamma_0$ being a new dimensionless function of only $\omegaup/\varepsilon_0^{1/d}$. This non-analytic factor is crucial to have a regular horizon cap as shown below. However, it does not affect the zeroth order equation of motion which takes the form of the Laplacian on a static black brane as discussed above. For any $\omegaup$, \eqref{Eq:Ansatz1} has the appearance of a trans-series in $\sigma^{-1}$ with $\omegaup$ characterizing the continuous instanton exponent. Note $p$ and $q$ are functions of $\omegaup$, $k_L$ and $k_T$ (and not the redefined momenta $\kappa_L$ and $\vec{\kappa}_T$), since we are integrating over $\omegaup$, $k_L$ and $k_T$, and utilizing the superposition principle to obtain the general solution with right behavior at the horizon cap. However $\mathcal{M}_n$ depend on $\kappa_L$ and $\vec{\kappa}_T$, and therefore they have non-trivial derivatives w.r.t. $\sigma$. The coefficients $p$ and $q$, which are functions of the ordinary momenta ($k_L$ and $\vec{k}_T$), and coefficients of the ingoing and outgoing modes respectively (see more below), are determined by imposing Dirichlet boundary conditions at the two boundaries.

The structure of $\mathcal{M}_n$ in \eqref{Eq:Mnc} is determined as follows. Note that at the zeroth order, i.e. at $n=0$, we obtain exactly the solutions of the static black brane as noted above. We can choose the basis of solutions which are ingoing and outgoing at the horizon, and satisfying the normalization conditions given by \eqref{Eq:in-out-static}. Explicitly, near the perturbative horizon $v = \varepsilon_0^{1/d}$ where the horizon cap is located, they take the forms\footnote{$\phi_{0,in} \rightarrow 1$ and $\left(\varepsilon_0^{-1/d}-v\right)^{-i\frac{2\omegaup }{d\varepsilon_0^{1/d}}}\phi_{0,out} \rightarrow 1$ as $v\rightarrow \varepsilon_0^{-1/d}$ are just normalizations.}
\begin{eqnarray}\label{Eq:NH0}
\phi_{0,in} &=& 1+ \sum_{k=1}^\infty p_{0,k} \left(\varepsilon_0^{-1/d} -v\right)^k, \nonumber\\ \phi_{0,out} &=&\left(\varepsilon_0^{-1/d}-v\right)^{i\frac{2\omegaup }{d\varepsilon_0^{1/d}}}\left( 1+ \sum_{k=1}^\infty q_{0,k} \left(\varepsilon_0^{-1/d} -v\right)^k\right),
\end{eqnarray}
respectively. Note we have used \eqref{Eq:beta-weyl} to set $\beta\omegaup/(2\pi) = 2\omegaup/(d\varepsilon_0^{1/d})$.
For $n\geq 1$, $\phi_{n,in}$ and $\phi_{n,out}$ represent corrections to these zeroth order solutions as discussed below. In \eqref{Eq:Mnc}, we have assumed that $\phi_{n,in}$ and $\phi_{n,out}$ have the same behavior at the horizon cap $v= v_h= \varepsilon_0^{-1/d}$ for $n\geq 1 $as in the case of the equilibrium (or equivalently at the zeroth order). This indeed turns out to be the case with appropriate residual gauge fixing as mentioned before and explicitly shown below.\footnote{In the following section, we show that preserving the near-horizon behavior to all orders is required for satisfying many consistency conditions, such as ensuring that the retarded correlation function is given by causal response.}

At higher orders, the equations determining $\phi_{n,in}$ and $\phi_{n,out}$ can be obtained from expanding $(\Box -m^2)\Phi = 0$ in the late proper time expansion after substituting $k_L$ and $\vec{k}_T$ in \eqref{Eq:Ansatz1} by the redefined momenta \eqref{Eq:kLkTredef},  and isolating the $\sigma^{-n}$ term. We obtain that $\phi_{n,in}$ and $\phi_{n,out}$ satisfies the linear inhomogeneous ordinary differential equations
\begin{equation}\label{Eq:nthordereoms}
    \mathcal{D} \phi_{n,in} = \mathcal{S}_{n,in}, \quad \mathcal{D} \phi_{n,out} = \mathcal{S}_{n,out}
\end{equation}
with $\mathcal{D}$ being the same operator \eqref{Eq:D}, which is simply that corresponding to the Klein-Gordon equation for the static black brane at temperature $\beta^{-1}$ given by \eqref{Eq:beta-weyl}, at all orders. The sources $\mathcal{S}_{n,in}$ and $\mathcal{S}_{n,out}$ are functions of $v$, $\omegaup$, $\kappa_L$ and $\vec{\kappa}_T$, and depend on  $\phi_{m,in}$ and $\phi_{m,out}$ respectively for $m<n$. Both of these sources also depend on the functions $a_{(i)}$, $k_{(i)}$ and $l_{(i)}$, which appear in the late proper time expansion \eqref{Eq:AsympGen2} of the background metric, with $i\leq n$. Note that the source is linear in the bulk field $\Phi$, and therefore splits into $\mathcal{S}_{n,in}$ and $\mathcal{S}_{n,out}$ at each order in the $\sigma^{-1}$ expansion for $n\geq 1$. Since $\phi_{n,in}$ for $n\geq 1$ corrects $\phi_{0,in}$, we include the particular solution with only $\phi_{m,in}$ (and $m<n$) appearing in the source term $\mathcal{S}_{n,in}$ in it. The particular solutions sourced by $\phi_{m,out}$ with $0\leq m<n$ which appear in $\mathcal{S}_{n,out}$ are added by definition to $\phi_{n,out}$ similarly. 

The regularity of the horizon cap implies that at first and higher orders in the proper time expansion, we should have
\begin{eqnarray}\label{Eq:NHn0}
\lim_{v\rightarrow\varepsilon_0^{-1/d}}\left(\varepsilon_0^{-1/d} -v\right)^{-i\frac{2\omegaup }{d\varepsilon_0^{1/d}}}\phi_{n,out}(v, \omegaup, \kappa_L, \vec{\kappa}_T) &=& -i\gamma_{n,out}(\omegaup,\kappa_L, \vec{\kappa}_T),\nonumber\\
\lim_{v\rightarrow\varepsilon_0^{-1/d}}\phi_{n,in}(v, \omegaup, \kappa_L, \vec{\kappa}_T) &=& -i\gamma_{n,in}(\omegaup,\kappa_L, \vec{\kappa}_T),
\end{eqnarray}
with $\gamma_{n,out}$s and $\gamma_{n,in}$s as new functions of $\omegaup$, $\kappa_L$ and $\vec{\kappa}_T$ for $n\geq1$. This will ensure that the analytical dependence of the field on $v$ at the horizon cap $v=v_h = \varepsilon_0^{-1/d}$ is the same to all orders in the proper time expansion. Note that at the zeroth order, the analogous conditions for $\phi_{0,in}$ and $\phi_{0,out}$ are simply given by $1$ on the RHS in both cases by choice (see \eqref{Eq:NH0}). We will show that the $\gamma_{n,out}$s and $\gamma_{n,in}$s can be determined uniquely for $n\geq 1$ via horizon cap regularity and field theory identities. 

Firstly, note that \eqref{Eq:NHn0} implies that  $\phi_{n,in} $ and $\phi_{n,out} $ can be written in the form
\begin{eqnarray}\label{Eq:inoutnthorder}
\phi_{n,in}  = \phi_{n(p),in}  - i \gamma_{n,in} \,\,\phi_{0,in}, \quad \phi_{n,out}  = \phi_{n(p),out}  - i \gamma_{n,out} \,\,\phi_{0,out}
\end{eqnarray}
for $n\geq 1$ such that $\phi_{n(p),in}$ and $\phi_{n(p),out}$ are the particular solutions of the inhomogeneous ordinary differential equations \eqref{Eq:nthordereoms}.  Both $\phi_{n(p),in}$ and $\phi_{n(p),out}$ are determined by the sources $\mathcal{S}_{n,in}$ and $\mathcal{S}_{n,out}$, and are proportional to the coefficients $p$ and $q$, respectively (i.e. they vanish when $p=q=0$). Near the horizon cap $v\approx \varepsilon_0^{1/d}$, we explicitly find that $\phi_{n(p),in}$ and $\phi_{n(p),out}$ behave as
\begin{eqnarray}\label{Eq:NHn1}
\lim_{v\rightarrow\varepsilon_0^{-1/d}}\left(\varepsilon_0^{-1/d} -v\right)^{-i\frac{2\omegaup }{d\varepsilon_0^{1/d}}}\phi_{n(p),out} &=& \mathcal{O}\left(\varepsilon_0^{-1/d} -v\right),\nonumber\\
\lim_{v\rightarrow\varepsilon_0^{-1/d}}\phi_{n(p),in} &=& \mathcal{O}\left(\varepsilon_0^{-1/d} -v\right),
\end{eqnarray}
respectively. Equivalently, 
\begin{eqnarray}\label{Eq:NHn2}
\phi_{n(p),out} &=& \left(\varepsilon_0^{-1/d} -v\right)^{i\frac{2\omegaup }{d\varepsilon_0^{1/d}}}\sum_{k=1}^\infty q_{n,k} \left(\varepsilon_0^{-1/d} -v\right)^k ,\nonumber\\
\phi_{n(p),in} &=&  \sum_{k=1}^\infty p_{n,k} \left(\varepsilon_0^{-1/d} -v\right)^k,
\end{eqnarray}
where the coefficients should be determined by the equations of motion. For the outgoing solution, we find that this behavior is possible only when the residual gauge parameter $\alpha_n$ appearing in the background metric and $\gamma_{n,out}$ are chosen appropriately to cancel double and single poles appearing in the equation of motion \eqref{Eq:nthordereoms} at the horizon cap for each $n\geq1$. Furthermore, $\alpha_n$s are simply numerical constants (as they are defined to be), and $\gamma_{n,out}$s are linear functions of $\omegaup/\varepsilon_0^{1/d}$ only. Thus the outgoing solutions appearing in $\mathcal{M}_n$ in our ansatz \eqref{Eq:Ansatz1} are determined uniquely. As discussed before, and will be explicitly shown again in the next section this completely determines the advanced propagator of the Bjorken flow. The nonequilibrium retarded propagator then is also determined uniquely, since even out of equilibrium, the advanced and retarded propagators are related by the exchange of the spatial and temporal arguments. Utilizing this, we can determine $\gamma_{n,in}$s uniquely as well for $n\geq1$ as will be shown in the next section. In this section, we will focus mainly on the outgoing mode. 

It is easy to see that \eqref{Eq:NHn1} (equivalently \eqref{Eq:NHn2}) implies that $\gamma_{n,in}$ and $\gamma_{n,out}$ are simply the coefficients of the homogeneous solutions of the equations of motion \eqref{Eq:nthordereoms} for $n\geq 1$. Therefore $\gamma_{n,in}$ and $\gamma_{n,out}$ appear in $\mathcal{S}_{m,in}$ and $\mathcal{S}_{m,out}$ respectively for $m>n$. We will illustrate by example how requiring the regularity condition \eqref{Eq:NHn1} (equivalently \eqref{Eq:NHn2}) at the $n$-th order determines $\gamma_{n-1,out}$ along with the gauge parameter $\alpha_n$ (which is a constant) recursively.

Unlike the case of the outgoing mode, the ingoing mode is always analytic at the horizon cap. Therefore, we need to use consistency conditions for the Schwinger-Keldysh correlation functions to determine $\gamma_{n,in}$s for $n\geq1$. However, $\gamma_0$ in the ansatz \eqref{Eq:Ansatz1} appears in both the ingoing and outgoing modes. Our construction passes a significant consistency test that the same function $\gamma_0(\omegaup/\varepsilon_0^{1/d})$ determines the homogeneous transients (sourceless solutions which are ingoing at the horizon) with the argument $\omegaup/\varepsilon_0^{1/d}$ taking values corresponding to the appropriately rescaled quasi-normal mode frequencies of the static black hole, as will be discussed in Section \ref{sec:cs}. This is remarkable as we determine the function $\gamma_0$ analytically by imposing the regularity condition \eqref{Eq:NHn2} on the outgoing mode at the horizon cap.

In what follows, we illustrate how we determine the outgoing mode and the residual gauge fixing uniquely in the case of $AdS_5$. Let us first see how we determine $\gamma_{0}$ and $\alpha_1$ at the first order in the proper time expansion. This requires the first-order correction to the background metric given by \eqref{Eq:a1l1k1} and \eqref{Eq:g}. We find that the equation of motion \eqref{Eq:nthordereoms} for $\phi_{1,out}$ explicitly takes the following form near the horizon cap $v\approx \varepsilon_0^{1/4}$ up to overall proportionality factors:
\begin{align} \label{poleqn}
    &\frac{1}{(\varepsilon_0^{1/4}v-1)^2}\frac{\omegaup}{4\varepsilon_0^{1/4}}(\alpha_1 +3)(\omegaup\varepsilon_0^{-1/4}+ 2i)\nonumber\\
    &+\frac{1}{(\varepsilon_0^{1/4}v-1)}\frac{i\omegaup}{\varepsilon_0^{1/4}}\Bigg(\frac{2 \kappa_L^2 + 2\kappa_T^2-m^2 - 3\omegaup^2 + 240\omegaup\varepsilon_0^{1/4} }{24\varepsilon_0^{1/2}}(\alpha_1+3)\nonumber\\&\qquad\qquad\qquad\qquad+\frac{(16+ 5\alpha_1)\omegaup^2- 4\gamma_0\omegaup\varepsilon_0^{1/4}}{\varepsilon_0^{1/2}}\Bigg) + \cdots,
\end{align}
with $\cdots$ denoting terms which are regular at $v = \varepsilon_0^{1/4}$ if $\phi_{1,out}$ is of the form \eqref{Eq:NHn2}. See Appendix \ref{sec:AdS5} for more details. We readily see that To have a solution of the desired form \eqref{Eq:NHn1} we must impose
\begin{equation} \label{alpha1}
    \alpha_1 = - 3, \quad \gamma_0 = \frac{\omegaup}{4\varepsilon_0^{1/4}}
\end{equation}
so that the double and single pole terms of the equation of motion appearing at the horizon cap vanish. The double pole term determines $\alpha_1$ and the single pole term determines $\gamma_0$. As claimed before, we find that $\gamma_0$ is indeed a simple linear function of $\omegaup/\varepsilon_0^{1/4}$ (rather just proportional to it) while the gauge parameter $\alpha_1$ is a numerical constant as it should be.  

At the second order in the proper time expansion, similarly $\alpha_2$ and $\gamma_1$ are determined by the vanishing of the double and single pole terms in the equation of motion for $\phi_{2,out}$ respectively. Here we have to utilize the explicit second-order correction to the background metric given in Appendix \ref{sec:BJ2}. Explicitly,
\begin{equation} \label{alpha2}
    \alpha_2 = \frac{1}{72}\left(20 + 9\pi -12 \ln 2\right), \quad \gamma_1 = \frac{1}{16}\left(4i -\frac{9\pi +4 - 24 \ln 2}{9}\frac{\omegaup}{\varepsilon_0^{1/4}}\right).
\end{equation}
We find once again $\alpha_2$ is just a numerical constant as it should be and $\gamma_1$ is a linear function of $\omegaup/\varepsilon_0^{1/4}$. 

It is indeed crucial that the $\gamma_{n,out}$s for $n\geq0$ are functions of $\omegaup$ only and are independent of $\kappa_L$ and $\vec{\kappa}_T$ which depend on $\sigma$. Otherwise, the central assumption \eqref{Eq:NHn0} (and thus \eqref{Eq:NHn2}) are not valid for the outgoing mode at the horizon cap, and should be corrected by log terms. The latter would have implied that the behavior near the horizon cap at first and higher orders is different from the zeroth order which is the same as in thermal equilibrium. Note that the ansatz for the ingoing mode in \eqref{Eq:NHn2} remains valid to all orders even if $\gamma_{n,in}$s depend on $\kappa_L$ and $\vec{\kappa}_T$. The coefficients $p_{n,k}$ in \eqref{Eq:NHn2} also involve derivatives of $\gamma_{m,in}$s w.r.t. $\kappa_L$ and $\vec{\kappa}_T$ with $m<n$. See Appendix \ref{sec:AdS5} for more details.

Finally, we have explicitly verified that the values of the residual gauge parameters $\alpha_1$ and $\alpha_2$ are such that the event horizon is pinned to the horizon cap $v=v_h = \varepsilon_0^{-1/d}$ at the first and second orders respectively for both $AdS_4$ and $AdS_5$. Note that the apparent horizon differs from the event horizon from second order onwards, so the evolving apparent horizon is behind the horizon cap. See Appendix \ref{sec:horizons} for details. We expect that this feature persists to all orders so that although the interior of the event horizon is excised, the full double-sheeted geometry with the horizon cap still covers the entire bulk regions which can send signals to the boundary.\footnote{Although we do not have a rigorous proof to all orders, we believe that this follows from the general result that the event horizon is generated by null geodesics which also determine the singularities arising in the equation of motion at the horizon cap.} This feature mirrors the causal nature \cite{Berges:2004yj} of the Schwinger-Dyson equations for the correlation functions in the field theory. We will discuss more about the consistency of this result in Sec \ref{sec:cs}.

The quadratic on-shell action for the bulk scalar field is the sum of three pieces, namely $S_{in-in}$ and $S_{out-out}$ which are quadratic in the ingoing and outgoing modes respectively, and the cross-term $S_{in-out}$. As in the thermal case discussed in Section \ref{sec:CGL}, $S_{in-in} =0$. The ingoing mode is analytic at the horizon and the contributions from the forward and backward arms of the radial contour cancel out. Once again this is required for consistency, as if we keep only the ingoing mode by setting $q=0$ in \eqref{Eq:Ansatz1}, then $J_1 = J_2$ and $W[J_1 = J_2] = 0$ for an arbitrary initial (non-thermal) state. (Recall $W$ is identified with $iS_{on-shell}$.) The cross term $S_{in-out}$ has a branch point at the horizon cap and the integration over the radial contour results in the two boundary terms like in the thermal case discussed in Section \ref{sec:CGL}. $S_{out-out}$ potentially has a single pole $(v_h- v)^{-1}$ divergence which we denote as $S_\epsilon$. Explicitly,
\begin{eqnarray}\label{Eq:SepsilonBJ}
    S_\epsilon &\propto& \int {\rm d}\omegaup \int {\rm d}k_L \int{\rm d}^{d-2}k_T \oint_{\epsilon}\, {\rm d}v \sqrt{-G}\Bigg(G^{vv}\partial_v \phi_{n,out}^*\partial_v \phi_{n,out}\qquad\qquad \nonumber\\&&+  G^{v\sigma}\left(\partial_v \phi_{n,out}^*\partial_\sigma \phi_{n,out}+ \partial_v \phi_{n,out}\partial_\sigma \phi_{n,out}^*\right)+\cdots\Bigg)
\end{eqnarray}
at the $n$-th order in the late proper time expansion. We have verified that $S_\epsilon = 0$ for the solution with the regular behavior at the horizon cap given by \eqref{Eq:NHn0}, obtained for the appropriate choices of $\alpha_n$ and $\gamma_{n-1,out}$ as discussed above. One can also check that terms like $\phi_{k,out}^*~\phi_{l,out}$ with $k+l\leq 2n$ also do not contribute to $S_\epsilon$. However, our arguments for $S_{out-out} = 0$ for the thermal case in Section \ref{sec:CGL} do not go through here since they rely on the KMS boundary condition. Nevertheless, since the pole at the horizon vanishes, $S_{out-out}$ is the sum of two boundary terms as well. 

{Since we preserve the horizon cap regularity at each order, the analytic continuation of the outgoing mode across the horizon cap at each order works exactly in the same way as in the case of thermal equilibrium, i.e. $\phi_{n,out}(\omegaup, v, \kappa_L, \kappa_T)$ picks up a factor of $e^{\beta\omegaup}$ as evident from \eqref{Eq:inoutnthorder} and \eqref{Eq:NHn2}. Repeating the argument in Section \ref{sec:CGL}, each boundary term in $S_{out-out}$ involves one $\phi_b(\omega, \kappa_L, \kappa_T)$ and another $\phi_b(-\omega, \kappa_L, \kappa_T)$ (with $b$ standing for the boundary value) or the corresponding boundary values of the radial derivatives. Also, the contribution from the forward and backward arms come with opposite signs. While $\phi_b(\omega, \kappa_L, \kappa_T)$ picks up a $e^{\beta\omega}$ factor via analytic continuation across the horizon cap, $\phi_b(-\omega, \kappa_L, \kappa_T)$ picks up a $e^{-\beta\omega}$ factor, and these multiply to unity. Therefore, the boundary term contributions from the two arms cancel out resulting in $S_{out-out} =0$.}

Finally, as in the thermal case, the on-shell action is simply the sum of two boundary terms (including the counter-terms for holographic renormalization) obtained from $S_{in-out}$. We can then readily differentiate this on-shell action to obtain the Schwinger-Keldysh correlation functions of the Bjorken flow.  

As shown in the following section, the boundary correlation functions obtained solely from $S_{in-out}$ ensure that the (nonequilibrium) retarded correlation function is given always by the linear causal response. {Furthermore, a major consistency check is that we reproduce the homogeneous transients as poles of the retarded Green's function in complexified $\omegaup$ not only at the leading order as computed {in \cite{Janik:2006gp}}, but also at the subleading orders as shown in Appendix \ref{sec:transients} and discussed further below.} We will discuss more consistency tests in Section \ref{sec:cs}.

\section{The real time out-of-equilibrium correlation functions}\label{sec:CFs}
In the previous section, we have shown that via the bulk diffeomorphism dual to the Weyl transformation we can obtain the solution of the bulk scalar field which has the same analytic thermal nature of the near-horizon modes to all orders in the (hydrodynamic) large proper time expansion. Furthermore, this solution is unique up to the coefficients that give the leading near-horizon behavior of the ingoing modes, and the on-shell action to all orders is the product of the ingoing and outgoing modes, as the terms which are quadratic in the ingoing modes or the outgoing modes cancel exactly between the radial arms corresponding to the forward and backward parts of the time contour due to the thermal nature of their leading near-horizon behavior. Here we will compute the Schwinger-Keldysh correlation functions from the on-shell action of the bulk scalar field in an appropriate late-time expansion. We will show that the correlation functions are consistent with the field theory identities which hold even for out-of-equilibrium states and one of them can be used to fix the coefficients that give the leading near-horizon behavior of the ingoing modes and thus determine the bulk scalar field solution uniquely. Furthermore, we will show that the all-order hydrodynamic correlation function has a matrix factorized form giving a bilocal generalization of the corresponding form of the thermal correlation functions.

In Sec. \ref{Sec:IdNeq}, we will discuss the general identities satisfied by the Schwinger-Keldysh correlation functions and discuss how they transform under Weyl transformations. In Sec. \ref{Sec:GenStr}, we will obtain the Schwinger-Keldysh correlation functions from the on-shell action and establish its matrix factorized form which is consistent with the field theory identities. We will also show that one of the latter identities can determine the bulk solution of the dual scalar field uniquely and therefore the Schwinger-Keldysh correlation functions. In Sec. \ref{Sec:PFSK}, we will obtain the correlation functions in the perfect fluid limit and show that it is thermal after space-time reparametrizations up to overall Weyl factors. In Sec. \ref{Sec:First-Order-Corr}, we will establish the systematic late-time expansion of the correlation functions. Finally, in Sec. \ref{sec:cs}, we will discuss consistency checks including the feature find that the retarded propagator given by the ingoing modes reproduces the normalizable bulk solutions at complex frequencies (which map to quasi-normal modes of the static black brane \cite{Janik:2006gp}) with vanishing sources to all orders although the relevant phase factors at first and higher orders are determined as functions of the frequency via the near-horizon behavior of the outgoing modes at real frequencies. The latter implies non-trivially that one can indeed obtain the normalizable bulk solutions dual to the nonequilibrium transients of the dual field theory to all orders in the large proper time expansion from the retarded correlation function generalizing how we can obtain the quasinormal modes (thermal transients) as poles of the thermal retarded correlation function.

\subsection{Some useful relations and their consequences}\label{Sec:IdNeq}
Some crucial identities are valid for the Schwinger-Keldysh correlation functions even in out-of-equilibrium states. The first such identity of interest is
\begin{eqnarray}\label{Eq:GRG11G12}
G_R(x_1, x_2) &=& -i \theta(x_1^0-x_2^0) {\rm Tr}(\rho [\hat{O}(x_1), \hat{O}(x_2)]) \nonumber\\
              &=& -i \left( {\rm Tr}(\rho\, {\rm T}(\hat{O}(x_1) \hat{O}(x_2))) - {\rm Tr}(\rho \hat{O}(x_2) \hat{O}(x_1))\right)\nonumber\\
              &=& G_{11}(x_1, x_2) - G_{12}(x_1, x_2).
\end{eqnarray}
Similarly, we can arrive at the identity    
\begin{eqnarray}\label{Eq:GAG11G21}
G_A(x_1, x_2) &=& i \theta(x_2^0-x_1^0) {\rm Tr}(\rho [\hat{O}(x_1), \hat{O}(x_2)]) \nonumber\\
              &=& -i \left( {\rm Tr}(\rho\, {\rm T}(\hat{O}(x_1) \hat{O}(x_2))) - {\rm Tr}(\rho \hat{O}(x_1) \hat{O}(x_2))\right)\nonumber\\
              &=& G_{11}(x_1, x_2) - G_{21}(x_1, x_2).
\end{eqnarray}
These identities give the retarded and advanced correlation functions in the usual Schwinger-Keldysh basis. The definitions of the correlation functions also imply that in any arbitrary state
\begin{equation}\label{Eq:SchKelSym}
    G_{11}(x_1,x_2) = G_{11}(x_2, x_1), \quad G_{12}(x_1,x_2) = G_{21}(x_2, x_1), \quad G_{22}(x_1,x_2) = G_{22}(x_2, x_1).
\end{equation}
Finally, it is obvious also that in any arbitrary (nonequilibrium) state
\begin{equation}\label{Eq:GRGAcon}
    G_R(x_1, x_2) = G_A(x_2, x_1).
\end{equation}
We will show that these identities imply the following for the horizon cap of the Bjorken flow. Firstly, will use \eqref{Eq:GRG11G12} to show that the retarded correlation function is always obtained from the ingoing mode, and \eqref{Eq:GAG11G21} to show that the advanced correlation function is always obtained from the outgoing mode. Second, due to \eqref{Eq:GRGAcon}, we can uniquely determine $\gamma_{n,in}$ for $n\geq 1$. These coefficients cannot be determined by the regularity condition \eqref{Eq:NHn0} of the horizon cap unlike $\gamma_{n,out}$ as discussed previously, but together with $\gamma_{n,out}$ they provide unique solution of the bulk scalar field for specified sources at the two boundaries via our ansatz \eqref{Eq:Ansatz1}. We will also see that by satisfying \eqref{Eq:GRGAcon} we can also ensure the validity of    \eqref{Eq:SchKelSym}.

Another important issue is the Weyl transformation. Consider a background metric $g_{\mu\nu}$ and its Weyl rescaled version $\Omega^2(x)g_{\mu\nu}$ such that $\Omega(x)$ is a non-vanishing function. Disregarding Weyl anomaly, the correlations functions of a scalar primary operator $\hat O$ of conformal dimension $\Delta_{O}$ in a conformal field theory living in these two background metrics would be related by
\begin{equation}
    {\rm Tr}(\tilde\rho \hat{O}(x_1) \cdots \hat{O}(x_n))_{\Omega^2 g_{\mu\nu}} =\Omega^{-\Delta_{O}}(x_1) \cdots\Omega^{-\Delta_{O}}(x_n) {\rm Tr}(\rho \hat{O}(x_1) \cdots \hat{O}(x_n))_{g_{\mu\nu}}
\end{equation}
where $\tilde\rho  = U^\dagger\rho U$ with $U$ denoting the unitary operator implementing the Weyl transformation. In the context of the holographic Bjorken flow, the states $\rho$ and $\tilde\rho$ would be described by the holographic geometries whose boundary metrics are the Milne metric and its Weyl rescaled version respectively, and the respective energy-momentum tensors are also appropriately Weyl transformed including the correct holographic Weyl anomaly. This has been described in detail already in Section \ref{sec2.1}.

We have seen that it is easier to implement the out-of-equilibrium horizon cap prescription in the Weyl transformed geometry (state) in which the temperature and entropy density become a constant at late proper time, and the Klein-Gordon equation in the bulk assumes the form of that in a static black brane at the leading order. It is convenient to compute the correlation functions in this background first, and then go back to the usual Bjorken flow background by Weyl transformation. In this case, we should use 
$$\Omega^{\Delta_{O}}(\tau_1)\Omega^{\Delta_{O}}(\tau_2)\widetilde{G}(x_1,x_2) $$
with $\widetilde{G}(x_1,x_2)$ being the correlation functions computed after Weyl rescaling to obtain the correlation functions in the usual Bjorken flow background with $\Omega$ given by \eqref{Eq:TimeRepa} and \eqref{Eq:Omega}. We disregard Weyl anomalies since they are not state-dependent (same as in vacuum).

Note that the full correlation functions to all orders in the late proper time expansion should depend on $\hat\zeta_1$ and $\hat\zeta_2$ only through the relative separation $\hat{\zeta}_1- \hat{\zeta}_2$ in rapidity due to the boost invariance of the background, and similarly only on $\vert \vec{x}_{\perp 1} -\vec{x}_{\perp 2}\vert $ due to translation and rotation symmetries of the background in the transverse spatial plane.

{We note that we can practically use the Weyl transformation to compute the correlation functions is because we are in the boost invariant hydrodynamic regime where the state is characterized only by a proper time-dependent temperature. A Weyl rescaling makes the temperature constant at late time leading us to set-up a perturbative derivative expansion controlled by a fixed dimensionful parameter. However, there will be contributions to the correlation functions in the form of a generalized trans-series which depends on the initial conditions, and are not constructed from hydrodynamic data. Such contributions which represent transients leading to hydrodynamization of correlation functions are discussed in Section \ref{sec:IC}. In this section, we discuss the Schwinger-Keldysh correlation functions in the hydrodynamic regime using a well-defined late proper time expansion.}

\subsection{General structure of the hydrodynamic correlation functions}\label{Sec:GenStr}
It is useful to define a new variable
\begin{equation}
    s(\sigma) = \frac{d-1}{d-2}\sigma - \tilde\gamma_0\varepsilon_0^{-1/d}\ln\left(\sigma/\tau_0\right)
\end{equation}
with $\tilde\gamma_0 = \gamma_0\varepsilon_0^{1/d}/\omegaup$ being a numerical constant (independent of $\omegaup$). We have already seen that the smoothness of the horizon cap requires that in the case of $AdS_5$, $\tilde\gamma_0 = 1/4$ (see \eqref{alpha1}).\footnote{In the case of $AdS_4$, $\tilde\gamma_0 = 2/3$.} In terms of this variable, we can readily see from \eqref{Eq:Ansatz1} that the non-normalizable mode of the scalar field in the gravitational background dual to the Bjorken flow takes the form:
\begin{eqnarray}
\begin{pmatrix}
J_1 (\sigma, \hat\zeta, \vec{x}_\perp)\\J_2 (\sigma, \hat\zeta, \vec{x}_\perp)
\end{pmatrix} &=& \int{\rm d}\omegaup {\rm d}k_L{\rm d}^{d-2}k_T \,\, e^{-i\omegaup s(\sigma)} e^{i\left(k_L \hat\zeta + \vec{k}_T\cdot \vec{x}_\perp\right)}\nonumber\\ &&
\sum_{n=0}^\infty \left(\varepsilon_0^{1/d}\sigma\right)^{-n}\mathcal{S}_n(\omegaup, \kappa_L, \vec{\kappa}_T)\cdot
\begin{pmatrix}
p(\omegaup, k_L, k_T)\\q(\omegaup, k_L, k_T)
\end{pmatrix}
\end{eqnarray}
and similarly, the expectation value of the dual operator (up to state-independent contact terms) is 
\begin{eqnarray}
\begin{pmatrix}
\langle O_1(\sigma, \hat\zeta, \vec{x}_\perp)\rangle  \\\langle O_2 \rangle (\sigma, \hat\zeta, \vec{x}_\perp)
\end{pmatrix} &=& (2\Delta_O -d)\int{\rm d}\omegaup {\rm d}k_L{\rm d}^{d-2}k_T \,\, e^{-i\omegaup s(\sigma)} e^{i\left(k_L \hat\zeta + \vec{k}_T\cdot \vec{x}_\perp\right)}\nonumber\\ &&
\sum_{n=0}^\infty \left(\varepsilon_0^{1/d}\sigma\right)^{-n}\mathcal{R}_n(\omegaup, \kappa_L, \vec{\kappa}_T)\cdot
\begin{pmatrix}
p(\omegaup, k_L, k_T)\\q(\omegaup, k_L, k_T)
\end{pmatrix}
\end{eqnarray}
where $\mathcal{S}_n$ and $\mathcal{R}_n$ can be defined from the asymptotic expansion of $\mathcal{M}_n$:
\begin{equation}
    \mathcal{M}_n(v,\omegaup, \kappa_L, \vec{\kappa}_T) = v^{d-\Delta_O}\left(\mathcal{S}_n(\omegaup, \kappa_L, \vec{\kappa}_T)+\cdots\right) + v^{\Delta_O}\left(\mathcal{R}_n(\omegaup, \kappa_L, \vec{\kappa}_T)+\cdots\right),
\end{equation}
and the labels $1$ and $2$ stand for the sheets of the bulk space-time ending on the forward and backward arms of the Schwinger-Keldysh contour respectively. 

It is also useful to define
\begin{eqnarray}
   \mathcal{R}(\sigma,\omegaup, \kappa_L, \vec{\kappa}_T) &=& \sum_{n=0}^\infty \left(\varepsilon_0^{1/d}\sigma\right)^{-n}\mathcal{R}_n(\omegaup, \kappa_L, \vec{\kappa}_T), \nonumber\\ \mathcal{S}(\sigma,\omegaup, \kappa_L, \vec{\kappa}_T) &=& \sum_{n=0}^\infty \left(\varepsilon_0^{1/d}\sigma\right)^{-n}\mathcal{S}_n(\omegaup, \kappa_L, \vec{\kappa}_T).
\end{eqnarray}
Clearly  \eqref{Eq:Mnc} leads to
\begin{eqnarray} \label{SR:matrix}
   \mathcal{S}(\sigma,\omegaup, \kappa_L, \vec{\kappa}_T) &=&\begin{pmatrix}
   a(\sigma,\omegaup, \kappa_L, \vec{\kappa}_T) \,&\, b(\sigma,\omegaup, \kappa_L, \vec{\kappa}_T) \\
   a(\sigma,\omegaup, \kappa_L, \vec{\kappa}_T) \,&\, e^{\beta\omegaup}b(\sigma,\omegaup, \kappa_L, \vec{\kappa}_T)
   \end{pmatrix},\nonumber\\
    \mathcal{R}(\sigma,\omegaup, \kappa_L, \vec{\kappa}_T) &=&\begin{pmatrix}
   A(\sigma,\omegaup, \kappa_L, \vec{\kappa}_T) \,&\, B(\sigma,\omegaup, \kappa_L, \vec{\kappa}_T) \\
   A(\sigma,\omegaup, \kappa_L, \vec{\kappa}_T) \,&\, e^{\beta\omegaup}B(\sigma,\omegaup, \kappa_L, \vec{\kappa}_T)
   \end{pmatrix},
\end{eqnarray}
where 
\begin{eqnarray}
   \phi_{in}(v,\sigma,\omegaup, \kappa_L, \vec{\kappa}_T) &=& \sum_{n=0}^\infty \left(\varepsilon_0^{1/d}\sigma\right)^{-n}\phi_{n,in}(v,\omegaup, \kappa_L, \vec{\kappa}_T), \nonumber\\ \phi_{out}(v,\sigma,\omegaup, \kappa_L, \vec{\kappa}_T) &=& \sum_{n=0}^\infty \left(\varepsilon_0^{1/d}\sigma\right)^{-n}\phi_{n,out}(v, \omegaup, \kappa_L, \vec{\kappa}_T),
\end{eqnarray}
gives us $a(\sigma,\omegaup, \kappa_L, \vec{\kappa}_T)$, $A(\sigma,\omegaup, \kappa_L, \vec{\kappa}_T)$, $b(\sigma,\omegaup, \kappa_L, \vec{\kappa}_T)$ and $B(\sigma,\omegaup, \kappa_L, \vec{\kappa}_T)$ from their following near boundary expansions 
\begin{eqnarray}
   \phi_{in}(v, \sigma,\omegaup, \kappa_L, \vec{\kappa}_T) &=& v^{d-\Delta_O}\left(a(\sigma,\omegaup, \kappa_L, \vec{\kappa}_T)+\cdots\right) + v^{\Delta_O}\left(A(\sigma,\omegaup, \kappa_L, \vec{\kappa}_T)+\cdots\right), \nonumber\\ 
   \phi_{out}(v, \sigma,\omegaup, \kappa_L, \vec{\kappa}_T) &=& v^{d-\Delta_O}\left(b(\sigma,\omegaup, \kappa_L, \vec{\kappa}_T)+\cdots\right) + v^{\Delta_O}\left(B(\sigma,\omegaup, \kappa_L, \vec{\kappa}_T)+\cdots\right).
\end{eqnarray}
It should also be obvious that if we define $a_n$, $A_n$, $b_n$ and $B_n$ via
\begin{eqnarray}
   \phi_{n,in}(v,\omegaup, \kappa_L, \vec{\kappa}_T) &=& v^{d-\Delta_O}\left(a_n(\omegaup, \kappa_L, \vec{\kappa}_T)+\cdots\right) + v^{\Delta_O}\left(A_n(\omegaup, \kappa_L, \vec{\kappa}_T)+\cdots\right), \nonumber\\ 
   \phi_{n,out}(v, \sigma,\omegaup, \kappa_L, \vec{\kappa}_T) &=& v^{d-\Delta_O}\left(b_n(\omegaup, \kappa_L, \vec{\kappa}_T)+\cdots\right) + v^{\Delta_O}\left(B_n(\omegaup, \kappa_L, \vec{\kappa}_T)+\cdots\right).
\end{eqnarray}
then
\begin{eqnarray}
   a(\sigma,\omegaup, \kappa_L, \vec{\kappa}_T) &=& \sum_{n=0}^\infty \left(\varepsilon_0^{1/d}\sigma\right)^{-n}a_n(\omegaup, \kappa_L, \vec{\kappa}_T), \nonumber\\
   A(\sigma,\omegaup, \kappa_L, \vec{\kappa}_T) &=& \sum_{n=0}^\infty \left(\varepsilon_0^{1/d}\sigma\right)^{-n}A_n(\omegaup, \kappa_L, \vec{\kappa}_T),\nonumber\\
   b(\sigma,\omegaup, \kappa_L, \vec{\kappa}_T) &=& \sum_{n=0}^\infty \left(\varepsilon_0^{1/d}\sigma\right)^{-n}b_n(\omegaup, \kappa_L, \vec{\kappa}_T), \nonumber\\ 
   B(\sigma,\omegaup, \kappa_L, \vec{\kappa}_T) &=& \sum_{n=0}^\infty \left(\varepsilon_0^{1/d}\sigma\right)^{-n}B_n(\omegaup, \kappa_L, \vec{\kappa}_T).
\end{eqnarray}
The correlation function can be extracted simply from the on-shell action, which as shown in the previous section, is the sum of the two boundary terms obtained from $S_{in-out}$. The computations are similar to the case of the thermal equilibrium discussed before. The general structure of the correlation function with $\{a,b\} = \{1,2\}$ standing for the forward and backward arms of the Schwinger-Keldysh contour (using \eqref{Eq:sqrt-tilde-g}) is (see Appendix \ref{sec:Deriv} for more details)
\begin{eqnarray}\label{Eq:StructureCentral}
   \widetilde G_{ab}(\sigma_1, \sigma_2, \hat{\zeta}_1- \hat{\zeta}_2 , \vert\vec{x}_{\perp 1} - \vec{x}_{\perp 2}\vert)&=&
\frac{1}{\sqrt{-\widetilde{g}_1}\sqrt{-\widetilde{g}_2}}   \frac{\delta^2 S_{on-shell}[J_1, J_2]}{\delta J_a(\sigma_1, \hat\zeta_1, \vec{x}_{\perp1})\delta J_b(\sigma_2, \hat\zeta_2, \vec{x}_{\perp2})}(-)^{a+b}
   \nonumber\\ &=& \frac{(d-2)^2}{(d-1)^2}
   \int{\rm d}\omegaup {\rm d}k_L{\rm d}^{d-2}k_T \, e^{-i\omegaup(s(\sigma_1)-s(\sigma_2))}\nonumber\\&& e^{i k_L(\hat{\zeta}_1- \hat{\zeta}_2)+i\vec{k}_T\cdot (\vec{x}_{\perp 1} - \vec{x}_{\perp 2})}
   \widehat{G}_{ab}(\sigma_1,\sigma_2,\omegaup,k_L, \vec{k}_T), 
\end{eqnarray}
where (compare with the thermal case \eqref{Eq:hatG} -- recall $\sigma_3$ is the Pauli matrix)
\begin{eqnarray}\label{Eq:StructureCentral-1}
\widehat{G}(\sigma_1,\sigma_2,\omegaup,k_L, \vec{k}_T) &=&  \frac{2\Delta_O-d}{2}\Bigg(s'(\sigma_2)\,\sigma_3~\cdot \mathcal{R}\left(\sigma_1, \omegaup,\kappa_{L1}, \vec{\kappa}_{T1}\right) \cdot~ \mathcal{S}^{-1}\left( \sigma_2,\omegaup,\kappa_{L2}, \vec{\kappa}_{T2}\right) \nonumber\\&& +  ({\rm transpose}, \,\, \sigma_1 \leftrightarrow \sigma_2,\omegaup \rightarrow -\omegaup , \kappa_{L1} \leftrightarrow - \kappa_{L2} , \vec{\kappa}_{T1} \leftrightarrow -\vec{\kappa}_{T2})\Bigg) \nonumber \\
\end{eqnarray}
with
\begin{equation}
    \kappa_{L1} =k_L\frac{\tau_0}{\sigma_1}, \quad \kappa_{L2} = k_L\frac{\tau_0}{\sigma_2}, \quad \vec{\kappa}_{T1}=\vec{k}_T\left(\frac{\tau_0}{\sigma_1}\right)^{-1/(d-2)}, \quad  \vec{\kappa}_{T2}=\vec{k}_T\left(\frac{\tau_0}{\sigma_2}\right)^{-1/(d-2)}.
\end{equation}
The transpose above denotes matrix transposition, $\leftrightarrow$ denotes exchange operation and $\rightarrow$ replacement. The second term in \eqref{Eq:StructureCentral-1} is produced by the symmetrization due to the differentiation in \eqref{Eq:StructureCentral}.

Explicitly, (compare with the thermal case given by Eq \eqref{Eq:Onshell-Action-JO-redux})
\begin{eqnarray}\label{Eq:StructureCentral-2}
   \widehat{G}_{11}(\sigma_1,\sigma_2,\omegaup,k_L, \vec{k}_T) &\equiv& \widehat{G}_R(\sigma_1,\sigma_2,\omegaup,k_L, \vec{k}_T)(1+ n(\omegaup))- \widehat{G}_A(\sigma_1,\sigma_2,\omegaup,k_L, \vec{k}_T)n(\omegaup),\nonumber\\
   \widehat{G}_{12}(\sigma_1,\sigma_2,\omegaup,k_L, \vec{k}_T) &\equiv& \left(\widehat{G}_R(\sigma_1,\sigma_2,\omegaup,k_L, \vec{k}_T)- \widehat{G}_A(\sigma_1,\sigma_2,\omegaup,k_L, \vec{k}_T)\right)n(\omegaup),\\\nonumber
   \widehat{G}_{21}(\sigma_1,\sigma_2,\omegaup,k_L, \vec{k}_T) &\equiv& \left(\widehat{G}_R(\sigma_1,\sigma_2,\omegaup,k_L, \vec{k}_T)- \widehat{G}_A(\sigma_1,\sigma_2,\omegaup,k_L, \vec{k}_T)\right)(1+n(\omegaup)),\\\nonumber
   \widehat{G}_{22}(\sigma_1,\sigma_2,\omegaup,k_L, \vec{k}_T) &\equiv& \widehat{G}_R(\sigma_1,\sigma_2,\omegaup,k_L, \vec{k}_T)n(\omegaup)- \widehat{G}_A(\sigma_1,\sigma_2,\omegaup,k_L, \vec{k}_T)(1+n(\omegaup)),
\end{eqnarray}
with $n(\omegaup) = 1/(e^{\beta\omegaup}-1)$ is the Bose-Einstein distribution and
\begin{eqnarray}\label{Eq:StructureCentral-3}
\widehat{G}_R(\sigma_1,\sigma_2,\omegaup,k_L, \vec{k}_T)  &\equiv&\frac{(d-2)^2}{(d-1)^2}(2\Delta_O-d)\frac{s'(\sigma_1)+s'(\sigma_2)}{2}\nonumber\\&& \frac{A\left(\sigma_1,\omegaup,\kappa_{L1}, \vec{\kappa}_{T1}\right) }{a\left(\sigma_2,\omegaup,\kappa_{L2}, \vec{\kappa}_{T2}\right) },\nonumber\\ \widehat{G}_A(\sigma_1,\sigma_2,\omegaup,k_L, \vec{k}_T)  &\equiv& \frac{(d-2)^2}{(d-1)^2}(2\Delta_O-d)\frac{s'(\sigma_1)+s'(\sigma_2)}{2}\nonumber\\&&\frac{B\left(\sigma_1,\omegaup,\kappa_{L1}, \vec{\kappa}_{T1}\right) }{b\left(\sigma_2,\omegaup,\kappa_{L2}, \vec{\kappa}_{T2}\right) }.
\end{eqnarray}
Above $\equiv$ denotes equality up to terms which vanish after the frequency and momentum integration in \eqref{Eq:StructureCentral} and $\beta$ is given by \eqref{Eq:beta-weyl}.

In order that \eqref{Eq:StructureCentral-2} and \eqref{Eq:StructureCentral-3} follow from \eqref{Eq:StructureCentral-1}, and also for the general identities \eqref{Eq:GRGAcon} and \eqref{Eq:SchKelSym} to be satisfied, we should have
\begin{equation}
    \frac{A\left(\sigma_1,\omegaup,\kappa_{L1}, \vec{\kappa}_{T1}\right) }{a\left(\sigma_2,\omegaup,\kappa_{L2}, \vec{\kappa}_{T2}\right) }\equiv\frac{B\left(\sigma_2,-\omegaup,-\kappa_{L1},- \vec{\kappa}_{T1}\right) }{b\left(\sigma_1,-\omegaup,-\kappa_{L2}, -\vec{\kappa}_{T2}\right) },
\end{equation}
i.e.
\begin{equation}
    \widehat{G}_R\left(\sigma_1,\sigma_2,\omegaup,\kappa_{L1}, \vec{\kappa}_{T1}\right) \equiv \widehat{G}_A\left(\sigma_2,\sigma_1,-\omegaup,-\kappa_{L2}, -\vec{\kappa}_{T2}\right).
\end{equation}
As mentioned, we will show in Section \ref{Sec:First-Order-Corr} that the above can be satisfied by appropriate choices of $\gamma_{n,in}s$ at each order in the late proper time expansion where we go to large values of the average reparametrized proper time $\overline{\sigma} = (\sigma_1 + \sigma_2)/2$ with fixed difference $\sigma_r = \sigma_1 - \sigma_2$. Thus we see that the correlation functions of the hydrodynamic Bjorken flow has a hidden and simple bilocal thermal structure to all orders in the late proper time expansion. We will show this satisfies crucial consistency tests.

Using \eqref{Eq:StructureCentral-2} and \eqref{Eq:StructureCentral-3}, and the identities \eqref{Eq:GRG11G12} and \eqref{Eq:GAG11G21} which are valid out-of-equilibrium, we readily see that the actual retarded and advanced propagators of the Weyl transformed Bjorken flow is
\begin{eqnarray}\label{Eq:GR-GA-all-orders}
 \widetilde G_R(\sigma_1, \sigma_2, \hat{\zeta}_1- \hat{\zeta}_2 , \vert\vec{x}_{\perp 1} - \vec{x}_{\perp 2}\vert)&=&\frac{(d-2)^2}{(d-1)^2}(2\Delta_O-d)\frac{s'(\sigma_1)+s'(\sigma_2)}{2}\nonumber\\&&\int{\rm d}\omegaup {\rm d}k_L{\rm d}^{d-2}k_T \, e^{-i\omegaup(s(\sigma_1)-s(\sigma_2))}\nonumber\\&& e^{i k_L(\hat{\zeta}_1- \hat{\zeta}_2)+i\vec{k}_T\cdot (\vec{x}_{\perp 1} - \vec{x}_{\perp 2})}
    \frac{A\left(\sigma_1,\omegaup,\kappa_{L1}, \vec{\kappa}_{T1}\right) }{a\left(\sigma_2,\omegaup,\kappa_{L2}, \vec{\kappa}_{T2}\right) },\nonumber\\
 \widetilde G_A(\sigma_1, \sigma_2, \hat{\zeta}_1- \hat{\zeta}_2 , \vert\vec{x}_{\perp 1} - \vec{x}_{\perp 2}\vert)&=&\frac{(d-2)^2}{(d-1)^2}(2\Delta_O-d)\frac{s'(\sigma_1)+s'(\sigma_2)}{2}\nonumber\\&&\int{\rm d}\omegaup {\rm d}k_L{\rm d}^{d-2}k_T \, e^{-i\omegaup(s(\sigma_1)-s(\sigma_2))}\nonumber\\&& e^{i k_L(\hat{\zeta}_1- \hat{\zeta}_2)+i\vec{k}_T\cdot (\vec{x}_{\perp 1} - \vec{x}_{\perp 2})}
    \frac{B\left(\sigma_1,\omegaup,\kappa_{L1}, \vec{\kappa}_{T1}\right) }{b\left(\sigma_2,\omegaup,\kappa_{L2}, \vec{\kappa}_{T2}\right) }.
\end{eqnarray}

Therefore, it follows from \eqref{Eq:StructureCentral-3} that indeed the retarded propagator is obtained purely from the out-of-equilibrium ingoing mode and the advanced propagator is also obtained from the out-of-equilibrium outgoing mode as claimed before. Note that at the zeroth order, these results are automatic as it reduces to the thermal case described earlier.\footnote{The bilocal thermal structure of hydrodynamic correlation functions in holography was argued earlier in \cite{Mukhopadhyay:2012hv} via Wigner transform. However the arguments here were less rigorous and relied on the possibility of obtaining correlation functions utilizing only one copy of the nonequilibrium background where the interior of the perturbative horizon could be removed.}

Finally, to obtain the Schwinger-Keldysh correlation function of the Bjorken flow, we should implement the Weyl transformation. Using \eqref{Eq:Omega}, the Weyl transformation finally yields the correlation function of the Bjorken flow:
\begin{equation}
    G(\sigma_1, \sigma_2, \hat{\zeta}_1- \hat{\zeta}_2 , \vert\vec{x}_{\perp 1} - \vec{x}_{\perp 2}\vert) = \left(\frac{\tau_0}{\sigma_1}\right)^{\frac{\Delta_O}{d-2}}\left(\frac{\tau_0}{\sigma_2}\right)^{\frac{\Delta_O}{d-2}}\widetilde{G}(\sigma_1, \sigma_2, \hat{\zeta}_1- \hat{\zeta}_2 , \vert\vec{x}_{\perp 1} - \vec{x}_{\perp 2}\vert).
\end{equation}

\subsection{In the limit of the perfect fluid expansion}\label{Sec:PFSK}

At a very late proper time, the Bjorken flow is simply a perfect fluid expansion. For the Weyl-transformed Bjorken flow which reaches a specific final temperature, the scalar field in the dual gravitational geometry can be mapped to that in the static thermal black brane space-time at the leading order in the late proper time expansion. This naturally implies that the Schwinger-Keldysh correlation functions at late proper time should be related to the corresponding thermal correlation functions via appropriate space-time-reparametrizations. After undoing the Weyl transformation, we should obtain the Schwinger-Keldysh correlation functions of the perfect fluid expansion. 

To obtain the correlation function in the perfect fluid expansion, we first take the limit in $\widetilde{G}$ given by Eqs. \eqref{Eq:StructureCentral}-\eqref{Eq:StructureCentral-3}, where we take the average reparametrized proper time coordinate $\overline\sigma = (1/2)(\sigma_1 + \sigma_2)$ to infinity keeping the relative reparametrized proper time coordinate $\sigma_r = \sigma_1 - \sigma_2$ fixed, and also $\hat{\zeta}_1- \hat{\zeta}_2$ and $\vec{x}_{\perp 1} -\vec{x}_{\perp 2} $ fixed. In this limit, 
\begin{equation}
    \kappa_{L1}, \, \kappa_{L2} \rightarrow \overline{\kappa}_L, \quad \vec{\kappa}_{T1}, \, \vec{\kappa}_{T2} \rightarrow \overline{\vec{\kappa}}_{T}
\end{equation}
where
\begin{equation}
    \overline{\kappa}_L =k_L\frac{\tau_0}{\overline{\sigma}},  \quad  {\overline{\vec\kappa}}_{T}=\vec{k}_T\left(\frac{\tau_0}{\overline{\sigma}}\right)^{-1/(d-2)}.
\end{equation}
As shown in the previous section, both $\phi_{in}$ and $\phi_{out}$ assume the form in the static black brane geometry, and therefore
\begin{equation}
    \widehat G_R \rightarrow \frac{A_0(\omegaup, \overline{\kappa}_L,{\overline{\vec\kappa}}_{T})}{a_0(\omegaup, \overline{\kappa}_L,{\overline{\vec\kappa}}_{T})}, \quad  \widehat G_A \rightarrow \frac{B_0(\omegaup, \overline{\kappa}_L,{\overline{\vec\kappa}}_{T})}{b_0(\omegaup, \overline{\kappa}_L,{\overline{\vec\kappa}}_{T})}~,
\end{equation}
given by the corresponding static black brane results and therefore all Schwinger-Keldysh correlation functions after the momentum integrals shown in \eqref{Eq:StructureCentral} should assume the thermal form, i.e.
\begin{equation}
    \widetilde{G} \rightarrow {G}_{\beta} \left(\sigma_1 - \sigma_2, (\hat{\zeta}_1- \hat{\zeta}_2)\frac{\overline\sigma}{\tau_0}, \vert \vec{x}_{\perp 1} -\vec{x}_{\perp 2}\vert \left(\frac{\tau_0}{\overline{\sigma}}\right)^{\frac{1}{d-2}}\right)
\end{equation}
with $G_\beta$ denoting the thermal correlation functions. To obtain the above form, we change variables in the momentum integrals from $k_L$ and $\vec{k}_T$ to $\overline{\kappa}_L$ and $\overline{\vec{\kappa}}_{T}$. Note the Jacobian of this transformation is identity. We also need to change the integration variable $\omegaup$ to $\omega = \omegaup (d-1)/(d-2)$ which yields a Jacobian $(d-2)/(d-1)$. Furthermore, we have used 
\begin{equation}
    s'(\sigma_1), \, s'(\sigma_2) \rightarrow \frac{d-1}{d-2}.
\end{equation}
The factor $(d-2)^2/(d-1)^2$ in \eqref{Eq:StructureCentral} is canceled by one factor of $(d-1)/(d-2)$ each from the Jacobian and $s'$.

Finally, the Weyl transformation in the late proper time limit gives 
\begin{equation}
   {G} \rightarrow \left(\frac{\tau_0}{\overline{\sigma}}\right)^{\frac{2\Delta_O}{d-2}} {G}_{\beta} \left(\frac{d-1}{d-2}(\sigma_1 - \sigma_2), (\hat{\zeta}_1- \hat{\zeta}_2)\frac{\overline\sigma}{\tau_0}, \vert \vec{x}_{\perp 1} -\vec{x}_{\perp 2}\vert \left(\frac{\tau_0}{\overline{\sigma}}\right)^{\frac{1}{d-2}}\right)
\end{equation}
the Schwinger-Keldysh correlation functions of the perfect fluid expansion. Since the static black brane has full rotational invariance in terms of the reparametrized space-time coordinates, we may also write 
\begin{equation}\label{Eq:Gbeta-pfn}
  {G} \rightarrow \left(\frac{\tau_0}{\overline{\sigma}}\right)^{\frac{2\Delta_O}{d-2}}{G}_{\beta} \left(\frac{d-1}{d-2}(\sigma_1 - \sigma_2), \sqrt{(\hat{\zeta}_1- \hat{\zeta}_2)^2\frac{\overline\sigma^2}{\tau_0^2} + \vert\vec{x}_{\perp 1} -\vec{x}_{\perp 2}\vert^2 \left(\frac{\tau_0}{\overline{\sigma}}\right)^{\frac{2}{d-2}}}\right)
\end{equation}
for the correlation functions in the limit of the late time perfect fluid expansion. The explicit analytic forms of the thermal holographic propagators is known, see \cite{Jana:2020vyx,Dodelson:2022yvn} as for instance.

It is obvious that we are actually resumming over the associated $\overline\sigma$ factors in the spatial factors $\hat{\zeta}_1- \hat{\zeta}_2$ and $\vec{x}_{\perp 1} -\vec{x}_{\perp 2}$. For brevity we will denote the variables as $\sigma_r = \sigma_1 - \sigma_2$ (as defined before) and
\begin{equation}
\widetilde{\zeta}_r = (\hat{\zeta}_1- \hat{\zeta}_2)\frac{\overline\sigma}{\tau_0}, \quad \widetilde{x_\perp}_r = \vert \vec{x}_{\perp 1} -\vec{x}_{\perp 2}\vert \left(\frac{\tau_0}{\overline{\sigma}}\right)^{\frac{1}{d-2}}.
\end{equation}
The late proper time expansion of the Schwinger-Keldysh correlation function then amounts to the following series
\begin{equation}\label{Eq:G-expansion}
    G(x_1, x_2) =\left(\frac{\tau_0}{\overline{\sigma}}\right)^{\frac{2\Delta_O}{d-2}} \sum_{n=0}^\infty \frac{1}{\overline\sigma^n \varepsilon_0^{n/d}}G_n\left(\sigma_r, \widetilde{\zeta}_r, \widetilde{x_\perp}_r\right)+\cdots
\end{equation}
with $\cdots$ denoting trans-series type completion about which we will discuss more below.

Since the choice of $\tau_0$ determines the effective final temperature $T= \beta^{-1}$ via \eqref{Eq:beta-weyl}, let's consider a scaling $\tau_0 \rightarrow \xi \tau_0$ to see if there is any ambiguity in the above result. Note that $G_\beta$ in \eqref{Eq:Gbeta-pfn} depends only on $T\sigma_r$ and $T\sqrt{\widetilde{\zeta}_r^2 +\widetilde{x_\perp}_r ^2}$ as it is a thermal correlation matrix of a CFT. Under this scaling, it is evident from \eqref{Eq:sigma-zeta-def} that $\sigma \rightarrow \xi^{\frac{1}{d-1}}\sigma$, and therefore $\sigma_r \rightarrow \xi^{\frac{1}{d-1}}\sigma_r $, $\overline{\sigma} \rightarrow \xi^{\frac{1}{d-1}}\overline{\sigma}$, $\widetilde{\zeta}_r \rightarrow \xi^{\frac{1}{d-1}}\widetilde{\zeta}_r $ and $\widetilde{x_\perp}_r \rightarrow \xi^{\frac{1}{d-1}}\widetilde{x_\perp}_r$. Also $\epsilon_0 \rightarrow \xi^{-\frac{d}{d-1}}\epsilon_0$ and $T  \rightarrow \xi^{-\frac{1}{d-1}}T$ as discussed earlier in Section \ref{sec:BjRev}. Together these imply that $G_\beta$ and also $G_n$ in \eqref{Eq:G-expansion} are invariant under $\tau_0 \rightarrow \xi \tau_0$ since $T\sigma_r$ and $T\widetilde{\zeta}_r$ and $T \widetilde{x_\perp}_r$ are invariant this scaling. However, the Weyl factor in \eqref{Eq:Gbeta-pf} scales as $\xi^{\frac{2\Delta_O}{d-1}}$ implying that the \textit{dimensionless correlation function} $\overline\sigma^{-2\Delta_0}G$ is invariant under the scaling of $\tau_0$.

\subsection{First and higher orders in the late proper time expansion}\label{Sec:First-Order-Corr}
From the form of the Schwinger-Keldysh correlation functions given by Eqs. \eqref{Eq:StructureCentral}-\eqref{Eq:StructureCentral-3} which are valid to all orders, we can readily go beyond the perfect fluid limit systematically and construct the late proper time expansion \eqref{Eq:G-expansion}. The contribution to the first-order correction comes from the following terms: 
\begin{enumerate}
    \item The phase factor,
    \begin{eqnarray}
e^{-i \omegaup (s(\sigma_1)- s(\sigma_2))} \rightarrow e^{- i \omegaup \sigma_r \frac{d-1}{d-2}} \left( 1 + i \omegaup \tilde{\gamma}_0 \varepsilon_0^{-1/d} \frac{\sigma_r}{\overline{\sigma}} + \mathcal{O}\left(\overline{\sigma}^{-2}\right)\right) ~.
\end{eqnarray} 
\item The Jacobian \begin{equation} s'(\sigma_1)~  \text{and}~  s'(\sigma_2) \rightarrow \frac{d-1}{d-2} - \frac{\tilde{\gamma}_0 }{\overline{\sigma}\varepsilon_0^{1/d}} + \mathcal{O}\left(\overline{\sigma}^{-2}\right)~. \end{equation}
\item Late-time expansion of  the matrices $\mathcal{R}$ and $\mathcal{S}$ given by \eqref{SR:matrix} in $\overline{\sigma}^{-1}$. This also includes  the $\kappa_L$ and $\vec{\kappa}_T$ dependence. For instance, consider any generic function $ f(\kappa_{L1}) $ and $ f(\vec{\kappa}_{T1})$,
\begin{eqnarray}
f(\kappa_{L1}) &=& f(\overline{\kappa}_L) - \partial_{\overline{\kappa}_L} f(\overline{\kappa}_L) \frac{\sigma_r \overline{\kappa}_L}{2 \overline{\sigma}}~, \nonumber \\
f(\vec{\kappa}_{T1}) &=& f\left({\overline{\vec\kappa}}_T\right) + \partial_{\overline{\vec{\kappa}}_T} f\left({\overline{\vec\kappa}}_T\right) \frac{\sigma_r \overline{\vec{\kappa}}_T}{2 (d-2) \overline{\sigma}}~,
\end{eqnarray}
and similarly for $ f(\kappa_{L2}) $ and $ f(\vec{\kappa}_{T2})$.
\end{enumerate}
From \eqref{Eq:StructureCentral-1}, we obtain 
\begin{eqnarray}
\widehat{G}_1 &=& \frac{(2 \Delta_0 -d)}{2} \frac{d-2}{d-1} \sigma_3 . \Big[ \mathcal{R}_0.~ \mathcal{S}_{~~~1}^{-1} +\mathcal{R}_1.~ \mathcal{S}_{~~~0}^{-1}  + \Big(\mathcal{R}_0.~ \partial_{\overline{\kappa}_L} \mathcal{S}_{~~~0}^{-1} - \partial_{\overline{\kappa}_L} \mathcal{R}_0.~ \mathcal{S}_{~~~0}^{-1}\Big) \frac{\sigma_r\overline{\kappa}_L}{2}  \nonumber \\ &-& \Big(\mathcal{R}_0.~ \partial_{\overline{\vec{\kappa}}_T} \mathcal{S}_{~~~0}^{-1} - \partial_{\overline{\vec{\kappa}}_T} \mathcal{R}_0.~ \mathcal{S}_{~~~0}^{-1}\Big) \frac{\sigma_r \overline{\vec{\kappa}}_T}{2(d-2)} - \tilde{\gamma}_0 \varepsilon_0^{-1/d}  \frac{d-2}{d-1} \mathcal{R}_0.~ \mathcal{S}_{~~~0}^{-1} \Big] \left(\omegaup,\overline{\kappa}_L,\overline{\vec{\kappa}}_T\right) \nonumber \\
&+& \left(\text{ transpose},~ \omegaup \rightarrow -\omegaup,~ \overline{\kappa}_L \rightarrow -\overline{\kappa}_L, ~\overline{\vec{\kappa}}_T \rightarrow -\overline{\vec{\kappa}}_T \right).
\end{eqnarray}
An easier way to read off the first order correction is to use \eqref{Eq:StructureCentral-2} and \eqref{Eq:StructureCentral-3}, and consider similar $\overline{\sigma}^{-1}$ expansion of  $ \widehat G_R $ and $ \widehat G_A $. Collecting first order terms from \eqref{Eq:StructureCentral-3}, as for instance, we obtain
\begin{eqnarray}\label{Eq:hatG-fs}
  \widehat G_{R,1} &=& \frac{(2 \Delta_0 -d)}{2} \frac{d-2}{d-1} \Bigg[\frac{2 A_1}{a_0} - \frac{2 A_0 a_1}{a_0^2} + \frac{ A_0 \sigma_r}{a_0^2} \Bigg( \frac{\overline{\vec{\kappa}}_T}{d-2}\partial_{\overline{\vec{\kappa}}_T}- \overline{\kappa}_L \partial_{\overline{\kappa}_L}\Bigg) a_0 \nonumber \\  &+& \frac{\sigma_r}{a_0} \Bigg(\frac{\overline{\vec{\kappa}}_T}{ (d-2)} \partial_{\overline{\vec{\kappa}}_T} - \overline{\kappa}_L \partial_{\overline{\kappa}_L} \Bigg) A_0 
 - \frac{ \tilde\gamma_0 \varepsilon_0^{-1/d} A_0 }{a_0}  \frac{d-2}{d-1}   \Bigg]\left(\omegaup,\overline{\kappa}_L, \overline{\vec{\kappa}}_T\right)
 \end{eqnarray}
 and
 \begin{eqnarray}
 \widehat G_{A,1} &=& \frac{(2 \Delta_0 -d)}{2} \frac{d-2}{d-1} \Bigg[\frac{2 B_1}{b_0} - \frac{2 B_0 b_1}{b_0^2} + \frac{ B_0 \sigma_r}{b_0^2} \Bigg( \frac{\overline{\vec{\kappa}}_T}{d-2}\partial_{\overline{\vec{\kappa}}_T}- \overline{\kappa}_L \partial_{\overline{\kappa}_L}\Bigg) b_0 \nonumber \\  &+& \frac{\sigma_r}{b_0} \Bigg(\frac{\overline{\vec{\kappa}}_T}{ (d-2)} \partial_{\overline{\vec{\kappa}}_T} -\overline{\kappa}_L \partial_{\overline{\kappa}_L} \Bigg) B_0 
 - \frac{ \tilde\gamma_0 \varepsilon_0^{-1/d}  B_0}{b_0}  \frac{d-2}{d-1}  \Bigg]\left(\omegaup,\overline{\kappa}_L, \overline{\vec{\kappa}}_T\right).
\end{eqnarray}
This allows us to finally determine $\gamma_{1,in}$ and similarly $\gamma_{n,in}$ for $n>1$ as follows. After integrating over the frequency and momentum integrals (i.e. doing integrals over $\omegaup$, $k_L$ and $\vec{k}_T$) as in  \eqref{Eq:GR-GA-all-orders}, we obtain the first order correction in the series expansion of the type \eqref{Eq:G-expansion}:
\begin{equation}
    G_R(x_1, x_2) = \sum_{n=0}^\infty \frac{1}{\overline\sigma^{n}\varepsilon_0^{n/d}} G_{R,n} \left(\sigma_r, \widetilde\zeta_r, \widetilde{x_{\perp r}}\right),
\end{equation}
and similarly for $G_A(x_1, x_2)$. For the identities \eqref{Eq:GRG11G12} and \eqref{Eq:GAG11G21} to be satisfied, we therefore need (note that $\widetilde{x_{\perp r}}$ is invariant under $x_1 \leftrightarrow x_2$)
\begin{equation}\label{GR=GA}
    G_{R,n} \left(-\sigma_r, -\widetilde\zeta_r, \widetilde{x_{\perp r}}\right) = G_{A,n}\left(\sigma_r, \widetilde\zeta_r, \widetilde{x_{\perp r}}\right)
\end{equation}
at each $n$, which can be ensured by appropriate choices of $\gamma_{n,in}(\omegaup, \kappa_L, \vec{\kappa}_T)$ at each order for $n\geq1$, since they determine both $A_n$ and $a_n$. In turn, this justifies the bilocal thermal structure \eqref{Eq:StructureCentral-2} and \eqref{Eq:StructureCentral-3} which implies via \eqref{Eq:GRG11G12} and \eqref{Eq:GAG11G21} that the retarded correlation originates purely from the ingoing mode and the advanced correlation function purely from the outgoing mode at all orders. We emphasize that although $\gamma_{n,out}$s are only functions of the frequency $\omegaup$, $\gamma_{n,in}$s are functions of both frequency and momenta.

With $\gamma_{n,in}$ determined we know the RHS of \eqref{Eq:hatG-fs} completely. Doing the frequency and momenta integrals shown in \eqref{Eq:StructureCentral}, we can obtain the first and similarly higher order corrections to all Schwinger-Keldysh propagators.

\subsection{Consistency checks}\label{sec:cs}


The general result for the hidden bilocal thermal structure of the hydrodynamic correlation functions given by Eqs. \eqref{Eq:StructureCentral}-\eqref{Eq:StructureCentral-3} to all orders satisfies a simple consistency check.  When $J_1 = J_2$, i.e. when the sources are the same at the two boundaries, we have only the ingoing solution which is analytic. The on-shell solution also vanishes. When $J_1 = J_2$, we have $$\langle O_1(x_1)\rangle = \int_{x_2} (G_{11}(x_1, x_2) J_1(x_2) - G_{12}(x_1, x_2) J_2(x_2)) = \int_{x_2} G_R(x_1,x_2) J_1(x_2),$$on the forward arm, and thus we see that the response in the forward arm is indeed given by the retarded correlation function, and this also follows from the bilocal thermal structure of the hydrodynamic correlation functions resulting in \eqref{Eq:GR-GA-all-orders} as shown above. We can also see that the analytic ingoing mode is indeed always related to the causal response from the full non-linear evolution (we just consider the forward arm here). There is a unique regular solution to Einstein's gravity minimally coupled to a scalar field corresponding to an initial condition for the bulk scalar field $\Phi(v,\zeta, \vec{x}_\perp)$ at a constant Eddington-Finkelstein time hypersurface  when the boundary source $J(\sigma,\zeta,\vec{x}_\perp)$ is specified for all time in the future. This solution can be obtained via a Chebyshev grid in the radial direction \cite{Chesler:2013lia} which implies analyticity at the horizon, and thus the causal evolution is built out of analytic ingoing modes. Our result then should follow in the linearized limit. The way to implement this explicitly was shown earlier in \cite{Banerjee:2016ray} which is thus reproduced by the horizon cap prescription.\footnote{See \cite{Banerjee:2012uq} for the nonequilibrium retarded correlation function in a hydrodynamic expansion, and \cite{Joshi:2017ump,Cartwright:2019opv} for AdS-Vaidya and states corresponding to quenches, etc. The latter uses the prescription of \cite{Banerjee:2016ray}.}

A more non-trivial consistency check involves the outgoing mode. Consider an outgoing mode with frequency $\omegaup$ and momenta $k_L$ and $\vec{k}_T$ in the Bjorken flow background. Evidently, from the discussion before, to all orders in the late proper time expansion, we have $$ J_2(\omegaup, k_L,\vec{k}_T) = e^{\beta\omegaup} J_1(\omegaup,k_L,\vec{k}_T) $$with $\beta$ defined by $\varepsilon_0$ as in \eqref{Eq:beta-weyl}.  Therefore, on the forward arm (where we choose to do the $k_L$ and $\vec{k}_T$ integrations first)
\begin{eqnarray*}
    \langle O_1(x_1)\rangle &=&\int {\rm d}\omegaup\int{\rm d}^d x_2 e^{-i\omegaup( s(\sigma_1)-s(\sigma_2))}(\widehat{G}_{11}(\omegaup,x_1, x_2) J_1(\omegaup,\vec{x}_2) - \widehat{G}_{12}(\omegaup,x_1, x_2) J_2(\omegaup,\vec{x}_2))\\ &=& \int {\rm d}\omegaup\int{\rm d}^d x_2 e^{-i\omegaup( s(\sigma_1)-s(\sigma_2))}(\widehat{G}_{11}(\omegaup,x_1, x_2)  - \widehat{G}_{12}(\omegaup,x_1, x_2) e^{\beta\omegaup}) J_1(\omegaup,\vec{x}_2) \\ &=& \int {\rm d}\omegaup\int{\rm d}^d x_2 e^{-i\omegaup( s(\sigma_1)-s(\sigma_2))}(\widehat{G}_{11}(\omegaup,x_1, x_2)  - \widehat{G}_{21}(\omegaup,x_1, x_2))J_1(\omegaup,\vec{x}_2) \\ &=&
    \int {\rm d}\omegaup\int{\rm d^d}x_2 e^{-i\omegaup( s(\sigma_1)-s(\sigma_2))}\widehat{G}_{A}(\omegaup,x_1, x_2)  J_1(\omegaup,\vec{x}_2)\\ &=& \int {\rm d}^d x_2 G_A(x_1,x_2) J_1(x_2)
\end{eqnarray*}
Above, we have used the result from \eqref{Eq:StructureCentral-2} that $\widehat{G}_{12}e^{\beta\omegaup} = \widehat{G}_{21}$ and also the identity \eqref{Eq:GAG11G21}. Thus indeed we see that the outgoing mode gives advanced response. Note for both $J_2 = e^{\beta\omegaup} J_1$ and $\widehat{G}_{12}e^{\beta\omegaup} = \widehat{G}_{21}$ (used above) to hold, we need absence of log terms at the horizon, which follows if $\gamma_{n,out}$s are functions of $\omegaup$ only. The outgoing solution which satisfies the regularity condition \eqref{Eq:NHn2} at the horizon cap indeed has this property with $\gamma_{n,out}$ determined from the cancellation of the double and single poles of the equation of motion being a linear function of $\omegaup$ and independent of $\kappa_L$ and $\vec{\kappa}_T$.

Furthermore, let us consider transients which are ingoing solutions with vanishing sources at the boundary. To be simplistic, let us consider the homogeneous transients first. Since we can map the leading order solution to the static black brane background, we should have
$$\omegaup = \omegaup_Q~,$$
where $\omegaup_Q$ corresponds to the homogeneous quasinormal mode frequencies of the static black brane, as noted by Janik and Peschanski. A non-trivial consistency test of our ansatz for the bulk scalar field is that the condition for the sources to vanish at the first subleading order is 
$$\gamma_0 = \gamma_0(\omegaup_Q),$$i.e. for $AdS_5$ we should have
$$ \gamma_0(\omegaup) =\frac{\omegaup_Q}{4\varepsilon_0^{1/4}}.$$The details of the numerical verification are shown in Appendix \ref{sec:transients}. These transients can be added to the solution with specific sources since the equation of the scalar field is linear and are needed for matching with arbitrary initial conditions for the bulk scalar field, as discussed below. The case of the inhomogeneous transients is complicated since the effective momenta $\kappa_L$ and $\vec{\kappa}_T$ also depend on the proper time, and will be examined in a later work.

Our key result that the horizon cap is pinned to the nonequilibrium event horizon is consistent with the feature in nonequilibrium quantum field theory that the evolution of the Schwinger-Keldysh correlation functions, which can be written in the form of the Schwinger-Dyson equations for the commutator and anti-commutator via functional derivatives of the two-particle irreducible effective action, is causal and uniquely determined once we give the initial conditions for these in the initial state \cite{Berges:2004yj}. The bulk analogue is that the evolution of the field configuration on the full complexified geometry with the horizon cap should be uniquely and causally determined for given initial conditions. For this to hold, the horizon cap should cover the space-time outside the event horizon, which is indeed the case.

\section{On initial conditions and seeing behind the event horizon}\label{sec:IC}
The discussion on initial conditions is subtle.\footnote{We thank Kostas Skenderis for discussions on these issues.} We only discuss this briefly here and hope to address it fully in the future. For the full complexified geometry, To obtain a unique solution for the bulk scalar field on both arms, we need to specify $\Phi_{1}(v, \zeta, \vec{x}_\perp)$ and $\Phi_{2}(v, \zeta, \vec{x}_\perp)$ at an initial time $\sigma_0$ on both arms, and also the sources $J_1(\sigma,\zeta, \vec{x}_\perp)$ and $J_2(\sigma,\zeta, \vec{x}_\perp)$ at the two boundaries for all times in the future. Our horizon cap prescription leads to unique solutions for the ingoing and outgoing modes corresponding to $J_1(\sigma,\zeta, \vec{x}_\perp)$ and $J_2(\sigma,\zeta, \vec{x}_\perp)$ in a large proper time expansion, however it is not guaranteed we can match with arbitrary initial conditions $\Phi_{1}(v, \zeta, \vec{x}_\perp)$ and $\Phi_{2}(v, \zeta, \vec{x}_\perp)$ at $\sigma = \sigma_0$. This is a conceptual problem although the matching with data for the initial density matrix is a separate issue from obtaining the correlation functions in the limit where the state hydrodynamizes and forgets most details of the initial conditions; we have concerned ourselves with the latter here. The two arms of the Schwinger-Keldysh contour specify the two in-states which have overlap, in principle, with arbitrary bulk field configurations at initial time, corresponding to $\Phi_{1}(v, \zeta, \vec{x}_\perp)$ and $\Phi_{2}(v, \zeta, \vec{x}_\perp)$ respectively. The full on-shell gravitational action beyond the hydrodynamic limit should yield the matrix element 
\begin{equation}\label{Eq:MEl}
\langle\Phi_1 \vert {T}_c\left(\exp\left(- i\oint J(x) \hat{O}(x)\right)\right)\vert \Phi_2 \rangle
\end{equation}
with ${T}_c$ denotes time-ordering in the closed time contour, and $\langle \Phi_1\vert$ and $\langle \Phi_2\vert$ denoting states in the dual field theory corresponding to the semi-classical bulk field configurations.\footnote{In the approach of Skenderis and van Rees \cite{Skenderis:2008dg}, the in-in states are defined using Euclidean path integrals. Even in this approach, it is unclear if one can glue arbitrary in-in states by bulk geometries. See \cite{Pedraza:2021fgp,Martinez:2021uqo} for some discussions on this issue.} 

This issue can be partly addressed by allowing for the ingoing transients discussed above. These do not change the sources at the two boundaries but they modify the initial conditions at the two slices. However, this is not enough because we have a pair of initial conditions, one each for each arm. The transients are analytic at the horizon cap and do not affect the initial conditions on the two arms independently. Although this needs to be investigated further, presently we may conclude that one may not be able to obtain semi-classical solutions corresponding to arbitrary in-in states meaning that decoherence, which suppresses some off-diagonal matrix elements, is built in the semi-classical gravity approximation. This issue is also relevant for the general approach of Skenderis and Balt van Rees \cite{Skenderis:2008dg}. We also note that there is possibility of adding other semi-classical complex saddles of the gravitational action which do not have a natural hydrodynamic limit. 

There is another independent route for matching with initial conditions. The late proper time expansion of the correlation functions \eqref{Eq:G-expansion} is divergent, and would require a trans-series completion which would naively be of the form (like multi-instanton series) 
\begin{eqnarray}
G(x_1, x_2) &=&
\prod_{\alpha =1}^\infty \left(\sum_{n_\alpha =0}^\infty \left(C_\alpha(\sigma_r, \widetilde{\zeta}_r, \widetilde{x_\perp}_r) \overline\sigma^{\gamma_\alpha(\sigma_r, \widetilde{\zeta}_r, \widetilde{x_\perp}_r)} e^{-\xi_\alpha(\sigma_r, \widetilde{\zeta}_r, \widetilde{x_\perp}_r) \overline\sigma}\right)^{n_\alpha}\right)\nonumber\\&&\Gamma_{n_1, n_2, \cdots}(\overline{\sigma},\sigma_r, \widetilde{\zeta}_r, \widetilde{x_\perp}_r) \quad {\rm with}\nonumber\\
\Gamma_{n_1, n_2, \cdots}(\overline{\sigma},\sigma_r, \widetilde{\zeta}_r, \widetilde{x_\perp}_r) &=& \sum_{k=0}^\infty \Gamma_{k; n_1, n_2, \cdots}(\sigma_r, \widetilde{\zeta}_r, \widetilde{x_\perp}_r)\overline{\sigma}^{-k}.
\end{eqnarray}
This is a generalization of the trans-series for $\epsilon(\sigma)$ for the Bjorken flow which completes the divergent asymptotic hydrodynamic series at large proper time \cite{Heller:2013fn,Heller:2015dha,Basar:2015ava}. Above $\Gamma_{n_\alpha =0}$ coincides with the matrix hydrodynamic power series \eqref{Eq:G-expansion} which we have explicitly computed here; while generally $\Gamma_{n_1, n_2, \cdots}$ is a similar matrix power series in $\overline\sigma^{-1}$ whose coefficients are related to those of \eqref{Eq:G-expansion} for each fixed $\sigma_r$, $\widetilde{\zeta}_r$ and $\widetilde{x_\perp}_r$. Physically, the functions $\xi_\alpha(\sigma_r, \widetilde{\zeta}_r, \widetilde{x_\perp}_r)$ and $\gamma_\alpha(\sigma_r, \widetilde{\zeta}_r, \widetilde{x_\perp}_r)$ govern the hydrodynamic relaxation of the Green's function for fixed separations of the reparametrized space-time arguments $x_1$ and $x_2$. It is possible that these functions are approximately constants. Crucially the Stokes data $C_\alpha(\sigma_r, \widetilde{\zeta}_r, \widetilde{x_\perp}_r)$ are determined by initial conditions. Instead of being constants as in the case of $\epsilon(\tau)$, the Stokes data are now functions.

The mathematical formulation of the trans-series is required to formulate a precise way to capture information about the initial state via Stokes data, and therefore the quantum fluctuations behind the event horizon. To match with the initial conditions, we should add the transients and therefore the Stokes data to hydrodynamic gravitational background as well. Furthermore, additional Stokes data for the evolving event horizon which can be captured by the time-dependent residual gauge transformation is needed too (recall that the residual gauge transformation itself is expressed in the large proper time expansion for which a trans-series completion is necessary like for $\epsilon(\sigma)$). One could interpret the latter feature as the horizon hair (more precisely a proper residual gauge transformation) playing a crucial role in decoding the interior of the event horizon.\footnote{See \cite{Raju:2020smc,Kibe:2021gtw} for recent reviews on current progress in resolving black hole information paradoxes with substantial discussion on the role played by \textit{hair} degrees of freedom at the horizon.}

\section{Conclusions and outlook}\label{sec:Con}

We believe that our method for computing the real time correlation functions of the hydrodynamic Bjorken flow can be generalized to generic situations where the state hydrodynamizes, meaning that the dynamics of one-point functions can be captured by an asymptotic series expansion which is generated by the hydrodynamic evolution of the temperature and velocity fields. The key would be to see if there exists Weyl transformations with non-trivial space-time dependence which can map the flow at late time to a configuration with constant temperature and entropy density although time translation symmetry may not be present. In this case, as demonstrated here for the Bjorken flow, the dual black hole would attain a horizon with constant surface gravity and area at late time, although a time-like Killing vector may be eternally absent. The event horizon's shape would fluctuate in space and time keeping the total area fixed when entropy production ceases in this limit. Nevertheless, with appropriate space-time reparametrizations, the horizon cap prescription can be made to work as in the case of the Bjorken flow provided at the leading order in the large (suitably reparametrized) time expansion the equation of motion of bulk fields can be mapped to that on a static black brane. We will pursue this direction in the future, particularly the Gubser flow \cite{Gubser:2010ze}.

Another important problem would be to understand the Borel resummation of the hydrodynamic series \eqref{Eq:G-expansion} and compute the Schwinger-Keldysh correlation functions of the holographic hydrodynamic attractor \cite{Heller:2013fn}. Furthermore, the late time thermal nature of the correlation functions in terms of the reparametrized space-time arguments could be of phenomenological relevance in heavy ion collisions especially with regards to the dynamics of heavy quarks and bound states, and jets in the expanding quark-gluon plasma \cite{Busza:2018rrf}.
 
Our prescription can be used to construct the quantum generalization of classical stochastic hydrodynamics (see \cite{Kovtun_2012,Liu:2018crr} for reviews) in holographic theories by considering the backreaction of the fluctuations on the background geometry systematically. This is especially important in the context of superfluid fluctuations since quantum dynamics is important at coherence time and length scales which are shorter than the scattering time and the mean free path respectively (see \cite{Bernard_2021} for an excellent related discussion)-- non-linearities can potentially cause novel non-trivial effects such as quantum corrections to the long time tails.\footnote{For a recent work on the role of classical fluctuations of the superfluid order parameter in modifying transport properties see  \cite{Grossi:2021gqi}.} 

{A more general goal is to develop a quantum generalization of the large deviation function \cite{TOUCHETTE20091} in classical nonequilibrium statistical mechanics, which is essentially a generalization of the equilibrium free energy to the hydrodynamic regime, assigning as for instance probability to a macroscopic hydrodynamic configuration which may not satisfy the hydrodynamic equations. The approach {of \cite{Jana:2020vyx}} which constructs an effective action at finite temperature from the complexified bulk space-time will be very relevant for such developments if it can be generalized to out-of-equilibrium situations at least in the hydrodynamic limit.}

More generally, we would like to use the horizon cap method to study holographic (evaporating) black holes interacting with heat baths or dynamical reservoirs, and understand the reconstruction of the islands \cite{Almheiri:2019hni,Kibe:2021gtw} (which include the black hole interior) from Hawking quanta. In such cases, semi-holographic formulations for open quantum systems (see \cite{Joshi:2019wgi,Mondkar:2021qsf} and also \cite{Jana:2020vyx} for instance) can provide useful models.

\newpage
\acknowledgments
We thank Kostas Skenderis and Balt van Rees for helpful discussions and sharing many insights on the issues discussed in Section \ref{sec:IC}. AM acknowledges support from the Ramanujan Fellowship and Early Career Research Award of the Science and Engineering Board of Department of Science and Technology of India, new faculty seed grant of IIT Madras, the institute of excellence schemes of the Ministry of Education of India and also IFCPAR/CEFIPRA funded project no. 6304-3. The research of AB is supported by IFCPAR/CEFIPRA funded project no. 6304-3.

\appendix
\section{Details of the five dimensional metric dual to the Bjorken flow upto second order in the late-time expansion}\label{sec:BJ2}
Consider the late-time expansion of Einstein's equations based on the ansatz \eqref{Eq:Metric-Bjorken-vsigma} along with (\ref{Eq:AsympGen2}). At zeroth order, the six components of Einstein equations read, 
\begin{eqnarray}
E_{vv} &:& 2 k_0'(x){}^2+4 k_0''(x)+l_0'(x){}^2+2 l_0''(x)  =0 ~, \nonumber\\ 
E_{v \sigma} &:& a_0(x) x^2 \left(4 k_0''(x)+l_0'(x){}^2+2 l_0''(x)\right)+ 2 x a_0(x) k_0'(x) \left(x l_0'(x)-6\right) + 3 x^2 a_0(x) k_0'(x){}^2  \nonumber \\ &-& 6 x a_0(x) l_0'(x)+24 a_0(x) + x a_0'(x) \left(2 x k_0'(x)+x l_0'(x)-6\right)-24 = 0~, \nonumber\\
E_{\sigma\sigma} &:& a_0(x) x^2 \left(4 k_0''(x)+l_0'(x){}^2+2 l_0''(x)\right)+ 2 x a_0(x) k_0'(x) \left(x l_0'(x)-6\right) + 3 x^2 a_0(x) k_0'(x){}^2  \nonumber \\ &-& 6 x a_0(x) l_0'(x)+24 a_0(x) + x a_0'(x) \left(2 x k_0'(x)+x l_0'(x)-6\right)-24 = 0 ~, \nonumber\\
E_{s_\perp s_\perp} &:& 2 x \left(a_0'(x) \left(x \left(k_0'(x)+l_0'(x)\right)-6\right)+x a_0''(x)\right) +24a_0(x) -24  \nonumber \\ &+& a_0(x) x \left(k_0'(x) \left(x l_0'(x)-6\right)+2 x \left(k_0''(x)+l_0''(x)\right)+x a_0'(x){}^2+x l_0'(x){}^2-6 l_0'(x)\right) = 0~, \nonumber\\
E_{{\zeta}\zeta} &:& a_0(x) \left(3 x^2 k_0'(x){}^2+4 x^2 k_0''(x)-12 x k_0'(x)+24\right)+ 4 x \left(x k_0'(x)-3\right) a_0'(x)+2 x^2 a_0''(x)-24 = 0~.\nonumber\\
\end{eqnarray}
where $ x \equiv  \varepsilon_0^{1/d}~v$. We solve these equations based on the conditions
$$a_0(x)= 1-x^4~,~~~~~~~~~~l_0(x)=0~,~~~~~~~~~~k_0(x)=0~,$$ which is motivated from the fact that at very late times, we have an $AdS_5$ black hole with constant horizon area and surface gravity. 

Once these initial conditions are imposed, then at each subleading order,  the six  equations can be repackaged to the following three equations 
\begin{flalign}
2 k_i''(x) +l_i''(x) &= S_{1,i}~, \nonumber \\ 
\left(x^4+3\right)( k_i'(x) -l_i'(x))+ x \left(x^4-1\right)  ( k_i''(x) -l_i''(x)) &= S_{2,i}~, \nonumber \\
\frac{3}{2} x^2 a_i''(x) -6 x a_i'(x)+6 a_i(x) -  2 x^5 \left(2 k_i'(x) +l_i'(x) \right)  &= S_{3,i}~, \end{flalign} 
where  $ S_{1,i}$, $ S_{2,i}$ and $ S_{3,i}$ are the  sources at the $i$-th order which depend on $a_j,l_j,k_j$ and their derivatives, for $j<i$. Now at each subleading order, the most general solution to these equations are given by, 
\begin{eqnarray} \label{geomsol}
2 k_i(x) + l_i(x) &=& \rho_i + \alpha_i x + P I_{1,i} ~,\nonumber \\
k_i(x) - l_i(x) &=& \xi_i - \frac{\kappa_i}{4} \log(1-x^4) + PI_{2,i}~, \nonumber\\
a_i(x) &=&  \beta_i x^4 + \alpha_i \frac{x(3+x^4)}{3} + P I_{3,i}~.
\end{eqnarray}
 where the Greek letters with subscript $ i$  correspond to the integration constants at $ i $-th order and  $PI_{1,i},PI_{2,i}$ and $PI_{3,i}$ are the particular solutions  determined by the sources $ S_{1,i}$, $ S_{2,i}$ and $ S_{3,i}$ respectively. The expressions for the $PI$s for $ i=1$ are simple and already mentioned in \eqref{Eq:a1l1k1}-\eqref{Eq:g}. For $ i=2$, we get
 \begin{eqnarray}
 PI_{1,2} &=& \frac{1}{12} \Big(8 x^2+6 \log \left(x^2+1\right)-\alpha _1^2 (x+2) x-4 \alpha _1 (x+2) x+  16 x+4 \log (x+1)+2 (2-6 x) \arctan{(x)}+4 \Big)~, \nonumber \\ 
PI_{2,2} &=&Re\Bigg[ \frac{1}{24} \Big(-24 x^2-\frac{16 x}{x^3+x^2+x+1}+16 x+(4+4 i) \arctan{(x)}^2-(1-i) \log ^2(i-x) \nonumber \\
&+& 4 \log ^2(x+1)-2 \log ^2(-i+x)-(3+i) \log ^2(i+x)-4 \log ^2\left(x^2+1\right)+(2+8 i) \arctan{\left(\frac{x+1}{1-x}\right)}\nonumber \\
&+& 2 i \pi  \arctan{(x)}-8 \arctan{(x)}+(2-8 i) \arctan{\left(\frac{x-1}{x+1}\right)}-\pi \log (2)+4 \pi  \log \left(1+e^{-2 i \arctan{(x)}}\right) \nonumber \\
&-& 16 \arctan{(x)} \log \left(1+e^{2 i \arctan{(x)}}\right)+8 \arctan{(x)} \log \left(1-i e^{2 i \arctan{(x)}}\right)+2 \pi  \log \left(1-i e^{2 i \arctan{(x)}}\right) \nonumber \\  
&+&6 \log (1-x)-(4+4 i) \log \left(\left(\frac{1}{2}+\frac{i}{2}\right) (i x+1)\right) \log (x+1) \nonumber \\  
&+&2 \log (x+1)+4 \log (x-1) \log \left(\left(\frac{1}{2}+\frac{i}{2}\right) (-i+x)\right)-(6+2 i) \log \left(\frac{1}{2} (i x+1)\right) \log (i+x) \nonumber \\  
&+& 4 \log (x-1) \log \left(\left(\frac{1}{2}-\frac{i}{2}\right) (i+x)\right)+(4+4 i) \log (x+1) \log \left(\left(-\frac{1}{2}-\frac{i}{2}\right) (i+x)\right) \nonumber \\  
&-& (2-2 i) \log (i-x) \log \left(-\frac{1}{2} i (i+x)\right)-4 \log (-i+x) \log \left(-\frac{1}{2} i (i+x)\right)  
+ (2-2 i) \log (i-x) \log \left(x^2+1\right)  \nonumber \\ &-& 4 \log (x-1) \log \left(x^2+1\right)+4 \log (-i+x) \log \left(x^2+1\right)+(6+2 i) \log (i+x) \log \left(x^2+1\right)  \nonumber \\ 
&+& 2 \pi  \log \left(x^2+1\right)-4 \log \left(x^2+1\right)+8 \log (x+1) \log \left(1-\left(\frac{1}{2}-\frac{i}{2}\right) (x+1)\right)  \nonumber \\
&-& 4 \pi \log \left(\sin \left(\arctan{(x)}+\frac{\pi }{4}\right)\right)+2 \pi  \log \left(\frac{x+1}{\sqrt{x^2+1}}\right)-(6-2 i) \text{Li}_2\left(\frac{i x}{2}+\frac{1}{2}\right)+8 i \text{Li}_2\left(-e^{2 i \arctan{(x)}}\right)  \nonumber \\ &-& 4 i \text{Li}_2\left(i e^{2 i \arctan{(x)}}\right)+4 \text{Li}_2\left(\left(-\frac{1}{2}+\frac{i}{2}\right) (x-1)\right)+4 \text{Li}_2\left(\left(-\frac{1}{2}-\frac{i}{2}\right) (x-1)\right)+(4+4 i) \text{Li}_2\left(\left(\frac{1}{2}+\frac{i}{2}\right) (x+1)\right)  \nonumber \\ &+&(4-4 i) \text{Li}_2\left(\left(\frac{1}{2}-\frac{i}{2}\right) (x+1)\right)-(6+2 i) \text{Li}_2\left(-\frac{1}{2} i (i+x)\right) + \Big(-\frac{8 x}{x^3+x^2+x+1}+8 x  \nonumber \\ &+& (2+4 i) \arctan{\left(\frac{x+1}{1-x}\right)}-4 \arctan{(x)}+(2-4 i) \arctan{\left(\frac{x-1}{x+1}\right)}\Big) \alpha _1-\frac{8}{x^3+x^2+x+1}\Big)\Bigg]~, \nonumber \\
PI_{3,2} &=& \frac{x}{108}  \Big(\alpha _1^2 \left(-9 x^5-6 x^4+3 x+6\right)-4 \alpha _1 \left(9 x^5+12 x^4-9 x-8\right) + 4 \Big(-9 x^5-9 \left(x^4+1\right) \arctan{(x)}+ x^3 \nonumber \\ 
 &+& 9 x^3 \log \left(x^2+1\right)+9 x+16\Big)\Big)~~,\nonumber\end{eqnarray}
where $\text{Li}_2(x) $ is the polylogarithmic function of order 2. The expressions for the particular solution  get increasingly complicated at higher orders. However, the simple structure of the homogeneous solutions remain the same at every order. The integration constants associated with the homogeneous solutions  can be fixed at every order in the following way:
 \begin{itemize}
 
 \item The fixation of $\alpha_i$ has been detailed in $(\ref{poleqn})-(\ref{alpha2})$. The same pattern repeats for the rest of $ \alpha_{i>2}$ .  
     \item The integration constants $ \beta_i$ are fixed by mapping the coefficients of $x^4 $ terms in (\ref{geomsol}) to  $ a_4(\tau)$, $ l_4(\tau)$ and $ k_4(\tau)$ of \eqref{renormalization:asymptotic} (or $ a_4(\sigma)$, $ l_4(\sigma)$ and $ k_4(\sigma)$ in the Weyl transformed coordinate) and then solving the constraint equations \eqref{Einstein:constraint}.
     \item The constants $\rho_i$ and $ \xi_i$ are fixed by
     demanding  that metric asymptotes to (\ref{Eq:Weylmetric}) at the boundary. 
     \item Finally, the integration constants $ \kappa_i$ are fixed  such that the dual geometry is regular at the horizon to all orders.
 \end{itemize}
 For $ i=1,2 $, this gives
\begin{eqnarray}
\beta_1 &=& \frac{2}{3}, \hspace*{0.5cm} \kappa_1 = 2, \hspace*{0.5cm} \xi_1 = 0, \hspace*{0.5cm} \rho_1= 0 \nonumber \\
\beta_2 &=& - \frac{1}{54}\left(11 + 6 \log(2) \right) , \hspace*{0.5cm} \kappa_2 = -\frac{1}{3} (2 \log (2)-3) , \hspace*{0.5cm}  \rho_2= -\frac{1}{3}\nonumber \\
\xi_2 &=& \frac{1}{288} i \Big(-24 \pi  \alpha _1+48 i C-(48-96 i) \text{Li}_2\left(\frac{1}{2}+\frac{i}{2}\right)+(48+96 i) \text{Li}_2\left(\frac{1}{2}-\frac{i}{2}\right) +(7+6 i) \pi ^2  \nonumber \\ 
&-& 96 i -48 \pi +72 i \log ^2(2)-(24-60 i) \pi  \log (2)+ 48 \pi  \log(2)+24 i \pi  \log (1-i) \Big)
\end{eqnarray}
where  $C$ is the Catalan's constant, $ C \approx 0.915966 $.

One can verify that these solutions reproduce the same late-time expansion of the stress tensor (setting $\varepsilon_0 =1$ and $\tau_0 =1$) 
\begin{eqnarray}
T_{\tau \tau} &\approx& \tau^{-4/3} - \frac{2}{3} \tau^{-2} + \frac{1+2\log(2)}{18} \tau^{-8/3} + \cdots, \nonumber \\
T_{\zeta \zeta} &\approx& \frac{1}{3} \tau^{2/3} - \frac{2}{3}+\frac{5 (1+2 \log (2))}{54} \tau^{-2/3} + \cdots, \nonumber \\ 
T_{x_\perp x_\perp} &\approx& \frac{1}{3} \tau^{-4/3} - \frac{(1+2 \log (2))}{54} \tau^{-8/3} + \cdots. \nonumber
\end{eqnarray}


\section{The event horizon and the apparent horizon of the Bjorken flow}\label{sec:horizons}
This appendix is devoted to a comparative discussion between the event horizon and the apparent horizon in the dynamical geometry dual to the Bjorken flow.
\subsection{Event horizon:}
Classically, the event horizon marks the null hypersurface from which no signal can come out, i.e, the outgoing null rays become tangential to the hypersurface. To determine its location in the geometry given by (\ref{Eq:Metric-Bjorken-vsigma}), consider radial null geodesics in this space-time whose equation is given by,
\begin{equation} \label{eventeq}
\frac{\partial v}{\partial \sigma} + \frac{v^2}{2} \frac{d-2}{d-1}g_{\sigma\sigma}=0 ~,~~~~~~\text{with}~~~~~~~g_{\sigma\sigma}= \frac{1}{v^2}\left(\frac{(d-1)^2}{(d-2)^2}A(v,\sigma)+\frac{2(d-1)v}{
(d-2)^2\sigma}\right) ~.
\end{equation}
For a static geometry, $v_E=cons.$ and the event horizon is simply given by the zero of $g_{\sigma\sigma}$. However, in a dynamical geometry, the event horizon will depend on $\sigma$. Consider the late-time expansion of the horizon of the form,
\begin{eqnarray}
v(\sigma) \equiv v_E(\sigma)= \sum_{i=0} v_{Ei}  ~ \sigma^{-i} 
\end{eqnarray}
Now to determine $ v_{Ei} $, we use the equation (\ref{eventeq}) along with the known late-time expansion of $A(v,\sigma)$,
\begin{equation}
A(v,\sigma) = 1- \varepsilon_0 ~ v^d + \frac{1}{\varepsilon_0^{1/d} \sigma} a_1(v) + \frac{1}{\varepsilon_0^{2/d} \sigma^2} a_2(v) + \cdots
\end{equation} 
At leading  order, we have
$$v_{E0} = 1/\varepsilon_0^{1/d}~ $$
in any space-time dimensions $d+1$. The sub-leading corrections to the event horizon depend on the residual gauge parameters\affiliation[]{}. For instance, the first and second sub-leading corrections in $ d= 4$\footnote{For $d=3$, the first sub-leading correction is
$ v_{E1} = \frac{ 5 (5+ 3 \alpha_1) } { 3 \varepsilon_0^{2/3} },$ which again depends on the residual parameter. However, these corrections do not have a universal form like the leading order.}  are, \begin{eqnarray}
 v_{E1} &=& \frac{ (3 + \alpha_1) } { 6 \varepsilon_0^{1/2} }~,\nonumber\\
  v_{E2} &=& \frac{72 {\alpha_2}  - 9 \pi - 20+ 12 \log (2)}{216 \varepsilon_0^{3/4}}~,
\end{eqnarray}
where $ \alpha_1$ and $ \alpha_2$ are the residual gauge parameters associated with the late-time expansion of  $ A(v,\sigma)$. Now the gauge parameters can be  uniquely fixed by demanding regularity of the horizon cap. In $d=4$, this gives (recall (\ref{alpha1}) and (\ref{alpha2}))  
\begin{eqnarray}\label{eventhorsublead}
  \alpha_1 &=& -3 ,\nonumber\\ \hspace{2mm}
   \alpha_2 &=& \frac{9 \pi +20-12 \log (2)}{72}.
\end{eqnarray}
These values, in turn fix the event horizon to $  1/\varepsilon_0^{1/d} $ upto the second subleading order. Thus, regularization of the horizon cap leads to vanishing of the sub-leading corrections $ v_{Ei}$ (for $ i>0$)  to the static event horizon. 
We expect this feature to remain true to all orders in the late-time expansion.
\subsection{Apparent horizon:}
The apparent horizon is a null hypersurface that acts as a boundary between the trapped and un-trapped regions, whose location is given by the product of the expansion parameters $ \theta_\pm$, i.e. $ \Theta = e^f \theta_+ \theta_-$, where the factor $e^f$ is defined below. The trapped region is characterized by $ \Theta > 0 $, where the light rays directed outward propagates inward, whereas in the un-trapped region the light rays directed outward propagates outward and is characterized by $\Theta < 0$. Therefore $ \Theta= 0 $ gives the location of the apparent horizon. 

Here we will adopt the dual-null formalism \cite{Kinoshita:2008dq, Hayward:1993wb,Hayward:2004fz,Figueras:2009iu} to study location of the apparent horizon, where one defines a pair of null hypersurfaces  $ \Sigma^\pm$ parameterised by scalars $ \zeta^\pm$, with associated one-forms $ n^\pm = -\rm d \zeta^\pm $. The null normal vector associated with these hypersurfaces are given by, 
\begin{eqnarray}
l^\mu_\pm = e^{-f} g^{\mu\nu} n^\mp_\nu~, \nonumber 
\end{eqnarray}
where $e^f $ is the normalization factor given by,
$$e^f = - g^{\mu \nu} n^-_ \mu n^+_ \nu ~.$$ 
Next we define the expansion parameters $ \theta_\pm$, 
\begin{equation}
\theta_\pm = \mathcal{L}_\pm Log(\mu)~, \nonumber 
\end{equation}
where $ \mu $ is the spatial volume element of the geometry in which the hypersurfaces are defined and $ \mathcal{L}_\pm $ is the lie derivative along the null normal vectors, $ l_\pm$.
Finally we introduce the invariant\footnote{This parameter $ \Theta$ is invariant under reparameterisation of the scalar $ \zeta^\pm \rightarrow \xi^\pm(\zeta^\pm)$ or interchange of $ \zeta^\pm \rightarrow \zeta^\mp$ as discussed in \cite{Hayward:1993wb}.} quantity $ \Theta = e^f \theta_+ \theta_- $, whose  zero gives the location of the apparent horizon.

In case of our geometry \eqref{Eq:Metric-Bjorken-vsigma}, we consider a pair of null hypersurface defined by constant values of the retarded and advanced radial null coordinates, whose normal one-forms are given by,
\begin{eqnarray} \label{1form_1}
 n^- = - \mathcal{N}^-{\rm d} \sigma,  \hspace{1cm} n^+ = - \mathcal{N}^+ \Big( \left(\frac{2 v}{\sigma} + (d-1)A(v,\sigma) \right)  {\rm d} \sigma + 2(d-2) {\rm d} v \Big), 
\end{eqnarray}
and the null normal vectors $ l^\mu_\pm$ along with the normalisation factor $ e^{f}$ and the volume element $\mu$ reads,
\begin{eqnarray}\label{1form_2}
e^f &=& \frac{2 (d-2)^2} {d-1} \mathcal{N}^+ \mathcal{N}^-  v^2 ~,\nonumber\\
l^\mu_+ &=& \left(\frac{1}{2 (d-2) \mathcal{N}^+ },0,0,\vec{0} \right) ~, \nonumber \\
l^\mu_- &=& \left(-\frac{ \sigma (d-1)   A(v,\sigma )+2 v}{2 (d-2) \mathcal{N}^- \sigma },1/\mathcal{N}^-,0,\vec{0} \right)~,\nonumber\\
\mu &=& \frac{\sigma+v}{v^3\sigma}\sqrt{e^{2K(v,\sigma)+L(v,\sigma)}}~.
\end{eqnarray}
 where $ \mathcal{N}^+ $ and $ \mathcal{N}^-$ are the overall normalizability factor that can be determined by the integrability condition $ {\rm d}({\rm d}n^\pm) = 0$. This ensures that the one-forms \eqref{1form_1} and \eqref{1form_2} are exact. However there is no need of computing them explicitly, as when we compute $ \Theta $ the contribution of $ \mathcal{N}^+ $ and $ \mathcal{N}^-$ from $ e^f$ will cancel with the ones coming from $ \theta_-$ and $ \theta_+$. 

\begin{figure}[h!]
    \centering
    \includegraphics[scale=0.5]{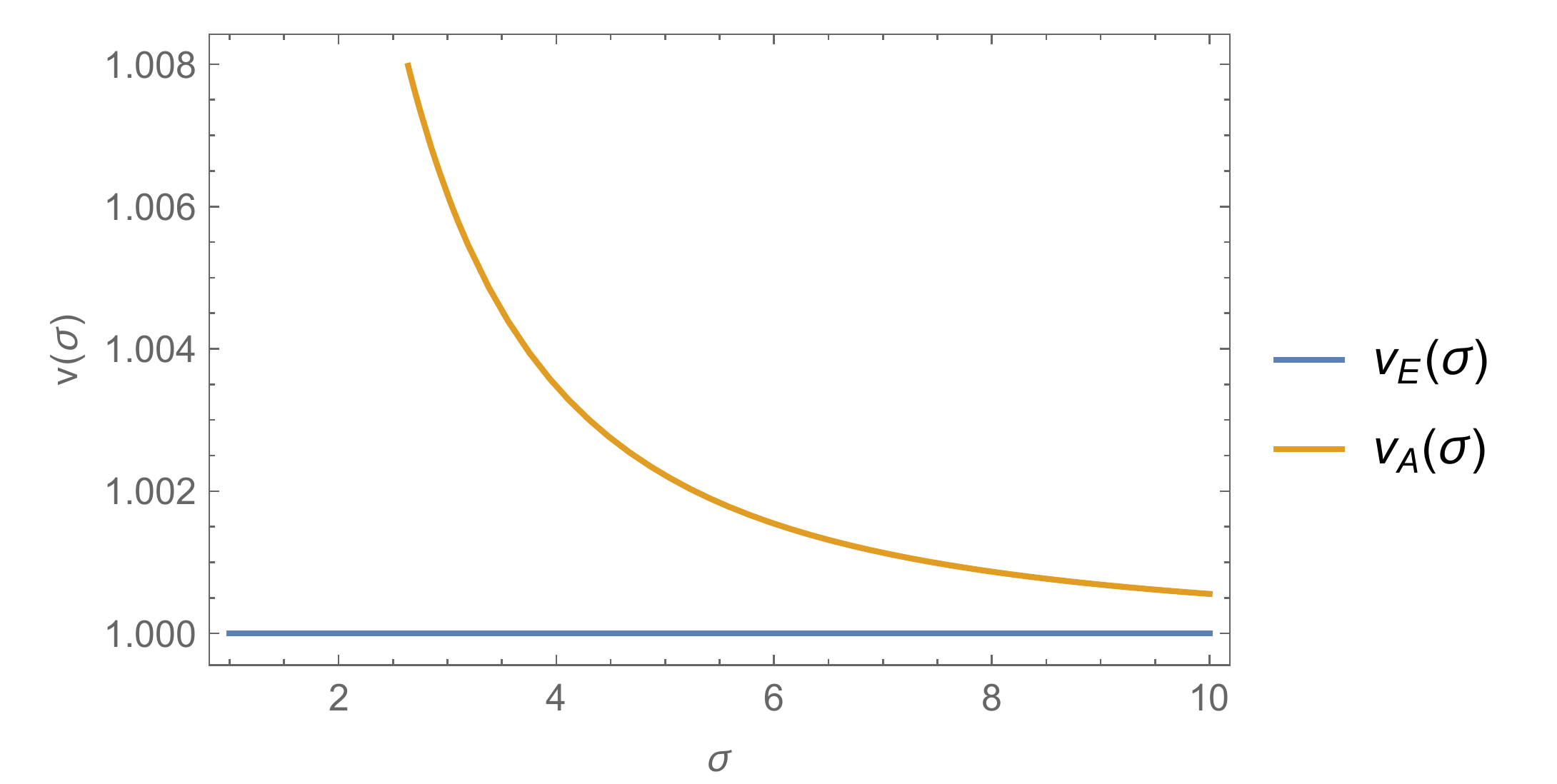}
    \caption{The figure shows the evolution of the event horizon(blue curve) and the apparent horizon(yellow curve) in the dynamical geometry dual to the Bjorken flow, upto second order in late-time expansion. Here $\varepsilon_0$ is set to 1 and the gauge parameters $ \alpha_1$ and $ \alpha_2$ are so chosen that the event horizon is fixed to 1. The apparent horizon always lies inside the event horizon (at greater $v$). }
    \label{fig:my_label}
\end{figure}

\paragraph{}
Now since the apparent horizon depends on $ \sigma$, one can consider a late-time expansion of the apparent horizon similar to the event horizon as, 
\begin{eqnarray}
v_A(\sigma) = \sum_{i=0} v_{Ai}~ \sigma^{-i} .\nonumber 
\end{eqnarray}
To determine the coefficients $ v_{Ai}$ we solve $\Theta =0 $ at every order. In case of any space-time dimensions  $ d+1 $, the leading order result is universal and turns out to be   $$ v_{A0} = 1/\varepsilon_0^{1/d}~,$$
which coincides with the event horizon. However,  the sub-leading corrections are dimension dependent. For $d=4$, the first and second sub-leading corrections to the leading terms are
\begin{eqnarray}
v_{A1} &=& \frac{3+\alpha_1}{6 \varepsilon_0^{1/2}} ~,\nonumber\\
v_{A2} &=& \frac{72 \alpha_2 -9 \pi -8+12 \log (2)}{216 ~ \varepsilon_0^{3/4}} ~.
\end{eqnarray}
Note that, for $\alpha_1=-3$ the first order correction to the apparent horizon vanishes similar to the case of the event horizon. However, at second order, the value of $\alpha_2$ (given by (\ref{alpha2})) for which the the correction to the event horizon vanishes, now renders $v_{A2}>0$. So the apparent horizon will lie inside the event horizon (see Figure \ref{fig:my_label}). Again, we expect this feature to remain true at higher orders as well.

\section{More details of the bulk scalar field in the $AdS_5$ Bjorken flow background}\label{sec:AdS5}

In this appendix, we will provide brief details of the solution of  Klein-Gordon equation at the leading  \eqref{Eq:ZerothKG} and first subleading  order \eqref{Eq:nthordereoms} (for $n=1$)  and validate the near-horizon behaviors \eqref{Eq:NH0} and \eqref{Eq:NHn1} for the same.  To be specific, we will consider the example of $ d=4$. 

At leading order, the homogenous Klein-Gordon equation \eqref{Eq:ZerothKG} simply takes the form of that in a static black brane geometry. So we can expand the solution in the standard basis provided by the ingoing and outgoing modes. Near the horizon, these modes admit the following expansions
\begin{eqnarray}
\phi_{0,in}(v,\omegaup,\kappa_L,\vec{\kappa}_T) =p_0\left[ 1 + (\varepsilon_0^{-1/4}-v) \frac{ \left(\kappa_L^2+ \vec{\kappa}_T^2 - m^2 \varepsilon_0^{1/2} +3 i \varepsilon_0^{-1/4} \omegaup \right)}{2 \left(2 \varepsilon_0^{1/4} - i \omegaup \right)} + \mathcal{O}((\varepsilon_0^{-1/4}-v)^2) +\cdots\right] 
 \nonumber \\
\end{eqnarray}
\begin{eqnarray}
(\varepsilon_0^{-1/4}-v)^{-\frac{i \omegaup}{2 \varepsilon_0^{-1/4}}}\phi_{0,out}(v,\omegaup,\kappa_L,\vec{\kappa}_T) = q_0\Bigg[1 + (\varepsilon_0^{-1/4}-v) \frac{ \left(2\kappa_L^2+ 2\vec{\kappa}_T^2 - 2 m^2 \varepsilon_0^{1/2} - 3 i \omegaup \right)}{4 \left(2 \varepsilon_0^{1/4} + i \omegaup \right)} \nonumber \\ + \mathcal{O}((\varepsilon_0^{-1/4}-v)^2) +\cdots \Bigg].
\end{eqnarray}
Now, without loss of generality, we can set $p_0=1$ and $q_0=1$. Then, in the limit $ v \rightarrow \varepsilon_0^{-1/4}$ we reproduce the behaviors given in
\eqref{Eq:NH0}.

At the first subleading order ($n=1$), the Klein-Gordon equation is given by \eqref{Eq:nthordereoms} with
\begin{eqnarray}
S_{1,in}= \tilde{\mathcal{S}}  \phi_{0,in}(v,\omegaup,\kappa_L, \vec{\kappa}_T) \nonumber 
\end{eqnarray}
\begin{eqnarray}
S_{1,out} &=&  \tilde{\mathcal{S}}  \phi_{0,out}(v,\omegaup,\kappa_L, \vec{\kappa}_T)
\end{eqnarray}
where, 
\begin{eqnarray}
\tilde{\mathcal{S}} &=& -\frac{v}{6 \varepsilon_0^{1/4}} \Bigg(\alpha_1 \varepsilon_0^{1/4} v \left(2 \kappa_L^2 v+2 \vec{\kappa}_T^2 v +3 i \omegaup \right)+2 v \left(2 {\kappa}_L^2-\vec{\kappa}_T^2\right) \tan ^{-1}\left(\varepsilon_0^{1/4} v\right) + 4 (\kappa_L^2 + \vec{\kappa}_T^2) \varepsilon_0^{1/4} v^2 \nonumber \\  &+& (2 \kappa_L^2 - \vec{\kappa}_T^2) v \Big(\log \left(\varepsilon_0^{1/2} v^2+1\right)+ 2 \log \left(\varepsilon_0^{1/4} v+1\right) \Big) +12 i \gamma_0  \varepsilon_0^{1/4}+6 i \varepsilon_0^{1/4} v \omegaup \Bigg) \nonumber \\
&-&\frac{1}{6} v \Bigg( (4 \varepsilon_0^{3/4} v^5  + 2 \alpha _1 \varepsilon_0 v^6   + 4 \varepsilon_0 v^6  + 2 \alpha _1 v^2  + 4 v^2 )\partial_v^2   + ( 2 \varepsilon_0 v^5   - \alpha _1 v   -8 i \gamma_0  v  + 4 \varepsilon_0^{3/4} v^4  + \alpha _1 \varepsilon_0 v^5 - 2 v ) \partial_v   \nonumber \\ &-& 12 \kappa_L \partial_{\kappa_L} + 6 \vec{\kappa}_T \partial_{\vec{\kappa}_T} - 4 \vec{\kappa}_T v \partial_{\vec{\kappa}_T} \partial_v  + 8 \kappa_L v  \partial_{{\kappa}_L} \partial_v  \Bigg)~.
\end{eqnarray}
The action of $ \tilde{\mathcal{S}}$  on $ \phi_{0,out}$ gives rise to singular terms having poles of order one and two at the horizon, as mentioned in \eqref{poleqn}. Upon removing these poles by suitably fixing $ \alpha_1$ and $ \gamma_0$ as in \eqref{alpha1},  the particular solution at the first subleading order for the outgoing mode  admits the near-horizon expansion,
\begin{eqnarray}
(\varepsilon_0^{-1/4}-v)^{-\frac{i \omegaup}{2 \varepsilon_0^{-1/4}}} \phi_{1(p),out}(\omegaup,v,\kappa_L,\vec{\kappa}_T) &=& \frac{(\varepsilon_0^{-1/4}-v)}{48 \varepsilon_0^{1/4} +24 i \omegaup }\Bigg(-2 \kappa_L^2 (\pi -7+\log (64))+\vec{\kappa}_T^2 (\pi +2+\log (64)) \nonumber \\ &-& 2 m^2 \varepsilon_0^{1/2}-4 i \varepsilon_0^{1/4} \omegaup -\omegaup ^2\Bigg) 
+ \mathcal{O}((\varepsilon_0^{-1/4}-v)^2) + \cdots 
\end{eqnarray}
However, for the  ingoing mode no such divergent terms appear and we simply have the following near horizon exapnsion
\begin{eqnarray}
\phi_{1(p),in}(\omegaup,v,\kappa_L,\vec{\kappa}_T) &=& \Bigg(\kappa_L^2 \left(-4 \varepsilon_0^{1/4} (\pi -7+\log (64))+2 i \omegaup  (\pi -3+\log (64))\right) \nonumber \\&+& \vec{\kappa}_T^2 \left(2 \varepsilon_0^{1/4} (\pi +2+\log (64))-i \omegaup  (\pi +6+\log (64))\right)  \nonumber \\ &-& 4 m^2 \varepsilon_0^{3/4}+2 i m^2 \varepsilon_0^{1/2} \omegaup+ 6 \varepsilon_0^{1/4} \omegaup ^2+12 i \varepsilon_0^{1/2} \omegaup \Bigg) \frac{(\varepsilon_0^{-1/4}-v)}{24 \left(2 \varepsilon_0^{1/4}-i \omegaup \right)^2} \nonumber \\ &+& \mathcal{O}((\varepsilon_0^{-1/4}-v)^2) + \cdots 
\end{eqnarray}
Clearly, in the limit $ v \rightarrow \varepsilon_0^{-1/4}$ , we reproduce \eqref{Eq:NHn1} for $d=4$ and $n=1$. Similar behavior persists at higher orders as well.

\section{Derivation of Eqs. \eqref{Eq:StructureCentral} }\label{sec:Deriv}
The source and vev of the dual operator at the boundary are given by
\begin{eqnarray} \label{sourcej}
J (\sigma_2, \hat\zeta_2, \vec{x}_{\perp2}) 
&=& \int{\rm d}\omegaup {\rm d}k_L{\rm d}^{d-2}k_T \,\, e^{-i\omegaup s(\sigma_2)} e^{i\left(k_L \hat\zeta_2 + \vec{k}_T\cdot \vec{x}_{\perp2}\right)}~~
 \mathcal{S}(\sigma_2,\omegaup, \kappa_L, \vec{\kappa}_T) \cdot Q(\omegaup,k_L,k_T), \label{source} \\
 \langle O (\sigma_1, \hat\zeta_1, \vec{x}_{\perp1})\rangle 
 &=& (2\Delta_O -d)\int{\rm d}\omegaup {\rm d}k_L{\rm d}^{d-2}k_T \,\, e^{-i\omegaup s(\sigma_1)} e^{i\left(k_L \hat\zeta_1 + \vec{k}_T\cdot \vec{x}_{\perp1}\right)}~~
\mathcal{R}(\sigma_1, \omegaup, \kappa_L, \vec{\kappa}_T)\cdot Q(\omegaup,k_L,k_T)~,\nonumber \label{vev}\\
\end{eqnarray}
where$$
J (\sigma, \hat\zeta, \vec{x}_\perp)=
\begin{pmatrix}
J_1 (\sigma, \hat\zeta, \vec{x}_\perp)\\J_2 (\sigma, \hat\zeta, \vec{x}_\perp)
\end{pmatrix},~~~
\langle O (\sigma, \hat\zeta, \vec{x}_\perp)\rangle= 
\begin{pmatrix}
\langle O_1 (\sigma, \hat\zeta, \vec{x}_\perp)\rangle\\\langle O (\sigma, \hat\zeta, \vec{x}_\perp)\rangle
\end{pmatrix},~~~Q(\omegaup, k_L, k_T)=
\begin{pmatrix}
p(\omegaup, k_L, k_T)\\q(\omegaup, k_L, k_T)
\end{pmatrix}.
$$
To extract the SK Green's function, we first invert  (\ref{source}) to get,
\begin{equation} \label{D3}
    Q(\omegaup,k_L,k_T)= \int{\rm d}\sigma_2 {\rm d}\hat\zeta_2{\rm d}^{d-2}x_{\perp2} \,\,s'(\sigma_2)~~ e^{i\omegaup s(\sigma_2)} e^{-i\left(k_L \hat\zeta_2 + \vec{k}_T\cdot \vec{x}_{\perp2}\right)}~~ \mathcal{S}^{-1}(\sigma_2,\omegaup, \kappa_L, \vec{\kappa}_T) \cdot J (\sigma_2, \hat\zeta_2, \vec{x}_{2\perp}).
\end{equation}
Finally, plugging it back to (\ref{vev}) and using linear response theory, 
\begin{eqnarray}
\langle O (\sigma_1, \hat{\zeta}_1, \vec{x}_{\perp1})\rangle 
 &=& \int{\rm d}\sigma_2 {\rm d}\hat{\zeta}_2{\rm d}^{d-2} \vec{x}_{\perp2} \,\, \widetilde{G}(\sigma_1,\sigma_2, \hat{\zeta}_1,\hat{\zeta}_2, \vec{x}_{\perp1},\vec{x}_{\perp2}) \,  J(\sigma_2, \hat{\zeta}_2, \vec{x}_{\perp2}) \nonumber 
\end{eqnarray}
we read off the SK Green's function
\begin{eqnarray} \label{D4}
\widetilde{G}(\sigma_1,\sigma_2, \hat{\zeta}_1-\hat{\zeta}_2, \vec{x}_{\perp1}-\vec{x}_{\perp2}) &=& 
   \int{\rm d}\omegaup {\rm d}k_L{\rm d}^{d-2}k_T \, e^{-i\omegaup(s(\sigma_1)-s(\sigma_2))}\nonumber\\&& e^{i k_L(\hat{\zeta}_1- \hat{\zeta}_2)+i\vec{k}_T\cdot (\vec{x}_{\perp 1} - \vec{x}_{\perp 2})}
   \widehat{G}(\sigma_1,\sigma_2,\omegaup,k_L, \vec{k}_T)
\end{eqnarray}
where,
\begin{eqnarray}
\widehat{G}(\sigma_1,\sigma_2,\omegaup,k_L, \vec{k}_T) &=&  \frac{1}{2}\Bigg(s'(\sigma_2)\,\sigma_3~\cdot \mathcal{R}\left(\sigma_1, \omegaup,\kappa_{L1}, \vec{\kappa}_{T1}\right) \cdot~ \mathcal{S}^{-1}\left( \sigma_2,\omegaup,\kappa_{L2}, \vec{\kappa}_{T2}\right) \nonumber\\&&\quad +  ({\rm transpose}, \,\, \sigma_1 \leftrightarrow \sigma_2,\omegaup \rightarrow -\omegaup , \kappa_{L1} \leftrightarrow - \kappa_{L2} , \vec{\kappa}_{T1} \leftrightarrow -\vec{\kappa}_{T2})\Bigg) \nonumber \\
\end{eqnarray}
with,
\begin{equation}
    \kappa_{L1} =k_L\frac{\tau_0}{\sigma_1}, \quad \kappa_{L2} = k_L\frac{\tau_0}{\sigma_2}, \quad \vec{\kappa}_{T1}=\vec{k}_T\left(\frac{\tau_0}{\sigma_1}\right)^{-1/(d-2)}, \quad  \vec{\kappa}_{T2}=\vec{k}_T\left(\frac{\tau_0}{\sigma_2}\right)^{-1/(d-2)}. \nonumber
\end{equation}
Firstly, note that the variable conjugate to the momentum $\omegaup$ is the reparametrized time $s(\sigma)$ whereas the boundary quantities depend on $\sigma$. This gives rise to the factor $s'(\sigma_2)$ in (\ref{D3}) as the Jacobian of the transformation $s(\sigma_2) \rightarrow \sigma_2$. Also note that, due to translational invariance along the spatial directions, the matrices $\mathcal{S}$ and $\mathcal{R}$ are independent of $\hat\zeta$ and $\vec{x}_\perp$. Hence, on carrying out the momentum integrals in (\ref{D4}), we would get $\widetilde{G}(\sigma_1,\sigma_2, \hat{\zeta}_1,\hat{\zeta}_2, \vec{x}_{\perp1},\vec{x}_{\perp2}) \equiv \widetilde{G}(\sigma_1,\sigma_2, \hat{\zeta}_1-\hat{\zeta}_2, \vec{x}_{\perp1}-\vec{x}_{\perp2})$.

\section{Transients and $\gamma_0$}\label{sec:transients} 

The quasinormal frequencies and transients capture the causal response of a black hole to  small perturbations. Typically, for a black hole these are determined using the spectral representation method. However, here we adapt a simpler approach for the computation of the same. 

For the homogeneous transients, consider the field ansatz \eqref{Eq:Ansatz1}  with only the ingoing modes of a massless scalar, i.e.
\begin{eqnarray}
\Phi (v, \sigma)
 &=& \int{\rm d}\omegaup \,\, e^{-i\frac{d-1}{d-2}\omegaup\sigma}\left(\frac{\sigma}{\tau_0}\right)^{i\gamma_0(\omegaup/\varepsilon_0^{1/d})}
\sum_{n=0}^\infty \left(\varepsilon_0^{1/d}\sigma\right)^{-n}\phi_{n,in}(v, \omegaup)  
p(\omegaup)~,
\end{eqnarray}
where $ \phi_{n,in}(v, \omegaup)$ 
obeys the same equations \eqref{Eq:ZerothKG} and  \eqref{Eq:nthordereoms} (and hence admit the  solutions \eqref{Eq:NH0} and \eqref{Eq:inoutnthorder}) with $\kappa_L,\vec{\kappa_T}=0$. The residual gauge parameters $\alpha_i$ appearing in the subleading solutions are fixed such that at every order, the event horizon is pinned at  $v_h = \varepsilon_0^{-1/4}$ with associated Hawking temperature $ T = \varepsilon_0^{1/4}/\pi $.  For concreteness, we will again consider $d=4$.

Recall that, at leading order, the ingoing mode(with zero spatial momentum) admits  the near-horizon expansion
\begin{eqnarray} \label{transient_soln_1}
\phi_{0,in}(v, \omegaup) = 1 + \frac{3 i \varepsilon_0^{1/4} \omegaup}{2 \left(2 \varepsilon_0^{1/4} -i \omegaup \right)} (\varepsilon_0^{-1/4} - v) + \mathcal{O}((\varepsilon_0^{-1/4} - v)^2) + \cdots
\end{eqnarray}
To compute the transients, we introduce the dimensionless decay rate $ \lambda = \omegaup/\pi T$ and also scale the horizon to set $ v_h = 1 $. In terms of the  dimensionless parameter, the solution \eqref{transient_soln_1} when expanded near the boundary $v=0$ reads, 
\begin{eqnarray}
\phi_{0,in}(v=0, \lambda) \equiv a_0(\lambda)= 1 - \frac{3 \lambda}{2 \left(2i + \lambda \right)} + \cdots
\end{eqnarray}
The QNM corresponds to 
 $ a_0(\lambda_{qnm}) = 0$, implying that there is no contribution to the source from the leading order field. The lowest QNM frequency obtained in this method turns out to be $$ \lambda_{qnm} = -2.7668 i \pm 3.11945~, $$ which agrees with the one obtained from spectral representation method upto order $10^{-10}$.  
 
Similarly at subleading orders  $n>0 $, the sources can be made to vanish, i.e. $ a_{n>0}(\lambda_{qnm}) = 0$ by suitably fixing  $ \gamma_0$ (for $n=1$)  and $ \gamma_{n,in}$ (for $n>1$) as functions of $ \lambda_{qnm}$. For example, corresponding to this lowest $ \lambda_{qnm}$,  $\gamma_0(\lambda_{qnm})$ obtained by setting $ a_1(\lambda_{qnm}) = 0$ at the first subleading order is, 
\begin{eqnarray}
\gamma_0(\lambda_{qnm}) = -0.68669 i \pm 0.779863 ~,
\end{eqnarray}
which turns out to be same as the one computed from the outgoing mode analysis \eqref{alpha1},
\begin{eqnarray}
\gamma_0 = \frac{\omegaup_{Q}}{4 \varepsilon_0^{1/4}} = \frac{\lambda_{qnm}}{4} = -0.68669 i \pm 0.779863~. 
\end{eqnarray}
These two results agree upto order $10^{-9}$ . 

See \cite{Casalderrey-Solana:2018uag} for a recent relevant work in the large $D$ limit.

\bibliography{gravSK.bib}

\begin{thebibliography}{69}%
\makeatletter
\providecommand \@ifxundefined [1]{%
 \@ifx{#1\undefined}
}%
\providecommand \@ifnum [1]{%
 \ifnum #1\expandafter \@firstoftwo
 \else \expandafter \@secondoftwo
 \fi
}%
\providecommand \@ifx [1]{%
 \ifx #1\expandafter \@firstoftwo
 \else \expandafter \@secondoftwo
 \fi
}%
\providecommand \natexlab [1]{#1}%
\providecommand \enquote  [1]{``#1''}%
\providecommand \bibnamefont  [1]{#1}%
\providecommand \bibfnamefont [1]{#1}%
\providecommand \citenamefont [1]{#1}%
\providecommand \href@noop [0]{\@secondoftwo}%
\providecommand \href [0]{\begingroup \@sanitize@url \@href}%
\providecommand \@href[1]{\@@startlink{#1}\@@href}%
\providecommand \@@href[1]{\endgroup#1\@@endlink}%
\providecommand \@sanitize@url [0]{\catcode `\\12\catcode `\$12\catcode
  `\&12\catcode `\#12\catcode `\^12\catcode `\_12\catcode `\%12\relax}%
\providecommand \@@startlink[1]{}%
\providecommand \@@endlink[0]{}%
\providecommand \url  [0]{\begingroup\@sanitize@url \@url }%
\providecommand \@url [1]{\endgroup\@href {#1}{\urlprefix }}%
\providecommand \urlprefix  [0]{URL }%
\providecommand \Eprint [0]{\href }%
\providecommand \doibase [0]{http://dx.doi.org/}%
\providecommand \selectlanguage [0]{\@gobble}%
\providecommand \bibinfo  [0]{\@secondoftwo}%
\providecommand \bibfield  [0]{\@secondoftwo}%
\providecommand \translation [1]{[#1]}%
\providecommand \BibitemOpen [0]{}%
\providecommand \bibitemStop [0]{}%
\providecommand \bibitemNoStop [0]{.\EOS\space}%
\providecommand \EOS [0]{\spacefactor3000\relax}%
\providecommand \BibitemShut  [1]{\csname bibitem#1\endcsname}%
\let\auto@bib@innerbib\@empty
\bibitem [{\citenamefont {Berges}(2004)}]{Berges:2004yj}%
  \BibitemOpen
  \bibfield  {author} {\bibinfo {author} {\bibfnamefont {Juergen}\ \bibnamefont
  {Berges}},\ }\bibfield  {title} {\enquote {\bibinfo {title} {{Introduction to
  nonequilibrium quantum field theory}},}\ }\href {\doibase 10.1063/1.1843591}
  {\bibfield  {journal} {\bibinfo  {journal} {AIP Conf. Proc.}\ }\textbf
  {\bibinfo {volume} {739}},\ \bibinfo {pages} {3--62} (\bibinfo {year}
  {2004})},\ \Eprint {http://arxiv.org/abs/hep-ph/0409233}
  {arXiv:hep-ph/0409233} \BibitemShut {NoStop}%
\bibitem [{\citenamefont {Sieberer}\ \emph {et~al.}(2016)\citenamefont
  {Sieberer}, \citenamefont {Buchhold},\ and\ \citenamefont
  {Diehl}}]{Sieberer_2016}%
  \BibitemOpen
  \bibfield  {author} {\bibinfo {author} {\bibfnamefont {L~M}\ \bibnamefont
  {Sieberer}}, \bibinfo {author} {\bibfnamefont {M}~\bibnamefont {Buchhold}}, \
  and\ \bibinfo {author} {\bibfnamefont {S}~\bibnamefont {Diehl}},\ }\bibfield
  {title} {\enquote {\bibinfo {title} {Keldysh field theory for driven open
  quantum systems},}\ }\href {\doibase 10.1088/0034-4885/79/9/096001}
  {\bibfield  {journal} {\bibinfo  {journal} {Reports on Progress in Physics}\
  }\textbf {\bibinfo {volume} {79}},\ \bibinfo {pages} {096001} (\bibinfo
  {year} {2016})}\BibitemShut {NoStop}%
\bibitem [{\citenamefont {Heyl}(2018)}]{Heyl:2017blm}%
  \BibitemOpen
  \bibfield  {author} {\bibinfo {author} {\bibfnamefont {Markus}\ \bibnamefont
  {Heyl}},\ }\bibfield  {title} {\enquote {\bibinfo {title} {{Dynamical quantum
  phase transitions: a review}},}\ }\href {\doibase 10.1088/1361-6633/aaaf9a}
  {\bibfield  {journal} {\bibinfo  {journal} {Rept. Prog. Phys.}\ }\textbf
  {\bibinfo {volume} {81}},\ \bibinfo {pages} {054001} (\bibinfo {year}
  {2018})},\ \Eprint {http://arxiv.org/abs/1709.07461} {arXiv:1709.07461
  [cond-mat.stat-mech]} \BibitemShut {NoStop}%
\bibitem [{\citenamefont {Mori}\ \emph {et~al.}(2018)\citenamefont {Mori},
  \citenamefont {Ikeda}, \citenamefont {Kaminishi},\ and\ \citenamefont
  {Ueda}}]{Mori:2017qhg}%
  \BibitemOpen
  \bibfield  {author} {\bibinfo {author} {\bibfnamefont {Takashi}\ \bibnamefont
  {Mori}}, \bibinfo {author} {\bibfnamefont {TatsuhikoN.}\ \bibnamefont
  {Ikeda}}, \bibinfo {author} {\bibfnamefont {Eriko}\ \bibnamefont
  {Kaminishi}}, \ and\ \bibinfo {author} {\bibfnamefont {Masahito}\
  \bibnamefont {Ueda}},\ }\bibfield  {title} {\enquote {\bibinfo {title}
  {{Thermalization and prethermalization in isolated quantum systems: a
  theoretical overview}},}\ }\href {\doibase 10.1088/1361-6455/aabcdf}
  {\bibfield  {journal} {\bibinfo  {journal} {J. Phys. B}\ }\textbf {\bibinfo
  {volume} {51}},\ \bibinfo {pages} {112001} (\bibinfo {year} {2018})},\
  \Eprint {http://arxiv.org/abs/1712.08790} {arXiv:1712.08790
  [cond-mat.stat-mech]} \BibitemShut {NoStop}%
\bibitem [{\citenamefont {Policastro}\ \emph {et~al.}(2001)\citenamefont
  {Policastro}, \citenamefont {Son},\ and\ \citenamefont
  {Starinets}}]{Policastro:2001yc}%
  \BibitemOpen
  \bibfield  {author} {\bibinfo {author} {\bibfnamefont {G.}~\bibnamefont
  {Policastro}}, \bibinfo {author} {\bibfnamefont {Dan~T.}\ \bibnamefont
  {Son}}, \ and\ \bibinfo {author} {\bibfnamefont {Andrei~O.}\ \bibnamefont
  {Starinets}},\ }\bibfield  {title} {\enquote {\bibinfo {title} {{The Shear
  viscosity of strongly coupled N=4 supersymmetric Yang-Mills plasma}},}\
  }\href {\doibase 10.1103/PhysRevLett.87.081601} {\bibfield  {journal}
  {\bibinfo  {journal} {Phys. Rev. Lett.}\ }\textbf {\bibinfo {volume} {87}},\
  \bibinfo {pages} {081601} (\bibinfo {year} {2001})},\ \Eprint
  {http://arxiv.org/abs/hep-th/0104066} {arXiv:hep-th/0104066} \BibitemShut
  {NoStop}%
\bibitem [{\citenamefont {Bhattacharyya}\ \emph {et~al.}(2008)\citenamefont
  {Bhattacharyya}, \citenamefont {Hubeny}, \citenamefont {Minwalla},\ and\
  \citenamefont {Rangamani}}]{Bhattacharyya:2007vjd}%
  \BibitemOpen
  \bibfield  {author} {\bibinfo {author} {\bibfnamefont {Sayantani}\
  \bibnamefont {Bhattacharyya}}, \bibinfo {author} {\bibfnamefont {Veronika~E}\
  \bibnamefont {Hubeny}}, \bibinfo {author} {\bibfnamefont {Shiraz}\
  \bibnamefont {Minwalla}}, \ and\ \bibinfo {author} {\bibfnamefont {Mukund}\
  \bibnamefont {Rangamani}},\ }\bibfield  {title} {\enquote {\bibinfo {title}
  {{Nonlinear Fluid Dynamics from Gravity}},}\ }\href {\doibase
  10.1088/1126-6708/2008/02/045} {\bibfield  {journal} {\bibinfo  {journal}
  {JHEP}\ }\textbf {\bibinfo {volume} {02}},\ \bibinfo {pages} {045} (\bibinfo
  {year} {2008})},\ \Eprint {http://arxiv.org/abs/0712.2456} {arXiv:0712.2456
  [hep-th]} \BibitemShut {NoStop}%
\bibitem [{\citenamefont {Baier}\ \emph {et~al.}(2008)\citenamefont {Baier},
  \citenamefont {Romatschke}, \citenamefont {Son}, \citenamefont {Starinets},\
  and\ \citenamefont {Stephanov}}]{Baier:2007ix}%
  \BibitemOpen
  \bibfield  {author} {\bibinfo {author} {\bibfnamefont {Rudolf}\ \bibnamefont
  {Baier}}, \bibinfo {author} {\bibfnamefont {Paul}\ \bibnamefont
  {Romatschke}}, \bibinfo {author} {\bibfnamefont {Dam~Thanh}\ \bibnamefont
  {Son}}, \bibinfo {author} {\bibfnamefont {Andrei~O.}\ \bibnamefont
  {Starinets}}, \ and\ \bibinfo {author} {\bibfnamefont {Mikhail~A.}\
  \bibnamefont {Stephanov}},\ }\bibfield  {title} {\enquote {\bibinfo {title}
  {{Relativistic viscous hydrodynamics, conformal invariance, and
  holography}},}\ }\href {\doibase 10.1088/1126-6708/2008/04/100} {\bibfield
  {journal} {\bibinfo  {journal} {JHEP}\ }\textbf {\bibinfo {volume} {04}},\
  \bibinfo {pages} {100} (\bibinfo {year} {2008})},\ \Eprint
  {http://arxiv.org/abs/0712.2451} {arXiv:0712.2451 [hep-th]} \BibitemShut
  {NoStop}%
\bibitem [{\citenamefont {Rangamani}(2009)}]{Rangamani:2009xk}%
  \BibitemOpen
  \bibfield  {author} {\bibinfo {author} {\bibfnamefont {Mukund}\ \bibnamefont
  {Rangamani}},\ }\bibfield  {title} {\enquote {\bibinfo {title} {{Gravity and
  Hydrodynamics: Lectures on the fluid-gravity correspondence}},}\ }\href
  {\doibase 10.1088/0264-9381/26/22/224003} {\bibfield  {journal} {\bibinfo
  {journal} {Class. Quant. Grav.}\ }\textbf {\bibinfo {volume} {26}},\ \bibinfo
  {pages} {224003} (\bibinfo {year} {2009})},\ \Eprint
  {http://arxiv.org/abs/0905.4352} {arXiv:0905.4352 [hep-th]} \BibitemShut
  {NoStop}%
\bibitem [{\citenamefont {Chesler}\ and\ \citenamefont
  {Yaffe}(2014)}]{Chesler:2013lia}%
  \BibitemOpen
  \bibfield  {author} {\bibinfo {author} {\bibfnamefont {Paul~M.}\ \bibnamefont
  {Chesler}}\ and\ \bibinfo {author} {\bibfnamefont {Laurence~G.}\ \bibnamefont
  {Yaffe}},\ }\bibfield  {title} {\enquote {\bibinfo {title} {{Numerical
  solution of gravitational dynamics in asymptotically anti-de Sitter
  spacetimes}},}\ }\href {\doibase 10.1007/JHEP07(2014)086} {\bibfield
  {journal} {\bibinfo  {journal} {JHEP}\ }\textbf {\bibinfo {volume} {07}},\
  \bibinfo {pages} {086} (\bibinfo {year} {2014})},\ \Eprint
  {http://arxiv.org/abs/1309.1439} {arXiv:1309.1439 [hep-th]} \BibitemShut
  {NoStop}%
\bibitem [{\citenamefont {Banerjee}\ \emph {et~al.}(2016)\citenamefont
  {Banerjee}, \citenamefont {Ishii}, \citenamefont {Joshi}, \citenamefont
  {Mukhopadhyay},\ and\ \citenamefont {Ramadevi}}]{Banerjee:2016ray}%
  \BibitemOpen
  \bibfield  {author} {\bibinfo {author} {\bibfnamefont {Souvik}\ \bibnamefont
  {Banerjee}}, \bibinfo {author} {\bibfnamefont {Takaaki}\ \bibnamefont
  {Ishii}}, \bibinfo {author} {\bibfnamefont {Lata~Kh}\ \bibnamefont {Joshi}},
  \bibinfo {author} {\bibfnamefont {Ayan}\ \bibnamefont {Mukhopadhyay}}, \ and\
  \bibinfo {author} {\bibfnamefont {P.}~\bibnamefont {Ramadevi}},\ }\bibfield
  {title} {\enquote {\bibinfo {title} {{Time-dependence of the holographic
  spectral function: Diverse routes to thermalisation}},}\ }\href {\doibase
  10.1007/JHEP08(2016)048} {\bibfield  {journal} {\bibinfo  {journal} {JHEP}\
  }\textbf {\bibinfo {volume} {08}},\ \bibinfo {pages} {048} (\bibinfo {year}
  {2016})},\ \Eprint {http://arxiv.org/abs/1603.06935} {arXiv:1603.06935
  [hep-th]} \BibitemShut {NoStop}%
\bibitem [{\citenamefont {Balasubramanian}\ \emph {et~al.}(2011)\citenamefont
  {Balasubramanian}, \citenamefont {Bernamonti}, \citenamefont {de~Boer},
  \citenamefont {Copland}, \citenamefont {Craps}, \citenamefont
  {Keski-Vakkuri}, \citenamefont {Muller}, \citenamefont {Schafer},
  \citenamefont {Shigemori},\ and\ \citenamefont
  {Staessens}}]{Balasubramanian:2011ur}%
  \BibitemOpen
  \bibfield  {author} {\bibinfo {author} {\bibfnamefont {V.}~\bibnamefont
  {Balasubramanian}}, \bibinfo {author} {\bibfnamefont {A.}~\bibnamefont
  {Bernamonti}}, \bibinfo {author} {\bibfnamefont {J.}~\bibnamefont {de~Boer}},
  \bibinfo {author} {\bibfnamefont {N.}~\bibnamefont {Copland}}, \bibinfo
  {author} {\bibfnamefont {B.}~\bibnamefont {Craps}}, \bibinfo {author}
  {\bibfnamefont {E.}~\bibnamefont {Keski-Vakkuri}}, \bibinfo {author}
  {\bibfnamefont {B.}~\bibnamefont {Muller}}, \bibinfo {author} {\bibfnamefont
  {A.}~\bibnamefont {Schafer}}, \bibinfo {author} {\bibfnamefont
  {M.}~\bibnamefont {Shigemori}}, \ and\ \bibinfo {author} {\bibfnamefont
  {W.}~\bibnamefont {Staessens}},\ }\bibfield  {title} {\enquote {\bibinfo
  {title} {{Holographic Thermalization}},}\ }\href {\doibase
  10.1103/PhysRevD.84.026010} {\bibfield  {journal} {\bibinfo  {journal} {Phys.
  Rev. D}\ }\textbf {\bibinfo {volume} {84}},\ \bibinfo {pages} {026010}
  (\bibinfo {year} {2011})},\ \Eprint {http://arxiv.org/abs/1103.2683}
  {arXiv:1103.2683 [hep-th]} \BibitemShut {NoStop}%
\bibitem [{\citenamefont {Le~Bellac}(1996)}]{le_bellac_1996}%
  \BibitemOpen
  \bibfield  {author} {\bibinfo {author} {\bibfnamefont {Michel}\ \bibnamefont
  {Le~Bellac}},\ }\href {\doibase 10.1017/CBO9780511721700} {\emph {\bibinfo
  {title} {Thermal Field Theory}}},\ Cambridge Monographs on Mathematical
  Physics\ (\bibinfo  {publisher} {Cambridge University Press},\ \bibinfo
  {year} {1996})\BibitemShut {NoStop}%
\bibitem [{\citenamefont {Touchette}(2009)}]{TOUCHETTE20091}%
  \BibitemOpen
  \bibfield  {author} {\bibinfo {author} {\bibfnamefont {Hugo}\ \bibnamefont
  {Touchette}},\ }\bibfield  {title} {\enquote {\bibinfo {title} {The large
  deviation approach to statistical mechanics},}\ }\href {\doibase
  https://doi.org/10.1016/j.physrep.2009.05.002} {\bibfield  {journal}
  {\bibinfo  {journal} {Physics Reports}\ }\textbf {\bibinfo {volume} {478}},\
  \bibinfo {pages} {1--69} (\bibinfo {year} {2009})}\BibitemShut {NoStop}%
\bibitem [{\citenamefont {Bernard}(2021)}]{Bernard_2021}%
  \BibitemOpen
  \bibfield  {author} {\bibinfo {author} {\bibfnamefont {Denis}\ \bibnamefont
  {Bernard}},\ }\bibfield  {title} {\enquote {\bibinfo {title} {Can the
  macroscopic fluctuation theory be quantized?}}\ }\href {\doibase
  10.1088/1751-8121/ac2597} {\bibfield  {journal} {\bibinfo  {journal} {Journal
  of Physics A: Mathematical and Theoretical}\ }\textbf {\bibinfo {volume}
  {54}},\ \bibinfo {pages} {433001} (\bibinfo {year} {2021})}\BibitemShut
  {NoStop}%
\bibitem [{\citenamefont {Son}\ and\ \citenamefont
  {Starinets}(2002)}]{Son:2002sd}%
  \BibitemOpen
  \bibfield  {author} {\bibinfo {author} {\bibfnamefont {Dam~T.}\ \bibnamefont
  {Son}}\ and\ \bibinfo {author} {\bibfnamefont {Andrei~O.}\ \bibnamefont
  {Starinets}},\ }\bibfield  {title} {\enquote {\bibinfo {title} {{Minkowski
  space correlators in AdS / CFT correspondence: Recipe and applications}},}\
  }\href {\doibase 10.1088/1126-6708/2002/09/042} {\bibfield  {journal}
  {\bibinfo  {journal} {JHEP}\ }\textbf {\bibinfo {volume} {09}},\ \bibinfo
  {pages} {042} (\bibinfo {year} {2002})},\ \Eprint
  {http://arxiv.org/abs/hep-th/0205051} {arXiv:hep-th/0205051} \BibitemShut
  {NoStop}%
\bibitem [{\citenamefont {Herzog}\ and\ \citenamefont
  {Son}(2003)}]{Herzog:2002pc}%
  \BibitemOpen
  \bibfield  {author} {\bibinfo {author} {\bibfnamefont {C.~P.}\ \bibnamefont
  {Herzog}}\ and\ \bibinfo {author} {\bibfnamefont {D.~T.}\ \bibnamefont
  {Son}},\ }\bibfield  {title} {\enquote {\bibinfo {title} {{Schwinger-Keldysh
  propagators from AdS/CFT correspondence}},}\ }\href {\doibase
  10.1088/1126-6708/2003/03/046} {\bibfield  {journal} {\bibinfo  {journal}
  {JHEP}\ }\textbf {\bibinfo {volume} {03}},\ \bibinfo {pages} {046} (\bibinfo
  {year} {2003})},\ \Eprint {http://arxiv.org/abs/hep-th/0212072}
  {arXiv:hep-th/0212072} \BibitemShut {NoStop}%
\bibitem [{\citenamefont {Skenderis}\ and\ \citenamefont {van
  Rees}(2008)}]{Skenderis:2008dh}%
  \BibitemOpen
  \bibfield  {author} {\bibinfo {author} {\bibfnamefont {Kostas}\ \bibnamefont
  {Skenderis}}\ and\ \bibinfo {author} {\bibfnamefont {Balt~C.}\ \bibnamefont
  {van Rees}},\ }\bibfield  {title} {\enquote {\bibinfo {title} {{Real-time
  gauge/gravity duality}},}\ }\href {\doibase 10.1103/PhysRevLett.101.081601}
  {\bibfield  {journal} {\bibinfo  {journal} {Phys. Rev. Lett.}\ }\textbf
  {\bibinfo {volume} {101}},\ \bibinfo {pages} {081601} (\bibinfo {year}
  {2008})},\ \Eprint {http://arxiv.org/abs/0805.0150} {arXiv:0805.0150
  [hep-th]} \BibitemShut {NoStop}%
\bibitem [{\citenamefont {Skenderis}\ and\ \citenamefont {van
  Rees}(2009)}]{Skenderis:2008dg}%
  \BibitemOpen
  \bibfield  {author} {\bibinfo {author} {\bibfnamefont {Kostas}\ \bibnamefont
  {Skenderis}}\ and\ \bibinfo {author} {\bibfnamefont {Balt~C.}\ \bibnamefont
  {van Rees}},\ }\bibfield  {title} {\enquote {\bibinfo {title} {{Real-time
  gauge/gravity duality: Prescription, Renormalization and Examples}},}\ }\href
  {\doibase 10.1088/1126-6708/2009/05/085} {\bibfield  {journal} {\bibinfo
  {journal} {JHEP}\ }\textbf {\bibinfo {volume} {05}},\ \bibinfo {pages} {085}
  (\bibinfo {year} {2009})},\ \Eprint {http://arxiv.org/abs/0812.2909}
  {arXiv:0812.2909 [hep-th]} \BibitemShut {NoStop}%
\bibitem [{\citenamefont {Glorioso}\ \emph {et~al.}(2018)\citenamefont
  {Glorioso}, \citenamefont {Crossley},\ and\ \citenamefont
  {Liu}}]{Glorioso:2018mmw}%
  \BibitemOpen
  \bibfield  {author} {\bibinfo {author} {\bibfnamefont {Paolo}\ \bibnamefont
  {Glorioso}}, \bibinfo {author} {\bibfnamefont {Michael}\ \bibnamefont
  {Crossley}}, \ and\ \bibinfo {author} {\bibfnamefont {Hong}\ \bibnamefont
  {Liu}},\ }\bibfield  {title} {\enquote {\bibinfo {title} {{A prescription for
  holographic Schwinger-Keldysh contour in non-equilibrium systems}},}\
  }\href@noop {} {\  (\bibinfo {year} {2018})},\ \Eprint
  {http://arxiv.org/abs/1812.08785} {arXiv:1812.08785 [hep-th]} \BibitemShut
  {NoStop}%
\bibitem [{\citenamefont {Henningson}\ and\ \citenamefont
  {Skenderis}(1998)}]{Henningson:1998gx}%
  \BibitemOpen
  \bibfield  {author} {\bibinfo {author} {\bibfnamefont {M.}~\bibnamefont
  {Henningson}}\ and\ \bibinfo {author} {\bibfnamefont {K.}~\bibnamefont
  {Skenderis}},\ }\bibfield  {title} {\enquote {\bibinfo {title} {{The
  Holographic Weyl anomaly}},}\ }\href {\doibase 10.1088/1126-6708/1998/07/023}
  {\bibfield  {journal} {\bibinfo  {journal} {JHEP}\ }\textbf {\bibinfo
  {volume} {07}},\ \bibinfo {pages} {023} (\bibinfo {year} {1998})},\ \Eprint
  {http://arxiv.org/abs/hep-th/9806087} {arXiv:hep-th/9806087} \BibitemShut
  {NoStop}%
\bibitem [{\citenamefont {de~Haro}\ \emph {et~al.}(2001)\citenamefont
  {de~Haro}, \citenamefont {Solodukhin},\ and\ \citenamefont
  {Skenderis}}]{deHaro:2000vlm}%
  \BibitemOpen
  \bibfield  {author} {\bibinfo {author} {\bibfnamefont {Sebastian}\
  \bibnamefont {de~Haro}}, \bibinfo {author} {\bibfnamefont {Sergey~N.}\
  \bibnamefont {Solodukhin}}, \ and\ \bibinfo {author} {\bibfnamefont {Kostas}\
  \bibnamefont {Skenderis}},\ }\bibfield  {title} {\enquote {\bibinfo {title}
  {{Holographic reconstruction of space-time and renormalization in the AdS /
  CFT correspondence}},}\ }\href {\doibase 10.1007/s002200100381} {\bibfield
  {journal} {\bibinfo  {journal} {Commun. Math. Phys.}\ }\textbf {\bibinfo
  {volume} {217}},\ \bibinfo {pages} {595--622} (\bibinfo {year} {2001})},\
  \Eprint {http://arxiv.org/abs/hep-th/0002230} {arXiv:hep-th/0002230}
  \BibitemShut {NoStop}%
\bibitem [{\citenamefont {Heller}\ \emph {et~al.}(2013)\citenamefont {Heller},
  \citenamefont {Janik},\ and\ \citenamefont {Witaszczyk}}]{Heller:2013fn}%
  \BibitemOpen
  \bibfield  {author} {\bibinfo {author} {\bibfnamefont {Michal~P.}\
  \bibnamefont {Heller}}, \bibinfo {author} {\bibfnamefont {Romuald~A.}\
  \bibnamefont {Janik}}, \ and\ \bibinfo {author} {\bibfnamefont {Przemyslaw}\
  \bibnamefont {Witaszczyk}},\ }\bibfield  {title} {\enquote {\bibinfo {title}
  {{Hydrodynamic Gradient Expansion in Gauge Theory Plasmas}},}\ }\href
  {\doibase 10.1103/PhysRevLett.110.211602} {\bibfield  {journal} {\bibinfo
  {journal} {Phys. Rev. Lett.}\ }\textbf {\bibinfo {volume} {110}},\ \bibinfo
  {pages} {211602} (\bibinfo {year} {2013})},\ \Eprint
  {http://arxiv.org/abs/1302.0697} {arXiv:1302.0697 [hep-th]} \BibitemShut
  {NoStop}%
\bibitem [{\citenamefont {Janik}\ and\ \citenamefont
  {Peschanski}(2006{\natexlab{a}})}]{Janik:2006gp}%
  \BibitemOpen
  \bibfield  {author} {\bibinfo {author} {\bibfnamefont {Romuald~A.}\
  \bibnamefont {Janik}}\ and\ \bibinfo {author} {\bibfnamefont {Robert~B.}\
  \bibnamefont {Peschanski}},\ }\bibfield  {title} {\enquote {\bibinfo {title}
  {{Gauge/gravity duality and thermalization of a boost-invariant perfect
  fluid}},}\ }\href {\doibase 10.1103/PhysRevD.74.046007} {\bibfield  {journal}
  {\bibinfo  {journal} {Phys. Rev. D}\ }\textbf {\bibinfo {volume} {74}},\
  \bibinfo {pages} {046007} (\bibinfo {year} {2006}{\natexlab{a}})},\ \Eprint
  {http://arxiv.org/abs/hep-th/0606149} {arXiv:hep-th/0606149} \BibitemShut
  {NoStop}%
\bibitem [{\citenamefont {Ghosh}\ \emph {et~al.}(2021)\citenamefont {Ghosh},
  \citenamefont {Loganayagam}, \citenamefont {Prabhu}, \citenamefont
  {Rangamani}, \citenamefont {Sivakumar},\ and\ \citenamefont
  {Vishal}}]{Ghosh:2020lel}%
  \BibitemOpen
  \bibfield  {author} {\bibinfo {author} {\bibfnamefont {Jewel~K.}\
  \bibnamefont {Ghosh}}, \bibinfo {author} {\bibfnamefont {R.}~\bibnamefont
  {Loganayagam}}, \bibinfo {author} {\bibfnamefont {Siddharth~G.}\ \bibnamefont
  {Prabhu}}, \bibinfo {author} {\bibfnamefont {Mukund}\ \bibnamefont
  {Rangamani}}, \bibinfo {author} {\bibfnamefont {Akhil}\ \bibnamefont
  {Sivakumar}}, \ and\ \bibinfo {author} {\bibfnamefont {V.}~\bibnamefont
  {Vishal}},\ }\bibfield  {title} {\enquote {\bibinfo {title} {{Effective field
  theory of stochastic diffusion from gravity}},}\ }\href {\doibase
  10.1007/JHEP05(2021)130} {\bibfield  {journal} {\bibinfo  {journal} {JHEP}\
  }\textbf {\bibinfo {volume} {05}},\ \bibinfo {pages} {130} (\bibinfo {year}
  {2021})},\ \Eprint {http://arxiv.org/abs/2012.03999} {arXiv:2012.03999
  [hep-th]} \BibitemShut {NoStop}%
\bibitem [{\citenamefont {Crossley}\ \emph {et~al.}(2017)\citenamefont
  {Crossley}, \citenamefont {Glorioso},\ and\ \citenamefont
  {Liu}}]{Crossley:2015evo}%
  \BibitemOpen
  \bibfield  {author} {\bibinfo {author} {\bibfnamefont {Michael}\ \bibnamefont
  {Crossley}}, \bibinfo {author} {\bibfnamefont {Paolo}\ \bibnamefont
  {Glorioso}}, \ and\ \bibinfo {author} {\bibfnamefont {Hong}\ \bibnamefont
  {Liu}},\ }\bibfield  {title} {\enquote {\bibinfo {title} {{Effective field
  theory of dissipative fluids}},}\ }\href {\doibase 10.1007/JHEP09(2017)095}
  {\bibfield  {journal} {\bibinfo  {journal} {JHEP}\ }\textbf {\bibinfo
  {volume} {09}},\ \bibinfo {pages} {095} (\bibinfo {year} {2017})},\ \Eprint
  {http://arxiv.org/abs/1511.03646} {arXiv:1511.03646 [hep-th]} \BibitemShut
  {NoStop}%
\bibitem [{\citenamefont {Glorioso}\ \emph {et~al.}(2017)\citenamefont
  {Glorioso}, \citenamefont {Crossley},\ and\ \citenamefont
  {Liu}}]{Glorioso:2017fpd}%
  \BibitemOpen
  \bibfield  {author} {\bibinfo {author} {\bibfnamefont {Paolo}\ \bibnamefont
  {Glorioso}}, \bibinfo {author} {\bibfnamefont {Michael}\ \bibnamefont
  {Crossley}}, \ and\ \bibinfo {author} {\bibfnamefont {Hong}\ \bibnamefont
  {Liu}},\ }\bibfield  {title} {\enquote {\bibinfo {title} {{Effective field
  theory of dissipative fluids (II): classical limit, dynamical KMS symmetry
  and entropy current}},}\ }\href {\doibase 10.1007/JHEP09(2017)096} {\bibfield
   {journal} {\bibinfo  {journal} {JHEP}\ }\textbf {\bibinfo {volume} {09}},\
  \bibinfo {pages} {096} (\bibinfo {year} {2017})},\ \Eprint
  {http://arxiv.org/abs/1701.07817} {arXiv:1701.07817 [hep-th]} \BibitemShut
  {NoStop}%
\bibitem [{\citenamefont {Glorioso}\ \emph {et~al.}(2019)\citenamefont
  {Glorioso}, \citenamefont {Liu},\ and\ \citenamefont
  {Rajagopal}}]{Glorioso:2017lcn}%
  \BibitemOpen
  \bibfield  {author} {\bibinfo {author} {\bibfnamefont {Paolo}\ \bibnamefont
  {Glorioso}}, \bibinfo {author} {\bibfnamefont {Hong}\ \bibnamefont {Liu}}, \
  and\ \bibinfo {author} {\bibfnamefont {Srivatsan}\ \bibnamefont
  {Rajagopal}},\ }\bibfield  {title} {\enquote {\bibinfo {title} {{Global
  Anomalies, Discrete Symmetries, and Hydrodynamic Effective Actions}},}\
  }\href {\doibase 10.1007/JHEP01(2019)043} {\bibfield  {journal} {\bibinfo
  {journal} {JHEP}\ }\textbf {\bibinfo {volume} {01}},\ \bibinfo {pages} {043}
  (\bibinfo {year} {2019})},\ \Eprint {http://arxiv.org/abs/1710.03768}
  {arXiv:1710.03768 [hep-th]} \BibitemShut {NoStop}%
\bibitem [{\citenamefont {Jensen}\ \emph {et~al.}(2018)\citenamefont {Jensen},
  \citenamefont {Marjieh}, \citenamefont {Pinzani-Fokeeva},\ and\ \citenamefont
  {Yarom}}]{Jensen:2018hse}%
  \BibitemOpen
  \bibfield  {author} {\bibinfo {author} {\bibfnamefont {Kristan}\ \bibnamefont
  {Jensen}}, \bibinfo {author} {\bibfnamefont {Raja}\ \bibnamefont {Marjieh}},
  \bibinfo {author} {\bibfnamefont {Natalia}\ \bibnamefont {Pinzani-Fokeeva}},
  \ and\ \bibinfo {author} {\bibfnamefont {Amos}\ \bibnamefont {Yarom}},\
  }\bibfield  {title} {\enquote {\bibinfo {title} {{A panoply of
  Schwinger-Keldysh transport}},}\ }\href {\doibase
  10.21468/SciPostPhys.5.5.053} {\bibfield  {journal} {\bibinfo  {journal}
  {SciPost Phys.}\ }\textbf {\bibinfo {volume} {5}},\ \bibinfo {pages} {053}
  (\bibinfo {year} {2018})},\ \Eprint {http://arxiv.org/abs/1804.04654}
  {arXiv:1804.04654 [hep-th]} \BibitemShut {NoStop}%
\bibitem [{\citenamefont {Liu}\ and\ \citenamefont
  {Glorioso}(2018)}]{Liu:2018kfw}%
  \BibitemOpen
  \bibfield  {author} {\bibinfo {author} {\bibfnamefont {Hong}\ \bibnamefont
  {Liu}}\ and\ \bibinfo {author} {\bibfnamefont {Paolo}\ \bibnamefont
  {Glorioso}},\ }\bibfield  {title} {\enquote {\bibinfo {title} {{Lectures on
  non-equilibrium effective field theories and fluctuating hydrodynamics}},}\
  }\href {\doibase 10.22323/1.305.0008} {\bibfield  {journal} {\bibinfo
  {journal} {PoS}\ }\textbf {\bibinfo {volume} {TASI2017}},\ \bibinfo {pages}
  {008} (\bibinfo {year} {2018})},\ \Eprint {http://arxiv.org/abs/1805.09331}
  {arXiv:1805.09331 [hep-th]} \BibitemShut {NoStop}%
\bibitem [{\citenamefont {de~Boer}\ \emph {et~al.}(2019)\citenamefont
  {de~Boer}, \citenamefont {Heller},\ and\ \citenamefont
  {Pinzani-Fokeeva}}]{deBoer:2018qqm}%
  \BibitemOpen
  \bibfield  {author} {\bibinfo {author} {\bibfnamefont {Jan}\ \bibnamefont
  {de~Boer}}, \bibinfo {author} {\bibfnamefont {Michal~P.}\ \bibnamefont
  {Heller}}, \ and\ \bibinfo {author} {\bibfnamefont {Natalia}\ \bibnamefont
  {Pinzani-Fokeeva}},\ }\bibfield  {title} {\enquote {\bibinfo {title}
  {{Holographic Schwinger-Keldysh effective field theories}},}\ }\href
  {\doibase 10.1007/JHEP05(2019)188} {\bibfield  {journal} {\bibinfo  {journal}
  {JHEP}\ }\textbf {\bibinfo {volume} {05}},\ \bibinfo {pages} {188} (\bibinfo
  {year} {2019})},\ \Eprint {http://arxiv.org/abs/1812.06093} {arXiv:1812.06093
  [hep-th]} \BibitemShut {NoStop}%
\bibitem [{\citenamefont {Chakrabarty}\ \emph {et~al.}(2020)\citenamefont
  {Chakrabarty}, \citenamefont {Chakravarty}, \citenamefont {Chaudhuri},
  \citenamefont {Jana}, \citenamefont {Loganayagam},\ and\ \citenamefont
  {Sivakumar}}]{Chakrabarty:2019aeu}%
  \BibitemOpen
  \bibfield  {author} {\bibinfo {author} {\bibfnamefont {Bidisha}\ \bibnamefont
  {Chakrabarty}}, \bibinfo {author} {\bibfnamefont {Joydeep}\ \bibnamefont
  {Chakravarty}}, \bibinfo {author} {\bibfnamefont {Soumyadeep}\ \bibnamefont
  {Chaudhuri}}, \bibinfo {author} {\bibfnamefont {Chandan}\ \bibnamefont
  {Jana}}, \bibinfo {author} {\bibfnamefont {R.}~\bibnamefont {Loganayagam}}, \
  and\ \bibinfo {author} {\bibfnamefont {Akhil}\ \bibnamefont {Sivakumar}},\
  }\bibfield  {title} {\enquote {\bibinfo {title} {{Nonlinear Langevin dynamics
  via holography}},}\ }\href {\doibase 10.1007/JHEP01(2020)165} {\bibfield
  {journal} {\bibinfo  {journal} {JHEP}\ }\textbf {\bibinfo {volume} {01}},\
  \bibinfo {pages} {165} (\bibinfo {year} {2020})},\ \Eprint
  {http://arxiv.org/abs/1906.07762} {arXiv:1906.07762 [hep-th]} \BibitemShut
  {NoStop}%
\bibitem [{\citenamefont {He}\ \emph {et~al.}(2022{\natexlab{a}})\citenamefont
  {He}, \citenamefont {Loganayagam}, \citenamefont {Rangamani},\ and\
  \citenamefont {Virrueta}}]{He:2021jna}%
  \BibitemOpen
  \bibfield  {author} {\bibinfo {author} {\bibfnamefont {Temple}\ \bibnamefont
  {He}}, \bibinfo {author} {\bibfnamefont {R.}~\bibnamefont {Loganayagam}},
  \bibinfo {author} {\bibfnamefont {Mukund}\ \bibnamefont {Rangamani}}, \ and\
  \bibinfo {author} {\bibfnamefont {Julio}\ \bibnamefont {Virrueta}},\
  }\bibfield  {title} {\enquote {\bibinfo {title} {{An effective description of
  momentum diffusion in a charged plasma from holography}},}\ }\href {\doibase
  10.1007/JHEP01(2022)145} {\bibfield  {journal} {\bibinfo  {journal} {JHEP}\
  }\textbf {\bibinfo {volume} {01}},\ \bibinfo {pages} {145} (\bibinfo {year}
  {2022}{\natexlab{a}})},\ \Eprint {http://arxiv.org/abs/2108.03244}
  {arXiv:2108.03244 [hep-th]} \BibitemShut {NoStop}%
\bibitem [{\citenamefont {Jana}\ \emph {et~al.}(2020)\citenamefont {Jana},
  \citenamefont {Loganayagam},\ and\ \citenamefont {Rangamani}}]{Jana:2020vyx}%
  \BibitemOpen
  \bibfield  {author} {\bibinfo {author} {\bibfnamefont {Chandan}\ \bibnamefont
  {Jana}}, \bibinfo {author} {\bibfnamefont {R.}~\bibnamefont {Loganayagam}}, \
  and\ \bibinfo {author} {\bibfnamefont {Mukund}\ \bibnamefont {Rangamani}},\
  }\bibfield  {title} {\enquote {\bibinfo {title} {{Open quantum systems and
  Schwinger-Keldysh holograms}},}\ }\href {\doibase 10.1007/JHEP07(2020)242}
  {\bibfield  {journal} {\bibinfo  {journal} {JHEP}\ }\textbf {\bibinfo
  {volume} {07}},\ \bibinfo {pages} {242} (\bibinfo {year} {2020})},\ \Eprint
  {http://arxiv.org/abs/2004.02888} {arXiv:2004.02888 [hep-th]} \BibitemShut
  {NoStop}%
\bibitem [{\citenamefont {He}\ \emph {et~al.}(2022{\natexlab{b}})\citenamefont
  {He}, \citenamefont {Loganayagam}, \citenamefont {Rangamani}, \citenamefont
  {Sivakumar},\ and\ \citenamefont {Virrueta}}]{He:2022jnc}%
  \BibitemOpen
  \bibfield  {author} {\bibinfo {author} {\bibfnamefont {Temple}\ \bibnamefont
  {He}}, \bibinfo {author} {\bibfnamefont {R.}~\bibnamefont {Loganayagam}},
  \bibinfo {author} {\bibfnamefont {Mukund}\ \bibnamefont {Rangamani}},
  \bibinfo {author} {\bibfnamefont {Akhil}\ \bibnamefont {Sivakumar}}, \ and\
  \bibinfo {author} {\bibfnamefont {Julio}\ \bibnamefont {Virrueta}},\
  }\bibfield  {title} {\enquote {\bibinfo {title} {{The timbre of Hawking
  gravitons: an effective description of energy transport from holography}},}\
  }\href {\doibase 10.1007/JHEP09(2022)092} {\bibfield  {journal} {\bibinfo
  {journal} {JHEP}\ }\textbf {\bibinfo {volume} {09}},\ \bibinfo {pages} {092}
  (\bibinfo {year} {2022}{\natexlab{b}})},\ \Eprint
  {http://arxiv.org/abs/2202.04079} {arXiv:2202.04079 [hep-th]} \BibitemShut
  {NoStop}%
\bibitem [{\citenamefont {van Rees}(2009)}]{vanRees:2009rw}%
  \BibitemOpen
  \bibfield  {author} {\bibinfo {author} {\bibfnamefont {Balt~C.}\ \bibnamefont
  {van Rees}},\ }\bibfield  {title} {\enquote {\bibinfo {title} {{Real-time
  gauge/gravity duality and ingoing boundary conditions}},}\ }\href {\doibase
  10.1016/j.nuclphysbps.2009.07.078} {\bibfield  {journal} {\bibinfo  {journal}
  {Nucl. Phys. B Proc. Suppl.}\ }\textbf {\bibinfo {volume} {192-193}},\
  \bibinfo {pages} {193--196} (\bibinfo {year} {2009})},\ \Eprint
  {http://arxiv.org/abs/0902.4010} {arXiv:0902.4010 [hep-th]} \BibitemShut
  {NoStop}%
\bibitem [{\citenamefont {Skenderis}(2002)}]{Skenderis:2002wp}%
  \BibitemOpen
  \bibfield  {author} {\bibinfo {author} {\bibfnamefont {Kostas}\ \bibnamefont
  {Skenderis}},\ }\bibfield  {title} {\enquote {\bibinfo {title} {{Lecture
  notes on holographic renormalization}},}\ }\href {\doibase
  10.1088/0264-9381/19/22/306} {\bibfield  {journal} {\bibinfo  {journal}
  {Class. Quant. Grav.}\ }\textbf {\bibinfo {volume} {19}},\ \bibinfo {pages}
  {5849--5876} (\bibinfo {year} {2002})},\ \Eprint
  {http://arxiv.org/abs/hep-th/0209067} {arXiv:hep-th/0209067} \BibitemShut
  {NoStop}%
\bibitem [{\citenamefont {Son}\ and\ \citenamefont
  {Teaney}(2009)}]{Son:2009vu}%
  \BibitemOpen
  \bibfield  {author} {\bibinfo {author} {\bibfnamefont {Dam~T.}\ \bibnamefont
  {Son}}\ and\ \bibinfo {author} {\bibfnamefont {Derek}\ \bibnamefont
  {Teaney}},\ }\bibfield  {title} {\enquote {\bibinfo {title} {{Thermal Noise
  and Stochastic Strings in AdS/CFT}},}\ }\href {\doibase
  10.1088/1126-6708/2009/07/021} {\bibfield  {journal} {\bibinfo  {journal}
  {JHEP}\ }\textbf {\bibinfo {volume} {07}},\ \bibinfo {pages} {021} (\bibinfo
  {year} {2009})},\ \Eprint {http://arxiv.org/abs/0901.2338} {arXiv:0901.2338
  [hep-th]} \BibitemShut {NoStop}%
\bibitem [{\citenamefont {Bjorken}(1983)}]{Bjorken:1982qr}%
  \BibitemOpen
  \bibfield  {author} {\bibinfo {author} {\bibfnamefont {J.~D.}\ \bibnamefont
  {Bjorken}},\ }\bibfield  {title} {\enquote {\bibinfo {title} {{Highly
  Relativistic Nucleus-Nucleus Collisions: The Central Rapidity Region}},}\
  }\href {\doibase 10.1103/PhysRevD.27.140} {\bibfield  {journal} {\bibinfo
  {journal} {Phys. Rev. D}\ }\textbf {\bibinfo {volume} {27}},\ \bibinfo
  {pages} {140--151} (\bibinfo {year} {1983})}\BibitemShut {NoStop}%
\bibitem [{\citenamefont {Jeon}\ and\ \citenamefont
  {Heinz}(2015)}]{Jeon:2015dfa}%
  \BibitemOpen
  \bibfield  {author} {\bibinfo {author} {\bibfnamefont {Sangyong}\
  \bibnamefont {Jeon}}\ and\ \bibinfo {author} {\bibfnamefont {Ulrich}\
  \bibnamefont {Heinz}},\ }\bibfield  {title} {\enquote {\bibinfo {title}
  {{Introduction to Hydrodynamics}},}\ }\href {\doibase
  10.1142/S0218301315300106} {\bibfield  {journal} {\bibinfo  {journal} {Int.
  J. Mod. Phys. E}\ }\textbf {\bibinfo {volume} {24}},\ \bibinfo {pages}
  {1530010} (\bibinfo {year} {2015})},\ \Eprint
  {http://arxiv.org/abs/1503.03931} {arXiv:1503.03931 [hep-ph]} \BibitemShut
  {NoStop}%
\bibitem [{\citenamefont {Shuryak}\ \emph {et~al.}(2007)\citenamefont
  {Shuryak}, \citenamefont {Sin},\ and\ \citenamefont
  {Zahed}}]{Shuryak:2005ia}%
  \BibitemOpen
  \bibfield  {author} {\bibinfo {author} {\bibfnamefont {Edward}\ \bibnamefont
  {Shuryak}}, \bibinfo {author} {\bibfnamefont {Sang-Jin}\ \bibnamefont {Sin}},
  \ and\ \bibinfo {author} {\bibfnamefont {Ismail}\ \bibnamefont {Zahed}},\
  }\bibfield  {title} {\enquote {\bibinfo {title} {{A Gravity dual of RHIC
  collisions}},}\ }\href {\doibase 10.3938/jkps.50.384} {\bibfield  {journal}
  {\bibinfo  {journal} {J. Korean Phys. Soc.}\ }\textbf {\bibinfo {volume}
  {50}},\ \bibinfo {pages} {384--397} (\bibinfo {year} {2007})},\ \Eprint
  {http://arxiv.org/abs/hep-th/0511199} {arXiv:hep-th/0511199} \BibitemShut
  {NoStop}%
\bibitem [{\citenamefont {Janik}\ and\ \citenamefont
  {Peschanski}(2006{\natexlab{b}})}]{Janik:2005zt}%
  \BibitemOpen
  \bibfield  {author} {\bibinfo {author} {\bibfnamefont {Romuald~A.}\
  \bibnamefont {Janik}}\ and\ \bibinfo {author} {\bibfnamefont {Robert~B.}\
  \bibnamefont {Peschanski}},\ }\bibfield  {title} {\enquote {\bibinfo {title}
  {{Asymptotic perfect fluid dynamics as a consequence of Ads/CFT}},}\ }\href
  {\doibase 10.1103/PhysRevD.73.045013} {\bibfield  {journal} {\bibinfo
  {journal} {Phys. Rev. D}\ }\textbf {\bibinfo {volume} {73}},\ \bibinfo
  {pages} {045013} (\bibinfo {year} {2006}{\natexlab{b}})},\ \Eprint
  {http://arxiv.org/abs/hep-th/0512162} {arXiv:hep-th/0512162} \BibitemShut
  {NoStop}%
\bibitem [{\citenamefont {Janik}(2007)}]{Janik:2006ft}%
  \BibitemOpen
  \bibfield  {author} {\bibinfo {author} {\bibfnamefont {Romuald~A.}\
  \bibnamefont {Janik}},\ }\bibfield  {title} {\enquote {\bibinfo {title}
  {{Viscous plasma evolution from gravity using AdS/CFT}},}\ }\href {\doibase
  10.1103/PhysRevLett.98.022302} {\bibfield  {journal} {\bibinfo  {journal}
  {Phys. Rev. Lett.}\ }\textbf {\bibinfo {volume} {98}},\ \bibinfo {pages}
  {022302} (\bibinfo {year} {2007})},\ \Eprint
  {http://arxiv.org/abs/hep-th/0610144} {arXiv:hep-th/0610144} \BibitemShut
  {NoStop}%
\bibitem [{\citenamefont {Kinoshita}\ \emph {et~al.}(2009)\citenamefont
  {Kinoshita}, \citenamefont {Mukohyama}, \citenamefont {Nakamura},\ and\
  \citenamefont {Oda}}]{Kinoshita:2008dq}%
  \BibitemOpen
  \bibfield  {author} {\bibinfo {author} {\bibfnamefont {Shunichiro}\
  \bibnamefont {Kinoshita}}, \bibinfo {author} {\bibfnamefont {Shinji}\
  \bibnamefont {Mukohyama}}, \bibinfo {author} {\bibfnamefont {Shin}\
  \bibnamefont {Nakamura}}, \ and\ \bibinfo {author} {\bibfnamefont {Kin-ya}\
  \bibnamefont {Oda}},\ }\bibfield  {title} {\enquote {\bibinfo {title} {{A
  Holographic Dual of Bjorken Flow}},}\ }\href {\doibase 10.1143/PTP.121.121}
  {\bibfield  {journal} {\bibinfo  {journal} {Prog. Theor. Phys.}\ }\textbf
  {\bibinfo {volume} {121}},\ \bibinfo {pages} {121--164} (\bibinfo {year}
  {2009})},\ \Eprint {http://arxiv.org/abs/0807.3797} {arXiv:0807.3797
  [hep-th]} \BibitemShut {NoStop}%
\bibitem [{\citenamefont {Beuf}\ \emph {et~al.}(2009)\citenamefont {Beuf},
  \citenamefont {Heller}, \citenamefont {Janik},\ and\ \citenamefont
  {Peschanski}}]{Beuf:2009cx}%
  \BibitemOpen
  \bibfield  {author} {\bibinfo {author} {\bibfnamefont {Guillaume}\
  \bibnamefont {Beuf}}, \bibinfo {author} {\bibfnamefont {Michal~P.}\
  \bibnamefont {Heller}}, \bibinfo {author} {\bibfnamefont {Romuald~A.}\
  \bibnamefont {Janik}}, \ and\ \bibinfo {author} {\bibfnamefont {Robi}\
  \bibnamefont {Peschanski}},\ }\bibfield  {title} {\enquote {\bibinfo {title}
  {{Boost-invariant early time dynamics from AdS/CFT}},}\ }\href {\doibase
  10.1088/1126-6708/2009/10/043} {\bibfield  {journal} {\bibinfo  {journal}
  {JHEP}\ }\textbf {\bibinfo {volume} {10}},\ \bibinfo {pages} {043} (\bibinfo
  {year} {2009})},\ \Eprint {http://arxiv.org/abs/0906.4423} {arXiv:0906.4423
  [hep-th]} \BibitemShut {NoStop}%
\bibitem [{\citenamefont {Heller}\ and\ \citenamefont
  {Spalinski}(2015)}]{Heller:2015dha}%
  \BibitemOpen
  \bibfield  {author} {\bibinfo {author} {\bibfnamefont {Michal~P.}\
  \bibnamefont {Heller}}\ and\ \bibinfo {author} {\bibfnamefont {Michal}\
  \bibnamefont {Spalinski}},\ }\bibfield  {title} {\enquote {\bibinfo {title}
  {{Hydrodynamics Beyond the Gradient Expansion: Resurgence and
  Resummation}},}\ }\href {\doibase 10.1103/PhysRevLett.115.072501} {\bibfield
  {journal} {\bibinfo  {journal} {Phys. Rev. Lett.}\ }\textbf {\bibinfo
  {volume} {115}},\ \bibinfo {pages} {072501} (\bibinfo {year} {2015})},\
  \Eprint {http://arxiv.org/abs/1503.07514} {arXiv:1503.07514 [hep-th]}
  \BibitemShut {NoStop}%
\bibitem [{\citenamefont {Soloviev}(2022)}]{Soloviev:2021lhs}%
  \BibitemOpen
  \bibfield  {author} {\bibinfo {author} {\bibfnamefont {Alexander}\
  \bibnamefont {Soloviev}},\ }\bibfield  {title} {\enquote {\bibinfo {title}
  {{Hydrodynamic attractors in heavy ion collisions: a review}},}\ }\href
  {\doibase 10.1140/epjc/s10052-022-10282-4} {\bibfield  {journal} {\bibinfo
  {journal} {Eur. Phys. J. C}\ }\textbf {\bibinfo {volume} {82}},\ \bibinfo
  {pages} {319} (\bibinfo {year} {2022})},\ \Eprint
  {http://arxiv.org/abs/2109.15081} {arXiv:2109.15081 [hep-th]} \BibitemShut
  {NoStop}%
\bibitem [{\citenamefont {Balasubramanian}\ and\ \citenamefont
  {Kraus}(1999)}]{Balasubramanian:1999re}%
  \BibitemOpen
  \bibfield  {author} {\bibinfo {author} {\bibfnamefont {Vijay}\ \bibnamefont
  {Balasubramanian}}\ and\ \bibinfo {author} {\bibfnamefont {Per}\ \bibnamefont
  {Kraus}},\ }\bibfield  {title} {\enquote {\bibinfo {title} {{A Stress tensor
  for Anti-de Sitter gravity}},}\ }\href {\doibase 10.1007/s002200050764}
  {\bibfield  {journal} {\bibinfo  {journal} {Commun. Math. Phys.}\ }\textbf
  {\bibinfo {volume} {208}},\ \bibinfo {pages} {413--428} (\bibinfo {year}
  {1999})},\ \Eprint {http://arxiv.org/abs/hep-th/9902121}
  {arXiv:hep-th/9902121} \BibitemShut {NoStop}%
\bibitem [{\citenamefont {Mukhopadhyay}(2013)}]{Mukhopadhyay:2012hv}%
  \BibitemOpen
  \bibfield  {author} {\bibinfo {author} {\bibfnamefont {Ayan}\ \bibnamefont
  {Mukhopadhyay}},\ }\bibfield  {title} {\enquote {\bibinfo {title}
  {{Nonequilibrium fluctuation-dissipation relation from holography}},}\ }\href
  {\doibase 10.1103/PhysRevD.87.066004} {\bibfield  {journal} {\bibinfo
  {journal} {Phys. Rev. D}\ }\textbf {\bibinfo {volume} {87}},\ \bibinfo
  {pages} {066004} (\bibinfo {year} {2013})},\ \Eprint
  {http://arxiv.org/abs/1206.3311} {arXiv:1206.3311 [hep-th]} \BibitemShut
  {NoStop}%
\bibitem [{\citenamefont {Dodelson}\ \emph {et~al.}(2022)\citenamefont
  {Dodelson}, \citenamefont {Grassi}, \citenamefont {Iossa}, \citenamefont
  {Panea~Lichtig},\ and\ \citenamefont {Zhiboedov}}]{Dodelson:2022yvn}%
  \BibitemOpen
  \bibfield  {author} {\bibinfo {author} {\bibfnamefont {Matthew}\ \bibnamefont
  {Dodelson}}, \bibinfo {author} {\bibfnamefont {Alba}\ \bibnamefont {Grassi}},
  \bibinfo {author} {\bibfnamefont {Cristoforo}\ \bibnamefont {Iossa}},
  \bibinfo {author} {\bibfnamefont {Daniel}\ \bibnamefont {Panea~Lichtig}}, \
  and\ \bibinfo {author} {\bibfnamefont {Alexander}\ \bibnamefont
  {Zhiboedov}},\ }\bibfield  {title} {\enquote {\bibinfo {title} {{Holographic
  thermal correlators from supersymmetric instantons}},}\ }\href@noop {} {\
  (\bibinfo {year} {2022})},\ \Eprint {http://arxiv.org/abs/2206.07720}
  {arXiv:2206.07720 [hep-th]} \BibitemShut {NoStop}%
\bibitem [{\citenamefont {Banerjee}\ \emph {et~al.}(2012)\citenamefont
  {Banerjee}, \citenamefont {Iyer},\ and\ \citenamefont
  {Mukhopadhyay}}]{Banerjee:2012uq}%
  \BibitemOpen
  \bibfield  {author} {\bibinfo {author} {\bibfnamefont {Souvik}\ \bibnamefont
  {Banerjee}}, \bibinfo {author} {\bibfnamefont {Ramakrishnan}\ \bibnamefont
  {Iyer}}, \ and\ \bibinfo {author} {\bibfnamefont {Ayan}\ \bibnamefont
  {Mukhopadhyay}},\ }\bibfield  {title} {\enquote {\bibinfo {title} {{The
  holographic spectral function in non-equilibrium states}},}\ }\href {\doibase
  10.1103/PhysRevD.85.106009} {\bibfield  {journal} {\bibinfo  {journal} {Phys.
  Rev. D}\ }\textbf {\bibinfo {volume} {85}},\ \bibinfo {pages} {106009}
  (\bibinfo {year} {2012})},\ \Eprint {http://arxiv.org/abs/1202.1521}
  {arXiv:1202.1521 [hep-th]} \BibitemShut {NoStop}%
\bibitem [{\citenamefont {Joshi}\ \emph {et~al.}(2017)\citenamefont {Joshi},
  \citenamefont {Mukhopadhyay}, \citenamefont {Preis},\ and\ \citenamefont
  {Ramadevi}}]{Joshi:2017ump}%
  \BibitemOpen
  \bibfield  {author} {\bibinfo {author} {\bibfnamefont {Lata~Kh}\ \bibnamefont
  {Joshi}}, \bibinfo {author} {\bibfnamefont {Ayan}\ \bibnamefont
  {Mukhopadhyay}}, \bibinfo {author} {\bibfnamefont {Florian}\ \bibnamefont
  {Preis}}, \ and\ \bibinfo {author} {\bibfnamefont {Pichai}\ \bibnamefont
  {Ramadevi}},\ }\bibfield  {title} {\enquote {\bibinfo {title} {{Exact time
  dependence of causal correlations and nonequilibrium density matrices in
  holographic systems}},}\ }\href {\doibase 10.1103/PhysRevD.96.106006}
  {\bibfield  {journal} {\bibinfo  {journal} {Phys. Rev. D}\ }\textbf {\bibinfo
  {volume} {96}},\ \bibinfo {pages} {106006} (\bibinfo {year} {2017})},\
  \Eprint {http://arxiv.org/abs/1704.02936} {arXiv:1704.02936 [hep-th]}
  \BibitemShut {NoStop}%
\bibitem [{\citenamefont {Cartwright}\ and\ \citenamefont
  {Kaminski}(2019)}]{Cartwright:2019opv}%
  \BibitemOpen
  \bibfield  {author} {\bibinfo {author} {\bibfnamefont {Casey}\ \bibnamefont
  {Cartwright}}\ and\ \bibinfo {author} {\bibfnamefont {Matthias}\ \bibnamefont
  {Kaminski}},\ }\bibfield  {title} {\enquote {\bibinfo {title} {{Correlations
  far from equilibrium in charged strongly coupled fluids subjected to a strong
  magnetic field}},}\ }\href {\doibase 10.1007/JHEP09(2019)072} {\bibfield
  {journal} {\bibinfo  {journal} {JHEP}\ }\textbf {\bibinfo {volume} {09}},\
  \bibinfo {pages} {072} (\bibinfo {year} {2019})},\ \Eprint
  {http://arxiv.org/abs/1904.11507} {arXiv:1904.11507 [hep-th]} \BibitemShut
  {NoStop}%
\bibitem [{\citenamefont {Pedraza}\ \emph {et~al.}(2022)\citenamefont
  {Pedraza}, \citenamefont {Russo}, \citenamefont {Svesko},\ and\ \citenamefont
  {Weller-Davies}}]{Pedraza:2021fgp}%
  \BibitemOpen
  \bibfield  {author} {\bibinfo {author} {\bibfnamefont {Juan~F.}\ \bibnamefont
  {Pedraza}}, \bibinfo {author} {\bibfnamefont {Andrea}\ \bibnamefont {Russo}},
  \bibinfo {author} {\bibfnamefont {Andrew}\ \bibnamefont {Svesko}}, \ and\
  \bibinfo {author} {\bibfnamefont {Zachary}\ \bibnamefont {Weller-Davies}},\
  }\bibfield  {title} {\enquote {\bibinfo {title} {{Sewing spacetime with
  Lorentzian threads: complexity and the emergence of time in quantum
  gravity}},}\ }\href {\doibase 10.1007/JHEP02(2022)093} {\bibfield  {journal}
  {\bibinfo  {journal} {JHEP}\ }\textbf {\bibinfo {volume} {02}},\ \bibinfo
  {pages} {093} (\bibinfo {year} {2022})},\ \Eprint
  {http://arxiv.org/abs/2106.12585} {arXiv:2106.12585 [hep-th]} \BibitemShut
  {NoStop}%
\bibitem [{\citenamefont {Mart\'\i{}nez}\ and\ \citenamefont
  {Silva}(2022)}]{Martinez:2021uqo}%
  \BibitemOpen
  \bibfield  {author} {\bibinfo {author} {\bibfnamefont {Pedro~Jorge}\
  \bibnamefont {Mart\'\i{}nez}}\ and\ \bibinfo {author} {\bibfnamefont
  {Guillermo~A.}\ \bibnamefont {Silva}},\ }\bibfield  {title} {\enquote
  {\bibinfo {title} {{Thermalization of holographic excited states}},}\ }\href
  {\doibase 10.1007/JHEP03(2022)003} {\bibfield  {journal} {\bibinfo  {journal}
  {JHEP}\ }\textbf {\bibinfo {volume} {03}},\ \bibinfo {pages} {003} (\bibinfo
  {year} {2022})},\ \Eprint {http://arxiv.org/abs/2110.07555} {arXiv:2110.07555
  [hep-th]} \BibitemShut {NoStop}%
\bibitem [{\citenamefont {Basar}\ and\ \citenamefont
  {Dunne}(2015)}]{Basar:2015ava}%
  \BibitemOpen
  \bibfield  {author} {\bibinfo {author} {\bibfnamefont {Gokce}\ \bibnamefont
  {Basar}}\ and\ \bibinfo {author} {\bibfnamefont {Gerald~V.}\ \bibnamefont
  {Dunne}},\ }\bibfield  {title} {\enquote {\bibinfo {title} {{Hydrodynamics,
  resurgence, and transasymptotics}},}\ }\href {\doibase
  10.1103/PhysRevD.92.125011} {\bibfield  {journal} {\bibinfo  {journal} {Phys.
  Rev. D}\ }\textbf {\bibinfo {volume} {92}},\ \bibinfo {pages} {125011}
  (\bibinfo {year} {2015})},\ \Eprint {http://arxiv.org/abs/1509.05046}
  {arXiv:1509.05046 [hep-th]} \BibitemShut {NoStop}%
\bibitem [{\citenamefont {Raju}(2022)}]{Raju:2020smc}%
  \BibitemOpen
  \bibfield  {author} {\bibinfo {author} {\bibfnamefont {Suvrat}\ \bibnamefont
  {Raju}},\ }\bibfield  {title} {\enquote {\bibinfo {title} {{Lessons from the
  information paradox}},}\ }\href {\doibase 10.1016/j.physrep.2021.10.001}
  {\bibfield  {journal} {\bibinfo  {journal} {Phys. Rept.}\ }\textbf {\bibinfo
  {volume} {943}},\ \bibinfo {pages} {1--80} (\bibinfo {year} {2022})},\
  \Eprint {http://arxiv.org/abs/2012.05770} {arXiv:2012.05770 [hep-th]}
  \BibitemShut {NoStop}%
\bibitem [{\citenamefont {Kibe}\ \emph {et~al.}(2022)\citenamefont {Kibe},
  \citenamefont {Mandayam},\ and\ \citenamefont {Mukhopadhyay}}]{Kibe:2021gtw}%
  \BibitemOpen
  \bibfield  {author} {\bibinfo {author} {\bibfnamefont {Tanay}\ \bibnamefont
  {Kibe}}, \bibinfo {author} {\bibfnamefont {Prabha}\ \bibnamefont {Mandayam}},
  \ and\ \bibinfo {author} {\bibfnamefont {Ayan}\ \bibnamefont
  {Mukhopadhyay}},\ }\bibfield  {title} {\enquote {\bibinfo {title}
  {{Holographic spacetime, black holes and quantum error correcting codes: A
  review}},}\ }\href {\doibase 10.1140/epjc/s10052-022-10382-1} {\bibfield
  {journal} {\bibinfo  {journal} {Eur. Phys. J. C}\ }\textbf {\bibinfo {volume}
  {82}},\ \bibinfo {pages} {463} (\bibinfo {year} {2022})},\ \Eprint
  {http://arxiv.org/abs/2110.14669} {arXiv:2110.14669 [hep-th]} \BibitemShut
  {NoStop}%
\bibitem [{\citenamefont {Gubser}(2010)}]{Gubser:2010ze}%
  \BibitemOpen
  \bibfield  {author} {\bibinfo {author} {\bibfnamefont {Steven~S.}\
  \bibnamefont {Gubser}},\ }\bibfield  {title} {\enquote {\bibinfo {title}
  {{Symmetry constraints on generalizations of Bjorken flow}},}\ }\href
  {\doibase 10.1103/PhysRevD.82.085027} {\bibfield  {journal} {\bibinfo
  {journal} {Phys. Rev. D}\ }\textbf {\bibinfo {volume} {82}},\ \bibinfo
  {pages} {085027} (\bibinfo {year} {2010})},\ \Eprint
  {http://arxiv.org/abs/1006.0006} {arXiv:1006.0006 [hep-th]} \BibitemShut
  {NoStop}%
\bibitem [{\citenamefont {Busza}\ \emph {et~al.}(2018)\citenamefont {Busza},
  \citenamefont {Rajagopal},\ and\ \citenamefont {van~der
  Schee}}]{Busza:2018rrf}%
  \BibitemOpen
  \bibfield  {author} {\bibinfo {author} {\bibfnamefont {Wit}\ \bibnamefont
  {Busza}}, \bibinfo {author} {\bibfnamefont {Krishna}\ \bibnamefont
  {Rajagopal}}, \ and\ \bibinfo {author} {\bibfnamefont {Wilke}\ \bibnamefont
  {van~der Schee}},\ }\bibfield  {title} {\enquote {\bibinfo {title} {{Heavy
  Ion Collisions: The Big Picture, and the Big Questions}},}\ }\href {\doibase
  10.1146/annurev-nucl-101917-020852} {\bibfield  {journal} {\bibinfo
  {journal} {Ann. Rev. Nucl. Part. Sci.}\ }\textbf {\bibinfo {volume} {68}},\
  \bibinfo {pages} {339--376} (\bibinfo {year} {2018})},\ \Eprint
  {http://arxiv.org/abs/1802.04801} {arXiv:1802.04801 [hep-ph]} \BibitemShut
  {NoStop}%
\bibitem [{\citenamefont {Kovtun}(2012)}]{Kovtun_2012}%
  \BibitemOpen
  \bibfield  {author} {\bibinfo {author} {\bibfnamefont {Pavel}\ \bibnamefont
  {Kovtun}},\ }\bibfield  {title} {\enquote {\bibinfo {title} {Lectures on
  hydrodynamic fluctuations in relativistic theories},}\ }\href {\doibase
  10.1088/1751-8113/45/47/473001} {\bibfield  {journal} {\bibinfo  {journal}
  {Journal of Physics A: Mathematical and Theoretical}\ }\textbf {\bibinfo
  {volume} {45}},\ \bibinfo {pages} {473001} (\bibinfo {year}
  {2012})}\BibitemShut {NoStop}%
\bibitem [{\citenamefont {Liu}\ and\ \citenamefont
  {Sonner}(2018)}]{Liu:2018crr}%
  \BibitemOpen
  \bibfield  {author} {\bibinfo {author} {\bibfnamefont {Hong}\ \bibnamefont
  {Liu}}\ and\ \bibinfo {author} {\bibfnamefont {Julian}\ \bibnamefont
  {Sonner}},\ }\bibfield  {title} {\enquote {\bibinfo {title} {{Holographic
  systems far from equilibrium: a review}},}\ }\href@noop {} {\  (\bibinfo
  {year} {2018})},\ \Eprint {http://arxiv.org/abs/1810.02367} {arXiv:1810.02367
  [hep-th]} \BibitemShut {NoStop}%
\bibitem [{\citenamefont {Grossi}\ \emph {et~al.}(2021)\citenamefont {Grossi},
  \citenamefont {Soloviev}, \citenamefont {Teaney},\ and\ \citenamefont
  {Yan}}]{Grossi:2021gqi}%
  \BibitemOpen
  \bibfield  {author} {\bibinfo {author} {\bibfnamefont {Eduardo}\ \bibnamefont
  {Grossi}}, \bibinfo {author} {\bibfnamefont {Alexander}\ \bibnamefont
  {Soloviev}}, \bibinfo {author} {\bibfnamefont {Derek}\ \bibnamefont
  {Teaney}}, \ and\ \bibinfo {author} {\bibfnamefont {Fanglida}\ \bibnamefont
  {Yan}},\ }\bibfield  {title} {\enquote {\bibinfo {title} {{Soft pions and
  transport near the chiral critical point}},}\ }\href {\doibase
  10.1103/PhysRevD.104.034025} {\bibfield  {journal} {\bibinfo  {journal}
  {Phys. Rev. D}\ }\textbf {\bibinfo {volume} {104}},\ \bibinfo {pages}
  {034025} (\bibinfo {year} {2021})},\ \Eprint
  {http://arxiv.org/abs/2101.10847} {arXiv:2101.10847 [nucl-th]} \BibitemShut
  {NoStop}%
\bibitem [{\citenamefont {Almheiri}\ \emph {et~al.}(2020)\citenamefont
  {Almheiri}, \citenamefont {Mahajan}, \citenamefont {Maldacena},\ and\
  \citenamefont {Zhao}}]{Almheiri:2019hni}%
  \BibitemOpen
  \bibfield  {author} {\bibinfo {author} {\bibfnamefont {Ahmed}\ \bibnamefont
  {Almheiri}}, \bibinfo {author} {\bibfnamefont {Raghu}\ \bibnamefont
  {Mahajan}}, \bibinfo {author} {\bibfnamefont {Juan}\ \bibnamefont
  {Maldacena}}, \ and\ \bibinfo {author} {\bibfnamefont {Ying}\ \bibnamefont
  {Zhao}},\ }\bibfield  {title} {\enquote {\bibinfo {title} {{The Page curve of
  Hawking radiation from semiclassical geometry}},}\ }\href {\doibase
  10.1007/JHEP03(2020)149} {\bibfield  {journal} {\bibinfo  {journal} {JHEP}\
  }\textbf {\bibinfo {volume} {03}},\ \bibinfo {pages} {149} (\bibinfo {year}
  {2020})},\ \Eprint {http://arxiv.org/abs/1908.10996} {arXiv:1908.10996
  [hep-th]} \BibitemShut {NoStop}%
\bibitem [{\citenamefont {Joshi}\ \emph {et~al.}(2020)\citenamefont {Joshi},
  \citenamefont {Mukhopadhyay},\ and\ \citenamefont
  {Soloviev}}]{Joshi:2019wgi}%
  \BibitemOpen
  \bibfield  {author} {\bibinfo {author} {\bibfnamefont {Lata~Kh}\ \bibnamefont
  {Joshi}}, \bibinfo {author} {\bibfnamefont {Ayan}\ \bibnamefont
  {Mukhopadhyay}}, \ and\ \bibinfo {author} {\bibfnamefont {Alexander}\
  \bibnamefont {Soloviev}},\ }\bibfield  {title} {\enquote {\bibinfo {title}
  {{Time-dependent $NAdS_2$ holography with applications}},}\ }\href {\doibase
  10.1103/PhysRevD.101.066001} {\bibfield  {journal} {\bibinfo  {journal}
  {Phys. Rev. D}\ }\textbf {\bibinfo {volume} {101}},\ \bibinfo {pages}
  {066001} (\bibinfo {year} {2020})},\ \Eprint
  {http://arxiv.org/abs/1901.08877} {arXiv:1901.08877 [hep-th]} \BibitemShut
  {NoStop}%
\bibitem [{\citenamefont {Mondkar}\ \emph {et~al.}(2021)\citenamefont
  {Mondkar}, \citenamefont {Mukhopadhyay}, \citenamefont {Rebhan},\ and\
  \citenamefont {Soloviev}}]{Mondkar:2021qsf}%
  \BibitemOpen
  \bibfield  {author} {\bibinfo {author} {\bibfnamefont {Sukrut}\ \bibnamefont
  {Mondkar}}, \bibinfo {author} {\bibfnamefont {Ayan}\ \bibnamefont
  {Mukhopadhyay}}, \bibinfo {author} {\bibfnamefont {Anton}\ \bibnamefont
  {Rebhan}}, \ and\ \bibinfo {author} {\bibfnamefont {Alexander}\ \bibnamefont
  {Soloviev}},\ }\bibfield  {title} {\enquote {\bibinfo {title} {{Quasinormal
  modes of a semi-holographic black brane and thermalization}},}\ }\href
  {\doibase 10.1007/JHEP11(2021)080} {\bibfield  {journal} {\bibinfo  {journal}
  {JHEP}\ }\textbf {\bibinfo {volume} {11}},\ \bibinfo {pages} {080} (\bibinfo
  {year} {2021})},\ \Eprint {http://arxiv.org/abs/2108.02788} {arXiv:2108.02788
  [hep-th]} \BibitemShut {NoStop}%
\bibitem [{\citenamefont {Hayward}(1994)}]{Hayward:1993wb}%
  \BibitemOpen
  \bibfield  {author} {\bibinfo {author} {\bibfnamefont {S.~A.}\ \bibnamefont
  {Hayward}},\ }\bibfield  {title} {\enquote {\bibinfo {title} {{General laws
  of black hole dynamics}},}\ }\href {\doibase 10.1103/PhysRevD.49.6467}
  {\bibfield  {journal} {\bibinfo  {journal} {Phys. Rev. D}\ }\textbf {\bibinfo
  {volume} {49}},\ \bibinfo {pages} {6467--6474} (\bibinfo {year}
  {1994})}\BibitemShut {NoStop}%
\bibitem [{\citenamefont {Hayward}(2004)}]{Hayward:2004fz}%
  \BibitemOpen
  \bibfield  {author} {\bibinfo {author} {\bibfnamefont {Sean~A.}\ \bibnamefont
  {Hayward}},\ }\bibfield  {title} {\enquote {\bibinfo {title} {{Energy and
  entropy conservation for dynamical black holes}},}\ }\href {\doibase
  10.1103/PhysRevD.70.104027} {\bibfield  {journal} {\bibinfo  {journal} {Phys.
  Rev. D}\ }\textbf {\bibinfo {volume} {70}},\ \bibinfo {pages} {104027}
  (\bibinfo {year} {2004})},\ \Eprint {http://arxiv.org/abs/gr-qc/0408008}
  {arXiv:gr-qc/0408008} \BibitemShut {NoStop}%
\bibitem [{\citenamefont {Figueras}\ \emph {et~al.}(2009)\citenamefont
  {Figueras}, \citenamefont {Hubeny}, \citenamefont {Rangamani},\ and\
  \citenamefont {Ross}}]{Figueras:2009iu}%
  \BibitemOpen
  \bibfield  {author} {\bibinfo {author} {\bibfnamefont {Pau}\ \bibnamefont
  {Figueras}}, \bibinfo {author} {\bibfnamefont {Veronika~E.}\ \bibnamefont
  {Hubeny}}, \bibinfo {author} {\bibfnamefont {Mukund}\ \bibnamefont
  {Rangamani}}, \ and\ \bibinfo {author} {\bibfnamefont {Simon~F.}\
  \bibnamefont {Ross}},\ }\bibfield  {title} {\enquote {\bibinfo {title}
  {{Dynamical black holes and expanding plasmas}},}\ }\href {\doibase
  10.1088/1126-6708/2009/04/137} {\bibfield  {journal} {\bibinfo  {journal}
  {JHEP}\ }\textbf {\bibinfo {volume} {04}},\ \bibinfo {pages} {137} (\bibinfo
  {year} {2009})},\ \Eprint {http://arxiv.org/abs/0902.4696} {arXiv:0902.4696
  [hep-th]} \BibitemShut {NoStop}%
\bibitem [{\citenamefont {Casalderrey-Solana}\ \emph
  {et~al.}(2019)\citenamefont {Casalderrey-Solana}, \citenamefont {Herzog},\
  and\ \citenamefont {Meiring}}]{Casalderrey-Solana:2018uag}%
  \BibitemOpen
  \bibfield  {author} {\bibinfo {author} {\bibfnamefont {Jorge}\ \bibnamefont
  {Casalderrey-Solana}}, \bibinfo {author} {\bibfnamefont {Christopher~P.}\
  \bibnamefont {Herzog}}, \ and\ \bibinfo {author} {\bibfnamefont {Ben}\
  \bibnamefont {Meiring}},\ }\bibfield  {title} {\enquote {\bibinfo {title}
  {{Holographic Bjorken Flow at Large-$D$}},}\ }\href {\doibase
  10.1007/JHEP01(2019)181} {\bibfield  {journal} {\bibinfo  {journal} {JHEP}\
  }\textbf {\bibinfo {volume} {01}},\ \bibinfo {pages} {181} (\bibinfo {year}
  {2019})},\ \Eprint {http://arxiv.org/abs/1810.02314} {arXiv:1810.02314
  [hep-th]} \BibitemShut {NoStop}%
\end{thebibliography}%

\end{document}